\documentclass[useAMS,usenatbib]{mn2e}
\usepackage{graphicx}
\usepackage[section]{placeins}

\usepackage{times} 
\usepackage{amsmath}
\usepackage{amssymb}

\protected\def\hiddenmath{$e\cos^2(I/2)$}

\begin{document}

\title[Resonance capture at arbitrary inclination]{Resonance capture at arbitrary inclination}

\author[F. Namouni and M. H. M. Morais]{F. Namouni$^{1}$\thanks{E-mail:
namouni@obs-nice.fr (FN) ; helena.morais@rc.unesp.br (MHMM)} and  M. H. M. Morais
$^{2}$\footnotemark[1]\\
$^{1}$Universit\'e de Nice, CNRS, Observatoire de la C\^ote d'Azur, CS 34229, 06304 Nice, France\\
$^{2}$Instituto de Geoci\^encias e Ci\^encias Exatas, Universidade Estadual Paulista (UNESP), Av. 24-A, 1515 13506-900 Rio Claro, SP, Brazil}

\date{Accepted 2014 October 19. Received 2014 October 17; in original form 2014 August 22}

\maketitle

\begin{abstract}
Resonance capture is studied numerically in the three-body problem for arbitrary inclinations.    Massless particles are set to drift from outside the 1:5 resonance with a Jupiter-mass planet thereby encountering the  web of the planet's diverse mean motion resonances. Randomly constructed samples explore parameter space for inclinations from 0 to 180$^\circ$ with 5$^\circ$ increments totalling  nearly $ 6\times 10^5$ numerical simulations. Thirty resonances internal and external to the planet's location are monitored. We find that retrograde resonances are unexpectedly more efficient at capture than prograde resonances and that resonance order is not necessarily a good indicator of capture efficiency at arbitrary inclination. Capture probability drops significantly at moderate sample eccentricity for initial inclinations in the range [$10^\circ$,$110^\circ$]. Orbit inversion is possible for initially circular orbits with inclinations in the range [$60^\circ$,$130^\circ$].  Capture in the 1:1 coorbital resonance occurs with great likelihood at large retrograde inclinations.  The planet's orbital eccentricity, if larger than 0.1, reduces  the capture probabilities through the action of the eccentric Kozai-Lidov mechanism. A capture asymmetry appears between inner and outer resonances as prograde orbits are preferentially trapped in inner resonances. The relative capture efficiency of retrograde resonance suggests that the dynamical lifetimes of  Damocloids and Centaurs on retrograde orbits must be significantly larger than those on prograde orbits implying that the recently identified asteroids in retrograde resonance, 2006 BZ8, 2008 SO218, 2009 QY6 and 1999 LE31(Morais and Namouni, 2013, Mon. Not. R. Astron. Soc. 436, L30)  may be among the oldest  small bodies that wander between the outer giant planets.
\end{abstract}


\section{Introduction}
Capture in mean motion resonance is a ubiquitous process in the evolution of planetary systems. It is evidenced widely by the  Solar System's minor bodies and planetary satellites as well as by the sizable fraction of exoplanet multiple systems in orbital resonance  ($20\%$). A new research avenue was opened recently with the discovery of Centaurs and Damocloids (2006 BZ8, 2008 SO218, 2009 QY6 and 1999 LE31)  in retrograde resonance  with Jupiter and Saturn raising the question of how resonant capture operates at arbitrary inclination in the solar system \citep{MoraisNamouni13b}.
Analytical estimates of mean motion resonance capture probability rely on Henrard's original derivation using Poincar\'e's Hamiltonian model of resonance with small perturbations \citep{Poincare02,Henrard1,Lemaitre84,BorderiesGoldreich84,Henrard2,Henrard3,Henrard4}. Such analytical models tend to assume single isolated resonances. This limitation is not worrisome whenever the orbital eccentricity and inclination are small and no coupling occurs among  different mean motion resonances and additionally with the Kozai-Lidov secular resonance \citep{Kozai62,Lidov62,GronchiMilani99}. For instance, linear capture probability theory has been applied successfully to the formation of planetary satellite resonances under the influence of tidal forces \citep{ssdbook}. Whenever the growth of eccentricity and inclination is significant, resonance capture in the three-body problem becomes complex.  The activation of multiple high order resonances near the nominal resonant location may or may not result is resonance overlap but nonetheless  influences the capture of the drifting third body by, for instance, temporarily trapping it in higher order resonances before its arrival to the main low order resonance (see the example in section 2). The secular evolution of eccentricity and inclination through the Kozai-Lidov  resonance makes capture more involved through the libration of the argument of pericenter as well as the chaotic inclination excursions when the planet's orbit is eccentric \citep{Katzetal11,LithwickNaoz11}.

To address the problem of mean motion resonant capture at arbitrary inclination, we seek the simplest dynamical model that allows us to access the fundamentals of resonant excitation and capture. This is the three-body problem of a solar-mass star,  a Jupiter-mass planet and a massless particle located outside the planet's orbit and forced to drift inwards thereby encountering the web of the  planet's mean motion resonances. Numerical simulations of the equations of motion of such a system with a large set of initial conditions are necessary to characterize the capture mechanism at arbitrary inclination. Although the Damocloids and Centaurs mentioned earlier are in retrograde resonance, we chose initial conditions that cover a wide rage of inclinations (from 0 to $180^\circ$ with $5^\circ$ increments) so as to ascertain the similarities and differences of prograde, retrograde and polar resonance capture. We remark however that our study does not aim to reflect the actual evolution  of Centaurs and Damocloids as these objects tend to hop between resonances owing to their close encounters with the outer solar system's planets that are responsible for their semi-major axis drift   \citep{MasakiKinoshita03,BaileyMalhotra09,VolkMalhotra13}. The present work must be understood as a first exploration of the fundamentals of capture in resonance at arbitrary inclination.
 
 In section 2, we briefly examine resonant capture for the special configurations of prograde and retrograde planar circular orbits and how their evolution may be modified by the addition of a small inclination. In section 3, we discuss the initial conditions and results for the statistical simulations of capture at arbitrary inclination for a planet with a circular orbit. Section 4 is devoted to the effect of planetary eccentricity on resonance capture. Our conclusions are given in section 5.

 \section{How inclination modifies planar resonance capture} 
 The common approach to studying resonance capture, especially analytically, is to consider a three-body system near a specific resonance and estimate under what conditions capture may occur. As our original motivation is resonance capture of drifting Damocloids and Centaurs possibly on retrograde orbits, we may not consider a specific resonance; instead we shall set up the minor body far away from the main strong planetary resonances and estimate capture likelihood. Resonance capture probability theory \citep[see for instance][]{ssdbook} is not of great help in this case for two primary reasons. First, the Poincar\'e model of resonance upon which standard capture theory is based is valid  for small eccentricity and inclination orbits and only for single isolated resonances. Moreover, the model includes only the smallest linear contributions from the secular potential and misses out on the Kozai-Lidov dynamics. Second, resonance capture of a drifting minor body is conditioned by its resonance crossing history that alters the particle's state before encountering the final resonance that traps the body permanently.  In order to understand how the presence of a large mutual inclination may affect capture we first compare the simplest configurations of quasi-planar prograde and retrograde orbits.
   
Consider a solar-mass star, a Jupiter-mass planet on a circular orbit with unit semi-major axis and a massless particle on a nearly coplanar   circular orbit (initial inclination $I_0\sim$ a few degrees) with semi-major axis $3$ outside the 1:5 resonance with the planet ($2.924 $).\footnote{{In the restricted three-body problem, dimensions are unnecessary as long as one knows the planet-star mass ratio and measures distances in units of the star-planet distance and time in units of planetary orbital periods.}} An imposed slow inward semi-major axis drift will make the particle's orbit sweep all  major resonances. Drift is modeled through a velocity-dependent drag force of the form $-k {\bf v}$ {per unit mass} leading to a semi-major axis function $a(t)=\exp(-2k t)$ with a characteristic drift time $(2k)^{-1}=10^{5}$ planetary orbital periods so that resonance crossing is adiabatic. {This specific drag force does not alter the eccentricity and inclination of the massless particle so that any variations in these elements are solely due to the gravitational interactions.}

As the  particle's prograde orbit converges towards the planet's (Figure 1, left column), capture occurs with unit probability for coplanar orbits at the first encountered first order resonance which is the 1:2 eccentricity resonance of resonant argument $\phi_1=2\lambda-\lambda_{\rm J}-\varpi$. Higher order resonances do not capture the particle because they all possess a stable libration point at zero eccentricity far outside  nominal resonance (for portraits of the resonances' Hamiltonians see \citet[][Chapter 8]{ssdbook}. For instance, the stable equilibrium at zero eccentricity of the 1:3 (second order) resonance bifurcates to an unstable equilibrium as the particle's semi-major axis drifts creating two stable libration equilibria inside the particle's initial Hamiltonian curve. Circulation therefore continues but the eccentricity is increased  at resonance passage. First order resonances do possess stable equilibria far outside nominal resonance but they do not occur at zero eccentricity. This favors capture for an initially circular orbit  as the particle's Hamiltonian curve is transported along the resonance's stable equilibrium adiabatically into permanent resonance.\footnote{The particle may be released from resonance when its eccentricity reaches unity and evolution is allowed to continue. Orbital plane reversal may occur leading to resonance exit \citep{YuTremaine01}. This situation will not be considered here.} Only when the eccentricity has increased sensibly ($\sim 0.5$) does the particle enter the second order $eI^2$ resonance with resonant argument  $\phi_2=2\lambda-\lambda_{\rm J}+\varpi-2\Omega$ while the first order resonance is maintained and the argument of pericenter is stationary.

Resonances for retrograde motion are weaker than the usual prograde resonances as particle-planet encounters occur at higher velocities for a shorter duration. Planar retrograde $p$:$q$ mean motion resonances other than the coorbital resonance possess a nearly identical\footnote{The only difference is the force amplitude. \cite{MoraisNamouni13a} explain how to obtain retrograde force amplitudes from the expansion of the classical prograde disturbing function.} dynamical structure to that of prograde resonances with order $|p+q|$ (a $p$:$q$ prograde resonance has order $|p-q|$). For details, see \citet{MoraisGiuppone12}. The previous arguments about prograde capture therefore explain why particles on initially circular planar retrograde orbits may not be captured in any mean motion resonance other than the 1:1 coorbital resonance during their drift from outside the 1:5 resonance (Figure 2, left column). For instance, the 1:2 planar retrograde resonance presents the same dynamical equilibria as the 1:4 (third order) prograde resonance thus precluding capture. The dynamical structure of the retrograde coorbital resonance was examined by \cite{MoraisNamouni13a}.  The particle shown in Figure 2 with $I_0=175^\circ$, remains in the coorbital resonance with the libration angle $\phi_1=\lambda^\star-\lambda_{\rm J}-2\varpi^\star$, where $\lambda^\star=\lambda-2\omega$ and $\varpi^\star=\varpi-2\omega$ are the pertinent angles for retrograde motion \citep[see][]{MoraisNamouni13a}. At no time does the prograde resonant angle $\phi_2=\lambda-\lambda_{\rm J}$ librate.

Adding more relative inclination to both the prograde and retrograde  orbits modifies the dynamics as inclination resonances  (whose order is larger than 1) as well as mixed eccentricity and inclination resonances are activated. Capture may occur with greater likelihood in higher order resonances. We illustrate the similarity and differences between prograde and retrograde capture by considering capture in the 1:3 prograde resonance and the 1:2 retrograde resonance as the latter (but not the former) represents the likely outcome for an initial relative inclination of $10^\circ$ (see next section).

For prograde motion (Figure 1, right panel), capture in the second order 1:3 $I^2$ resonance occurs as the nominal location is reached and the corresponding angle $\phi_3=3\lambda-\lambda_{\rm J}-2\Omega$ librates. As the inclination increases substantially ($\sim 40^\circ$), an additional resonant argument  $\phi_4=3\lambda-\lambda_{\rm J}-2\varpi$ enters libration and the argument of pericenter is stationary. When eccentricity has increased substantially, the particle exits the original resonance but remains in the 1:3 $e^2$ resonance. 

For retrograde motion, capture in the 1:2 resonance is favored over the 1:1 resonance as inclination is increased. This happens because the  mixed $eI^{\star 2}$ 1:2 resonance (where $I^\star = 180^\circ-I$) with the librating angle $\phi_3=2\lambda^\star-\lambda_{\rm J}-\varpi^\star+2\Omega$ is encountered before the coorbital resonance. As the former resonance has a structure similar to first order prograde resonances far outside the nominal resonant semi-major axis (see Appendix A), particles are systematically captured at its location much like prograde planar orbits are captured in the 1:2 resonance. Further evolution is however more complex. After capture in the $eI^{\star 2}$ resonance, the Kozai-Lidov resonance sets in as the particle's argument of pericentre executes librations around $\pm 90^\circ$.  The presence of the Kozai-Lidov resonance is evidenced by the temporary evolution change in eccentricity and inclination that display the characteristic large amplitude cycles which cease to occur as the particle exits the resonance. The Kozai-Lidov mechanism forces the particle to enter the 1:2 $e^3$ retrograde resonance with a librating angle $\phi_4=2\lambda^\star-\lambda_{\rm J}-3\varpi^\star$ and later to exit the $eI^{\star 2}$ resonance as well as the Kozai-Lidov resonance.  The particle  remains in the $e^3$ resonance until the eccentricity nears unity and collision with the star becomes inevitable.  The preference for a third order resonance (1:2 $e^3$) over a first order (and for some parameters second order) resonance (1:2 $eI^{\star 2}$) is an unexpected feature of retrograde resonant capture. We have checked that this three-stage capture process is independent of the initial semi-major axis and the characteristic drift time. In particular, a longer drifting time would only increase in proportion the relative duration of the various libration episodes.

So far, we have restricted our examples to initially circular orbits but adding eccentricity to the planet and/or the particle introduces more excitation in the dynamical system as higher order resonances gain more importance and the Kozai-Lidov resonance limits  capture through large amplitude inclination oscillations.  These aspects are examined in the next two sections. 

 \section{Capture statistics}
We ascertain capture likelihood at arbitrary inclination by enlarging the initial conditions of the previous set-up. We choose the planet's eccentricity  from the following values 0, 0.05, 0.10 and 0.25 set  to model the circular restricted three-body problem, Jupiter's eccentricity in the solar system and two values typical of the observed exoplanets (whose current median value is 0.21). For each planet  eccentricity, we fix the particle's inclination  from 0 to 180$^\circ$ with 5$^\circ$ increments. For each inclination and each eccentricity standard deviation  $\sigma_e=0.01$, $0.1$, $0.3$ and $0.7$, we generate  a statistical ensemble of 1008 particles with random uniform distributions for mean longitudes and nodes, as well as a Rayleigh distribution for eccentricity. The study thus totals $596\,736$ numerical simulations of resonant capture. 
Thirty resonances are monitored: 5:1, 4:1, 3:1, 5:2, 7:3, 2:1, 9:5, 7:4, 5:4, 8:5, 3:2, 7:5, 4:3, 5:4,  6:5, 1:1, 5:6, 4:5, 3:4, 5:7, 2:3, 5:8, 3:5, 4:7, 5:9, 1:2, 3:7, 2:5, 1:3 and 1:4. They were chosen as the main first, second, third and fourth order (prograde) resonances. As explained in the previous section resonance order for retrograde motion is higher at the same nominal locations. 

The integration timespan of each simulation is 2.5 times the drift time of $10^5$ planetary orbital periods. The choice is motivated by two reasons: one is to allow particles that were not captured in outer resonance to drift past the planet  into the inner region. If unperturbed, a particle will drift and stop at the inner semi major axis of 0.246. The second reason explains why the timespan is not much longer. If it were then captured particles would inevitably increase their orbital eccentricity toward unity, as semi major axis drift continues, and collide with the star or be ejected from the system. In the following, we examine the results of resonant capture  for a circular planet. The effect of a planetary eccentricity is discussed in the next section.

\subsection{Total capture fraction}
The fraction of particles captured in each 1008 sample expressed in percent is shown in the first panel of Figure 3 as a function of the initial inclination. Its breakdown for each specific resonance is given in the following panels displayed in the order of planet encounter (left to right, top to bottom). Outer resonances are shown in Figure 3 whereas inner resonances are shown in Figure 4.  The displayed fraction in percent for a given resonance is the ratio of particle number in that resonance to the total particle number captured in resonance from the 1008 sample.  Total and specific resonance fractions are calculated for the four eccentricity standard deviations $\sigma_e=0.01$, $0.1$, $0.3$ and $0.7$.  In all plots of resonant capture, the data points for initially planar retrograde orbits are not shown for the following reason ($I_0=180^\circ$). Capture in this case occurs  in the coorbital 1:1 resonance at 100\% fraction. However as the planar coorbital resonance has a strong collision singularity \citep[see][]{MoraisNamouni13a}, semi-major axis drift, that systematically increases the particle's eccentricity, inevitably leads to planet collision or system ejection  shortly after capture. This issue does not appear for planar prograde orbits as they are stopped before reaching the planet by the 1:2 resonance (see Figure 2). Collision singularity disappears as relative inclination becomes finite, much as in the prograde coorbital resonance \citep{Namouni99,Namounietal99}. Consequently, early collision and/or ejection are prevented. 

Total capture fraction shows a strong dependence on initial inclination. For $0\leq I_0\leq 10^\circ$ and $100^\circ\leq I_0\leq 170^\circ$, capture occurs nearly with unit probability most of it in the 1:2 resonance for prograde as well as retrograde motion. A sharp decrease in capture efficiency occurs for $10^\circ\leq I_0\leq 100^\circ$ in the  small eccentricity samples ($\sigma_e=0.01$) progressively disappearing as initial eccentricity standard deviation is increased.  The asymmetry of the inclination trough towards prograde motion is unexpected as retrograde resonances tend to be weaker than their prograde counterpart. Furthermore as the mean eccentricity is increased and the trough disappears, retrograde resonances are significantly more efficient at resonant capture than their prograde counterparts.

\subsection{Resonance strength}
Eleven resonances reach or exceed $\sim20$\% capture fraction  considering all samples. In order of decreasing capture maximum efficiency for a specific inclination, they are: 
1:2 (100\%), 1:1 (100\%), 1:4 (85\%), 2:5 (36\%), 5:2 (35\%), 1:3 (34\%), 9:5 (30\%), 1:5 (26\%), 2:3 (24\%), 2:1 (23\%), 7:3 (17\%). Inner resonances show a somewhat similar inclination dependence evidenced by  the presence of a prominent  and sharp capture peak about 30$^\circ$-wide and centered from $50^\circ$ to $70^\circ$ depending on the specific resonance. Outer resonances share the presence of mainly two capture peaks whose slopes and position may vary sensibly from one resonance to the other. Efficient capture for outer resonances may depend sharply on the sample's mean eccentricity. Whereas the 1:4 resonance capture peaks for $\sigma_e=0.01$, 1:5 is most efficient for $\sigma_e=0.70$ and the 2:3 resonance's largest capture fractions occur for $\sigma_e=0.30$ and span initial inclination values from $100^\circ$ to $140^\circ$. 

The coorbital 1:1 resonance has a unique dependence on initial inclination. Capture does not occur for planar prograde orbits and is certain for planar retrograde orbits albeit with a quick catastrophic outcome as explained earlier. Capture fraction exhibits secondary peaks centered around $80^\circ$, $110^\circ$ and $160^\circ$ at respectively $10\%$, $5\%$ and $30\%$. The first two peaks are independent of the mean eccentricity whereas the latter appears for large eccentricities.

\subsection{Final states}
To illustrate the final configurations of resonant capture, we show in Figures (5), (6) and (7) the final orbital elements as a function of the final semi-major axis for initial inclinations $I_0=10^\circ, \ 170^\circ$ and $110^\circ$.  Similar plots for all simulations (all initial inclinations and planet initial eccentricities) are accessible online as supplementary information material. Capture for $I_0=10^\circ$ occurs mainly in the 1:2 and 1:3 resonances. Other resonances may trap some particles as the initial mean eccentricity is increased. With the exception of one outlier particle with moderate initial eccentricity ($\sigma_e=0.10$), no orbital inversion occurs for small initial inclination prograde orbits. It is interesting to note the three family-like clusters that appear markedly for  $\sigma_e=0.01$ at $I_f\sim 5^\circ$  and $40^\circ$ (1:2 resonance) and at $I_f\sim 50^\circ$ (1:3 resonance). The first two clusters are associated with the Kozai-Lidov resonance evidenced by the concentration of particles near $\omega_f=0$ and $180^\circ$. The clustering disappears for large mean eccentricities along with capture efficiency. Small relative inclination retrograde orbits ($I_0=170^\circ$) behave differently in that a mean eccentricity increase favours a greater resonance capture diversity as evidence in Figure (6).  In addition to the 1:1 and 1:2 processes discussed in the previous section, the 1:3, 1:4 and 1:5 higher order resonances increase their efficiency for larger initial eccentricity (see Figure 3). The Kozai-Lidov resonance is present at some of the resonances whereas orbital inversion is marginal despite large inclination oscillations. 

Intermediate initial inclinations such as $I_0=110^\circ$ (Figure 7) show understandably the most diverse final states as eccentricity and inclination resonances operate on a more equal footing. A number of  features are of particular interest: first, the formation of family-like clusters some of which are related to  the Kozai-Lidov resonance. Second,  dominant orbital inversion occurs for particles trapped outside the planet's orbit ($a_f>1$). Third, capture in inner resonances occurs with a final inclination similar to its initial value $I_f\sim I_0$ through the Kozai-Lidov mechanism.  Four, the family-like clustering, although weaker, persists as the mean eccentricity is increased. 

Close examination of the final states for all initial inclinations (online material) indicates that orbital inversion occurs in the range $60^\circ\lesssim I_0\lesssim 130^\circ$ for $\sigma_e=0.01$. It occurs in a wider inclination range as the mean eccentricity is increased.

 \section{Trends for eccentric planets}
Adding an eccentricity to the planet's orbits activates the eccentric Kozai-Lidov mechanism \citep{Katzetal11,LithwickNaoz11} whereby particle orbit inversion becomes possible for initially moderate relative inclinations. The mechanism is similar to a secular resonance in nature and couples to mean motion resonance capture as explained in the previous section. Examination of the final states of samples with an eccentric planet (online material) shows that the Kozai-Lidov process dominates evolution even for small initial inclinations when the planet's eccentricity is larger than 0.10.  A capture asymmetry appears for initially prograde orbits in that outer resonances are globally less efficient at capture than inner resonances. This trend is illustrated in Figure (8) where the capture fractions of the 2:1, 1:1 and 1:2 resonances are shown for the various planet eccentricities. In general, the Kozai mechanism induces  large inclination oscillations resulting in a decrease in capture probability  as collision and ejection are more likely. This is evidenced in Figure (9) where the total capture probabilities are shown for various values of the planet's eccentricity.  The more drastic change occurs for nearly circular ensembles ($\sigma_e=0.01$) with a decrease of capture probability of nearly 50\% at large inclinations.  For larger particle eccentricities, the global capture fraction decreases moderately as the planet eccentricity is increased and  the general initial inclination dependence is maintained as  capture is more efficient for  larger initial inclinations. 

\section{Conclusion}
Our extended numerical simulations of capture in resonance at arbitrary inclination have shown that retrograde orbits are more easily captured in resonance than prograde orbits. This fact is unexpected as  retrograde resonances are intrinsically weaker than prograde ones. The strongest resonances by far are the 1:1 and 1:2 resonances but higher order resonances such as the 2:5 and 1:4 may achieve large capture probabilities respectively for prograde and  small eccentricity polar orbits. We found that as the planet's eccentricity is increased above 0.1, overall capture is diminished but retrograde orbits are more likely to be captured than prograde orbits. These conclusions are the first step towards understanding resonant capture in the solar system. In particular, the measured capture probability depends on the mass of the perturbing planet \citep{Wyatt03} that sets resonance width as well as on the nature of the semi major axis drift  that is not generally unidirectional {and eccentricity and inclination preserving }(such as the one we chose). To remedy these assumptions, the next natural step requires the inclusion of the four giant planets in the solar system. With four planets,  the resonance web is denser and more diverse, orbital drift results naturally from close encounters and resonance overlap, the Kozai-Lidov mechanism is all but disrupted and chaotic orbital evolution rules. However, although limited, our present results have some direct relevance to the dynamics of the Centaur population. 

Centaurs are small bodies on a long-term unstable  orbits that lie in the space between the outer planets. Their orbital instability is caused by the close encounters with the planets. As a result Centaurs have chaotic orbits that wander radially between the planets and may be captured then released from mean motion resonances \citep{BaileyMalhotra09,VolkMalhotra13}. The conclusions of our present study suggest that large inclination and in particular retrograde Centaurs and Damolcoids are more likely to be temporarily trapped in resonance than smaller inclination counterparts. As temporary resonance capture is one of the possible ways to postpone disruptive close encounters, it is reasonable to deduce that the relative efficiency of capture in retrograde resonance implies that the dynamical lifetimes Centaurs and Damocloids on retrograde orbits are larger than those with prograde orbits. This suggests that the identified asteroids in retrograde resonance, 2006 BZ8, 2008 SO218, 2009 QY6 and 1999 LE31 \citep{MoraisNamouni13b}, may  be among the dynamically oldest small bodies in the outer solar system.

\section*{Acknowledgments}
The authors thank an anonymous reviewer for useful comments. The numerical simulations in this work were performed at the Center for Intensive Computing  `M\'esocentre {\sc sigamm}'  hosted by the Observatoire de la C{\^o}te d'Azur.

\appendix \section{Hamiltonian model for the \hiddenmath \ 1:2 retrograde resonance}
In order to understand why circular slightly inclined retrograde orbits are captured first in the $e\cos^2 (I/2)$ resonance, we study its dynamics  using the method developed by Poincar\'e for his hamiltonian resonance model of Hecuba-type asteroids' orbits \citep{Poincare02,ssdbook}. The hamiltonian of the three-body problem of a planet on a circular orbit and a particle on a small eccentricity and inclination retrograde orbit may be written as:  

\begin{eqnarray}
H&=&-\frac{Gm}{2a}-\frac{Gm^\prime}{2a^\prime}+U_{\rm res}+U_{\rm sec},\\
U_{\rm res}&=&-\frac{Gm m^\prime}{a^\prime} \left[\alpha f_{\rm d}-\frac{1}{2\alpha}\right]\, e\cos^2 (I/2)\times \nonumber \\&& \times \cos(2\lambda^\star-\lambda^{\prime\star}-\varpi^\star+2\Omega^\star),\\
U_{\rm sec}&=&-\frac{Gmm^\prime f_{\rm s}}{a^\prime} \left( e^2-I^2\right),
\end{eqnarray}
where the mean longitude $\lambda^\star=M+\omega-\Omega$ and the longitude of pericenter $\varpi^\star=\omega-\Omega$ and the primary's mass is unity and much larger than $m$ and $m^\prime$. The perturbation amplitudes are given as  $f_{\rm d}=\alpha(1+\alpha\partial_\alpha/4)b^{(0)}_{3/2}$ and $f_{\rm s} =\alpha b^{(1)}_{3/2}/8$ where $\alpha$ is the semi-major axis ratio (chosen $<1$) and $b^{(j)}_{3/2}$ are standard Laplace coefficients. In choosing the correct retrograde resonant argument we applied the method derived by \cite{MoraisNamouni13a} to extract retrograde resonant arguments of the classical expansion of the disturbing function of prograde orbits. We note that the simple form of the secular potential that is identical to that of prograde motion is due to the circular orbit of the perturber. The secular potential for retrograde orbits is generally different from that of prograde orbits if both perturber and perturbed have eccentric orbits \citep{KinoshitaNakai88}.

The usual Poincar\'e variables   must be modified for retrograde orbits using the generating function $F=\lambda^\prime \Lambda^{\prime\star} +(\lambda + 2 z)\Lambda^\star+ (\gamma-2z) \Gamma^\star-z Z^\star$ as follows:
\begin{eqnarray}
\Lambda^{\prime\star}&=&\Lambda^\prime=m^\prime n^\prime a^{\prime 2}, \ \lambda^{\prime\star}=\lambda^\prime,\\
\Lambda^\star&=&\Lambda_1=mna^2, \ \lambda^\star=M+\omega-\Omega,\\
\Gamma^\star&=& \Gamma_1=mna^2\left[1-(1-e^2)^{1/2}\right], \ \gamma^\star=\Omega-\omega,\\
Z^\star&=& 2(\Lambda-\Gamma)-Z,\nonumber \\
&=&m na^2\left(1-e^2\right)^{1/2}\left(1+\cos I\right), \ z^\star=\Omega,
\end{eqnarray}
where $\lambda=M+\omega+\Omega$, $\gamma=-\omega-\Omega$ and $z=-\Omega$ are the usual Poincar\'e angles. The variables are chosen so that $Z^\star$ approaches zero as $I$ approaches $180^\circ$. Henceforth only the new Poincar\'e variables are used and the star sign is removed. The lack of various angles in the expression of the Hamiltonian lead to its reduction using the following action-angle variables:
\begin{eqnarray}
&&\theta_1= 2\lambda-\lambda^{\prime}+\gamma+2z, \ \ \Theta_1=\Gamma,\\
&&\theta_2= \lambda, \ \ \Theta_2=\Lambda-2\Gamma,\\
&&\theta_3= \lambda^{\prime}, \ \ \Theta_3=\Lambda^{\prime}+\Gamma,\\
&&\theta_4= \gamma^{\prime}, \ \ \Theta_4=\Gamma^{\prime},\\
&&\theta_5= z^{\prime}, \ \ \Theta_5=Z^{\prime},\\
&&\theta_6= z, \ \ \Theta_6=Z-2\Gamma.
\end{eqnarray}
Except $\Theta_1$, all actions are constant. In particular, $\Theta_6$ relates eccentricity and inclination variations.  We use all the constants to expand the zero order hamiltonian to second order in $\Theta_1$ or equivalently $\Gamma$ (i.e. 4th order in $e$) and find:
\begin{eqnarray}
\frac{m}{2a}-\frac{m^\prime}{2a^\prime}&=&\left(2a^{-\frac{3}{2}}-a^{\prime-\frac{3}{2}}\right)\Gamma  - \left(\frac{6}{ma^2}+\frac{3}{2m^\prime a^{\prime 2}}\right)\Gamma^2. \nonumber \\
&&
\end{eqnarray}
The secular potential may be written up to a constant term as:
\begin{eqnarray}
H_{\rm sec}&=& \Gamma \dot\gamma_{\rm sec}+Z\dot z_{\rm sec} +\Lambda\dot\lambda_{\rm sec},\\
&=& \Gamma\left(-\dot\varpi_{\rm sec}+2\dot \Omega_{\rm sec}+2 \dot\lambda_{\rm sec}\right).
\end{eqnarray}
The  Hamiltonian may therefore be written as:
\begin{eqnarray}
H&=&\left[2(n+\dot\lambda_{\rm sec})-n^\prime -\dot\varpi_{\rm sec}+2\dot \Omega_{\rm sec}\right]\Gamma  \nonumber \\
&& - \frac{3}{2}\left(\frac{4}{ma^2}+\frac{1}{m^\prime a^{\prime 2}}\right)\Gamma^2 - \epsilon \Gamma^\frac{1}{2} \left(\Theta_6+2\Gamma\right)\ \cos\theta\nonumber \\
&&\epsilon=\frac{Gm m^\prime}{2^{1/2}a^\prime} \left[\alpha f_{\rm d}-\frac{1}{2\alpha}\right],
\end{eqnarray}
where the index of $\theta_1$ was dropped. We write the Hamiltoian as $H=\alpha_1 \Gamma-\alpha_2 \Gamma^2-\alpha_3 \Gamma^{1/2}(\Theta_6+2\Gamma)\cos \theta$ then scale $\Gamma$ by a factor $\beta$ and require that the last two terms have the same coefficient leading to $\beta=\alpha_2^2/\alpha_3^2$. Rescaling  the Hamiltonian using  $-\beta$ we end up with:
\begin{eqnarray}
H&=&\delta \Gamma +\Gamma^2+ \Gamma^{1/2}(T+2\Gamma)\cos\theta, \\
&& \delta=-\frac{\alpha_1\alpha_2}{\alpha_3^2}, \ \ T=\frac{\alpha_2^2\Theta_6}{\alpha_3^2}.
\end{eqnarray}
With this choice of sign for $\delta$, the perturbed body is outside (inside) the nominal resonance radius if $\delta>0$ ($\delta<0$). We determine the critical points of the Hamiltonian using cartesian coordinates $x=\Gamma^{1/2}\cos\theta$ and  $y=\Gamma^{1/2}\sin\theta$ and rewriting the Hamiltonain as:
\begin{eqnarray}
H&=& \delta (x^2+y^2)+ (x^2+y^2)^2+ x \left[T+2(x^2+y^2)\right].\nonumber \label{ham}\\
\end{eqnarray}
The critical points are obtained through:
\begin{eqnarray}
\partial_xH&=& 2\delta x+4x(x^2+y^2)+T+6 x^2+2y^2=0,\\
\partial_yH&=& y[\delta +2(x^2+y^2)+2x]=0.
\end{eqnarray}
For $y=0$,
\begin{equation}
4x^3+6 x^2+2\delta x+T=0\ \ \mbox{(CP1)}.
\end{equation}
Depending on the constant $T$, there can be several equilibria configurations (see below).   For $\delta +2(x^2+y^2)+2x=0$,
\begin{equation}
x=(T-\delta)/2 \ \ \mbox{and}\ \ y^2=-T/2-(T-\delta)^2/4\ \ \mbox{(CP2)}.\nonumber
\end{equation}  
The last pair of critical points (CP2) exists if $T<0$, $\delta<1/2$ and $1-\delta-(1-2\delta)^{1/2}<-T<1-\delta+(1-2\delta)^{1/2}$. They are the two unstable points that separate the three stable equilibria (CP1) when this stable configuration is possible. We plot the $x$-position (a proxy for eccentricity at equilibrium) of the critical points as a function of $\delta$ (a proxy for the distance to exact resonance)  in Figures (10) and (11) and the corresponding Hamiltonian contour plots in Figure (12) for two values of the parameter $T$ ($0.1$ and $-0.1$). For nearly circular nearly coplanar retrograde orbits, $T$ is a proxy for $(\pi-I)^2-2e^2$ where $I$ is measured in radians. For  $T>0$, the structure far outside exact resonance  is similar to that of a first order resonance \citep{ssdbook} and apocentric libration occurs at $x_0\sim -T/2\delta$ (panel $T=0.1$, $\delta=1.6$) where $\theta$ oscillates around $180^\circ$. As exact resonance is approached there is a spatial domain before exact resonance ($0.74\lesssim \delta \lesssim 1.2$) where two apocentric librations are possible.  In this domain the structure is similar to that of second order resonance (panel $T=0.1$, $\delta=1$). At exact resonance theres is only one libration equilibrium (panel $T=0.1$, $\delta=0$). Far inside resonance (panel $T=0.1$, $\delta=-1.6$), the structure is similar to that of a first order resonance with two libration centers (apocentric and pericentric). 

For $T<0$, libration far outside resonance is pericentric ($\theta$ oscillates around 0). The spatial domain where $0.35\lesssim \delta \lesssim 1.05$ has a structure similar to second order resonances (panel $T=-0.1$, $\delta=0.8$) and exact resonance is centered around a domain $-0.35\lesssim \delta \lesssim 0.35$ (panel $T=-0.1$, $\delta=0$)  where three libration centers are possible (two apocentric and one pericentric). Far inside resonance (panel $T=-0.1$, $\delta=-1.6$), the structure is similar to that of $T>0$ except that both librations are apocentric. 

We note that none of the equilibrium points of the $e\cos^2 (I/2)$ 1:2 retrograde resonance is located at $x=0$. In particular, initially circular orbits located far outside resonance are bound to be captured with unit probability as they get carried along the apocentric equilibrium point in Figure (10) ($\delta \gg 1$) during their semi-major axis drift. However as eccentricity is increased by resonant capture additional resonances need to be considered including the secular Kozai-Lidov mechanism. This limits the use of the hamiltonian model to the first stages of resonant capture. 

\begin{figure*}
\begin{center}
\includegraphics[width=140mm]{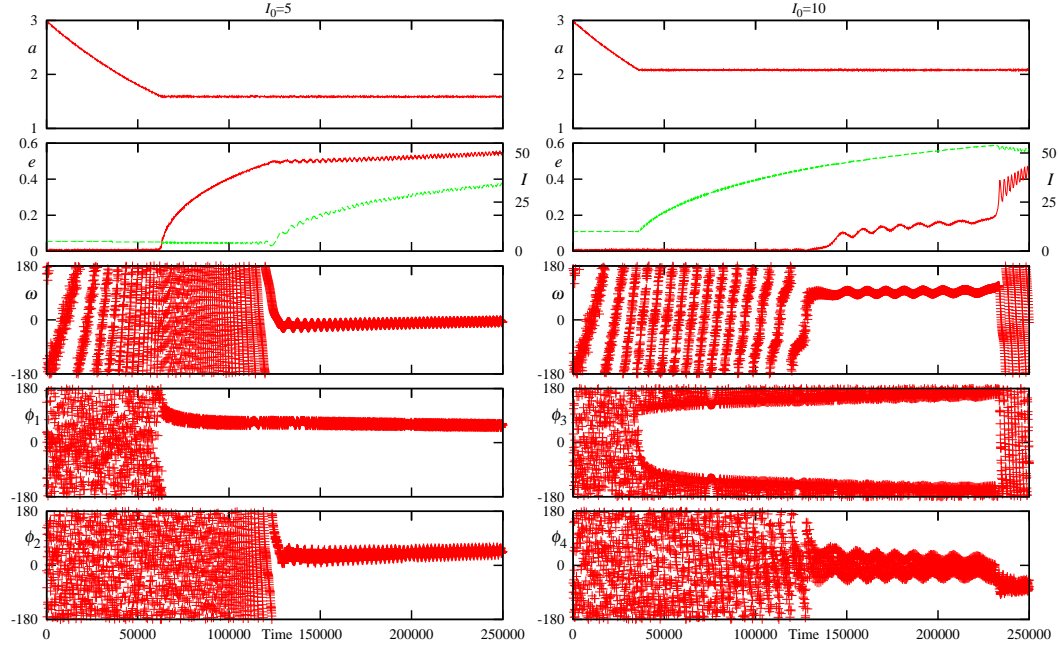}\\[-15mm]
\caption{Evolution towards capture for prograde circular orbits with initial inclinations $5^\circ$ and $10^\circ$. The eccentricity (inclination) is shown in solid red (dashed green). The angles are $\phi_1=2\lambda-\lambda_{\rm J}-\varpi$, $\phi_2=2\lambda-\lambda_{\rm J}+\varpi-2\Omega$, $\phi_3=3\lambda-\lambda_{\rm J}-2\Omega$ , $\phi_4=3\lambda-\lambda_{\rm J}-2\varpi$}
\end{center}
\end{figure*}

\newpage

\begin{figure*}
\begin{center}
\includegraphics[width=140mm]{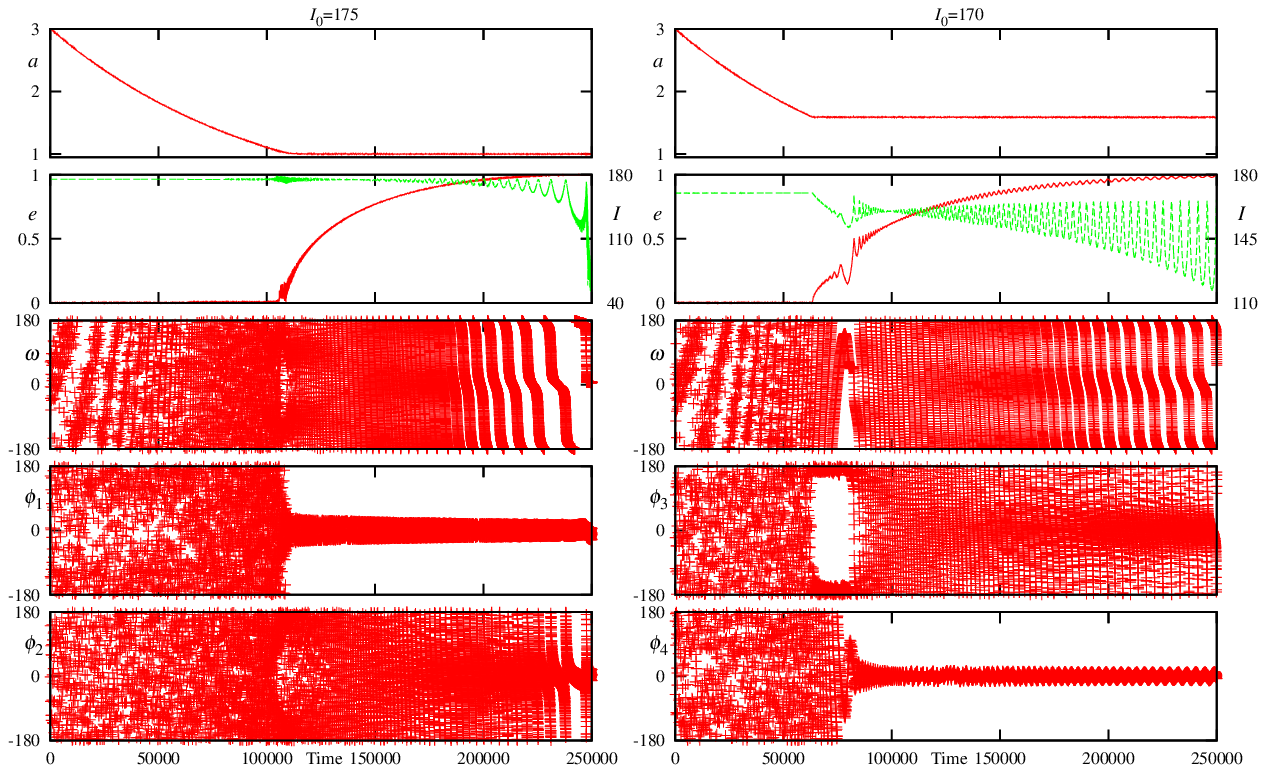}\\[-15mm]
\caption{Evolution towards capture for retrograde circular orbits with initial inclinations $175^\circ$ and $170^\circ$. The eccentricity (inclination) is shown in solid red (dashed green). The angles are $\phi_1=\lambda^\star-\lambda_{\rm J}-2\varpi^\star$, $\phi_2=\lambda-\lambda_{\rm J}$, $\phi_3=2\lambda^\star-\lambda_{\rm J}-\varpi^\star+2\Omega$ , $\phi_4=2\lambda^\star-\lambda_{\rm J}-3\varpi^\star$.}
\end{center}
\end{figure*}
\newpage
\begin{figure*}
\begin{center}
\includegraphics[width=60mm]{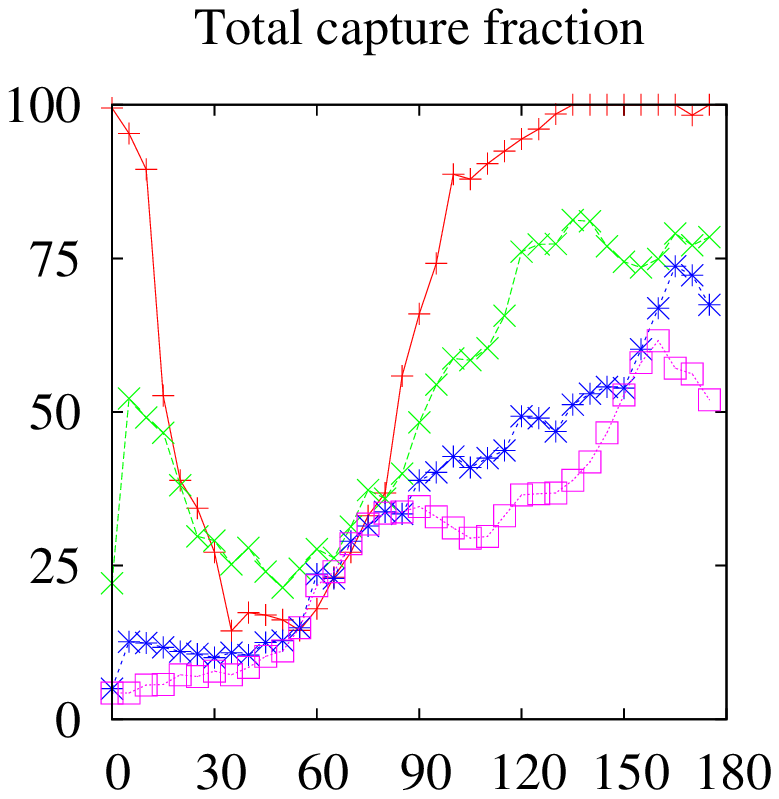}\hspace{-23mm}
\includegraphics[width=60mm]{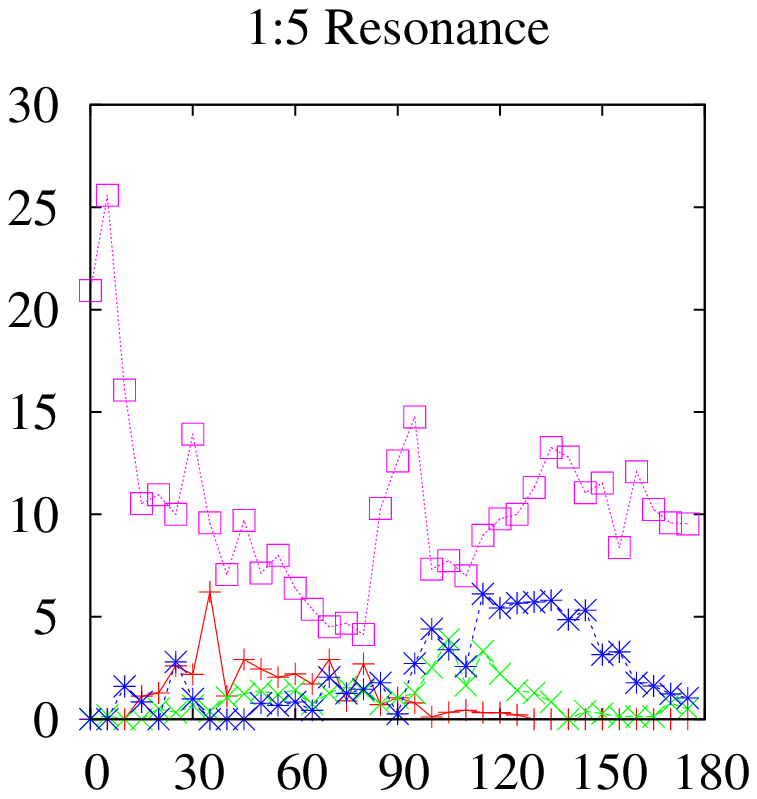}\hspace{-23mm}
\includegraphics[width=60mm]{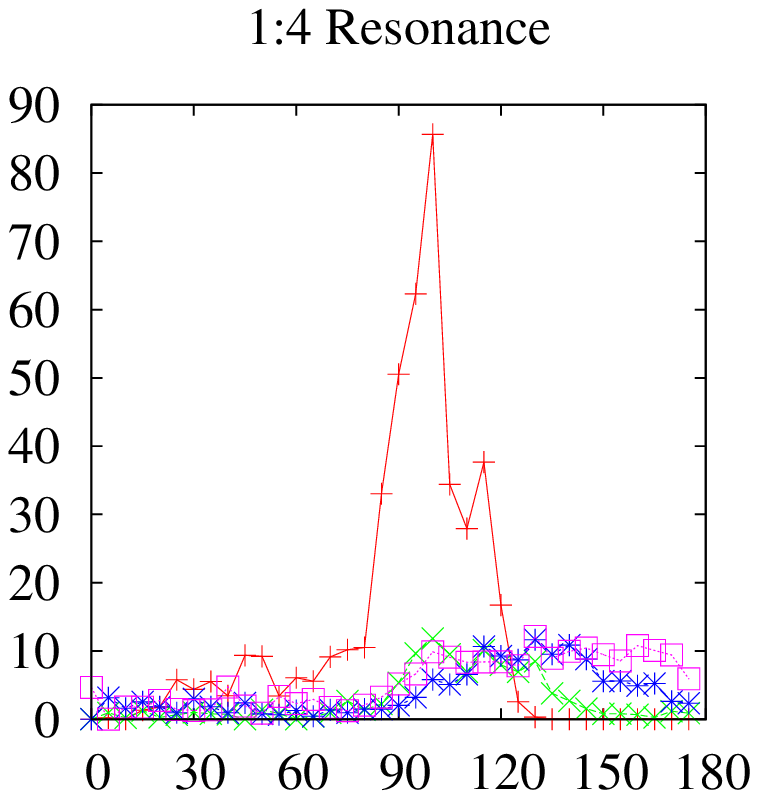}\hspace{-23mm}
\includegraphics[width=60mm]{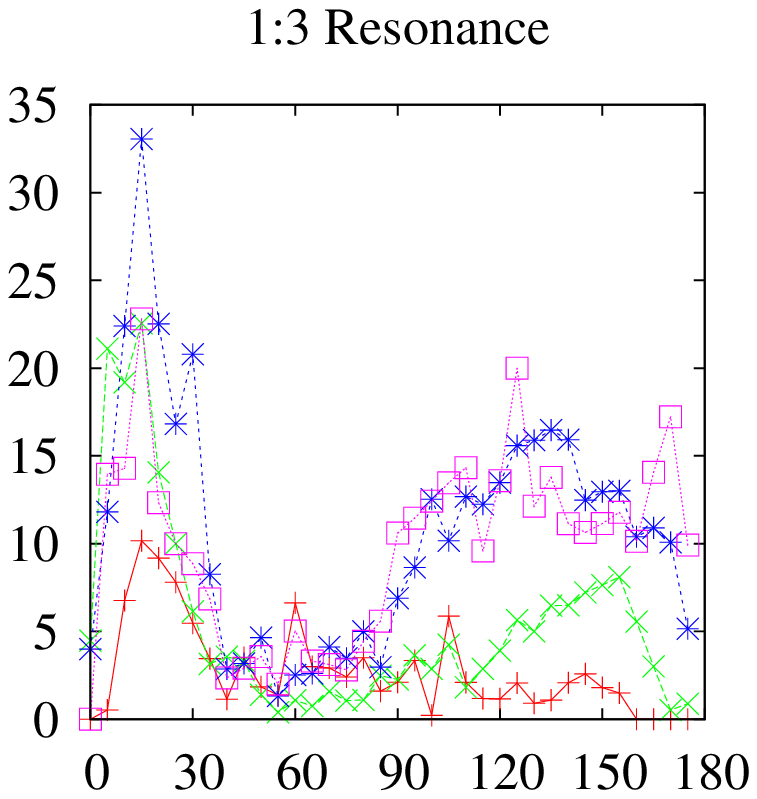}\\
\includegraphics[width=60mm]{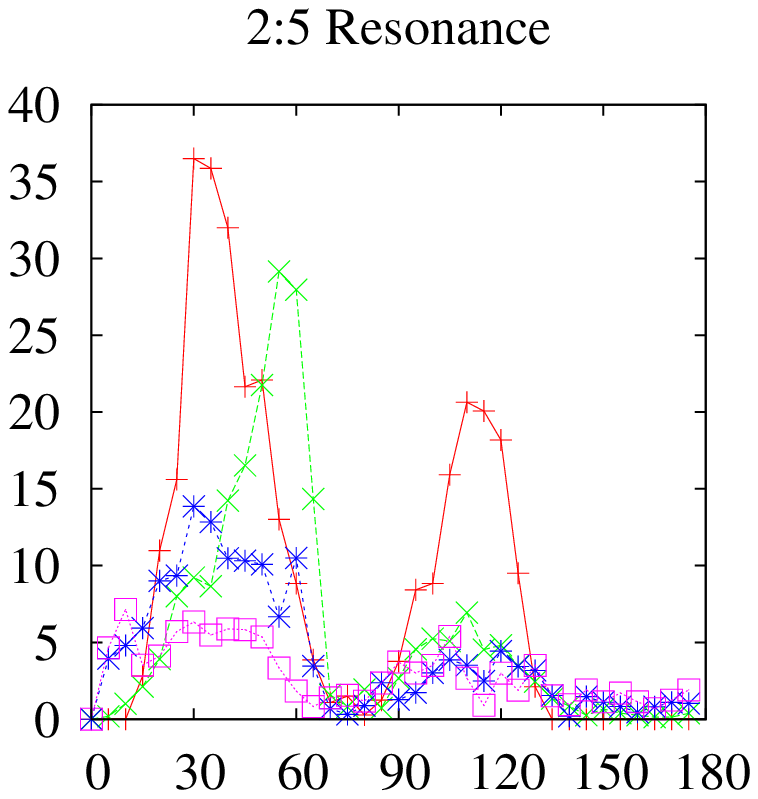}\hspace{-23mm}
\includegraphics[width=60mm]{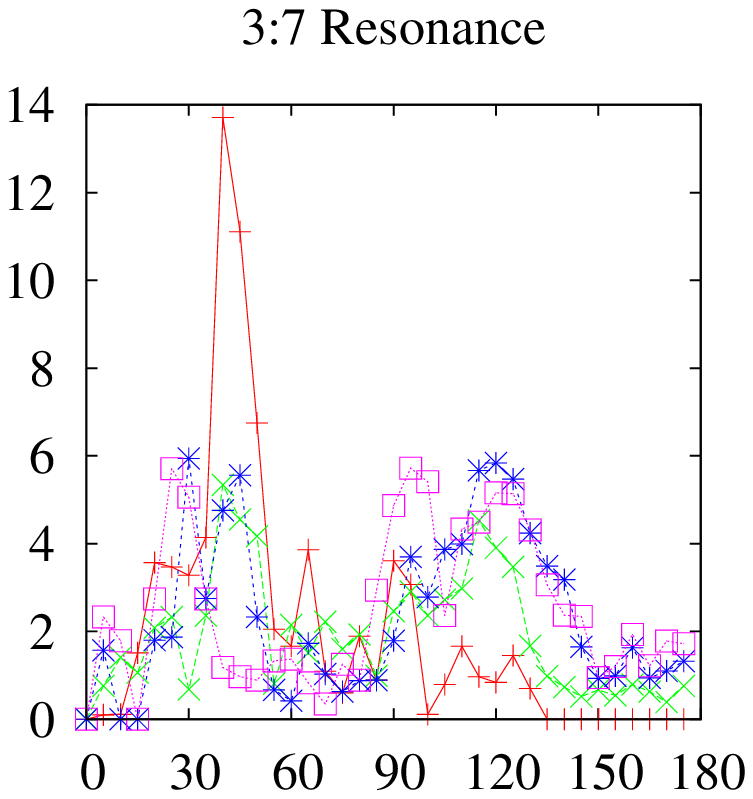}\hspace{-23mm}
\includegraphics[width=60mm]{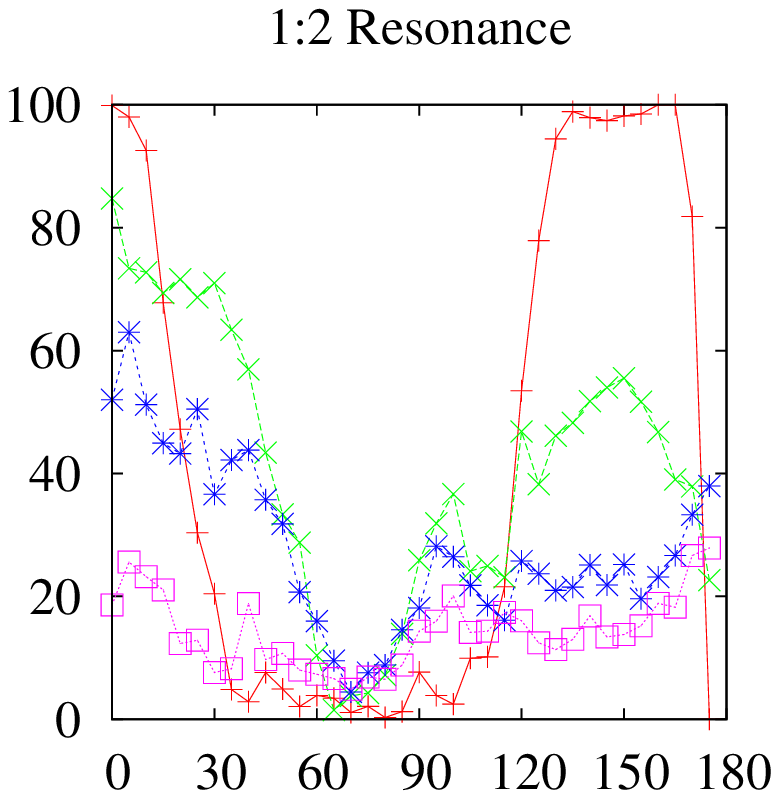}\hspace{-23mm}
\includegraphics[width=60mm]{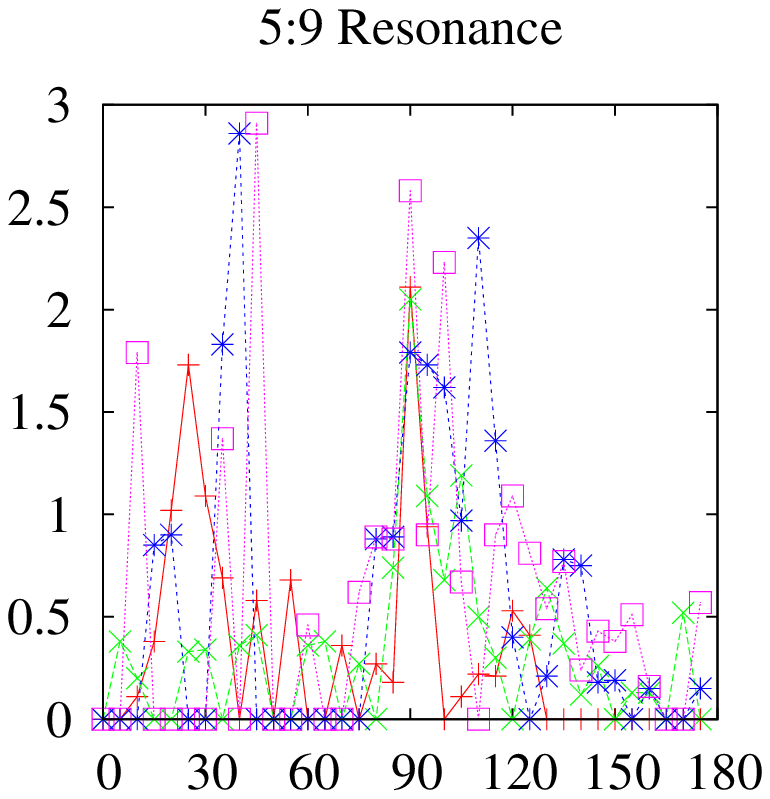}\\
\includegraphics[width=60mm]{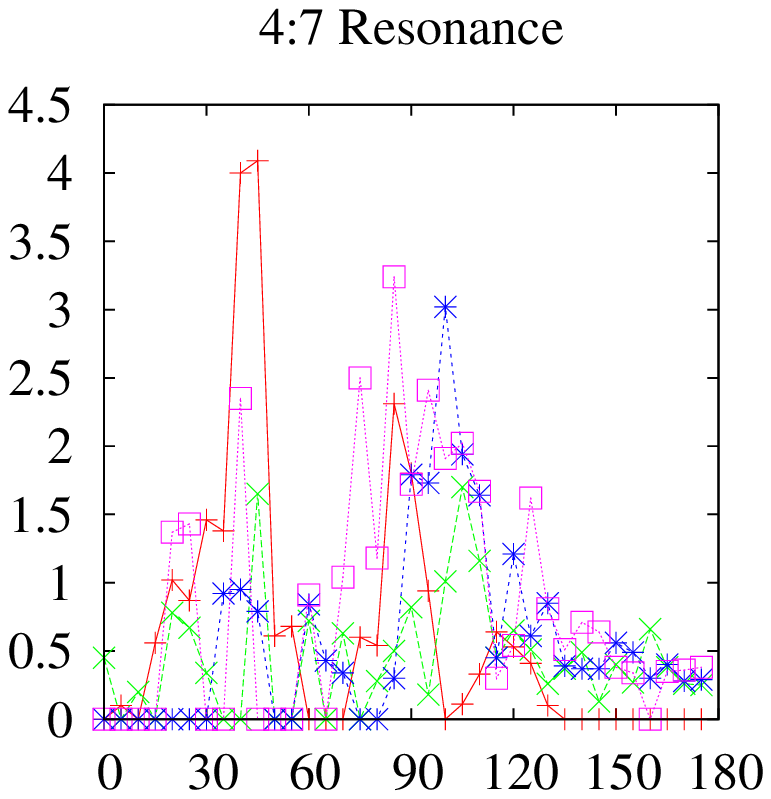}\hspace{-23mm}
\includegraphics[width=60mm]{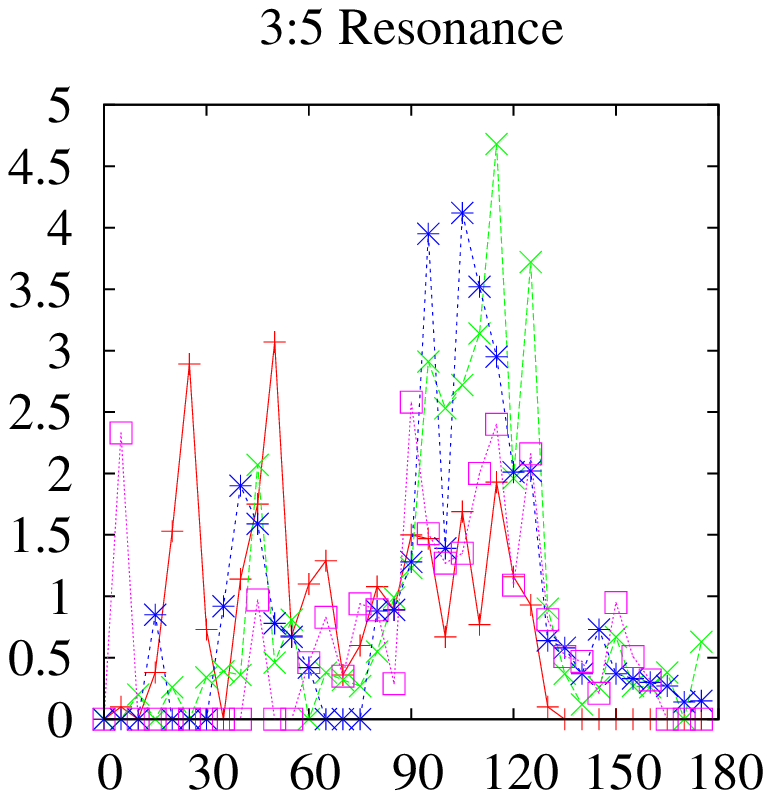}\hspace{-23mm}
\includegraphics[width=60mm]{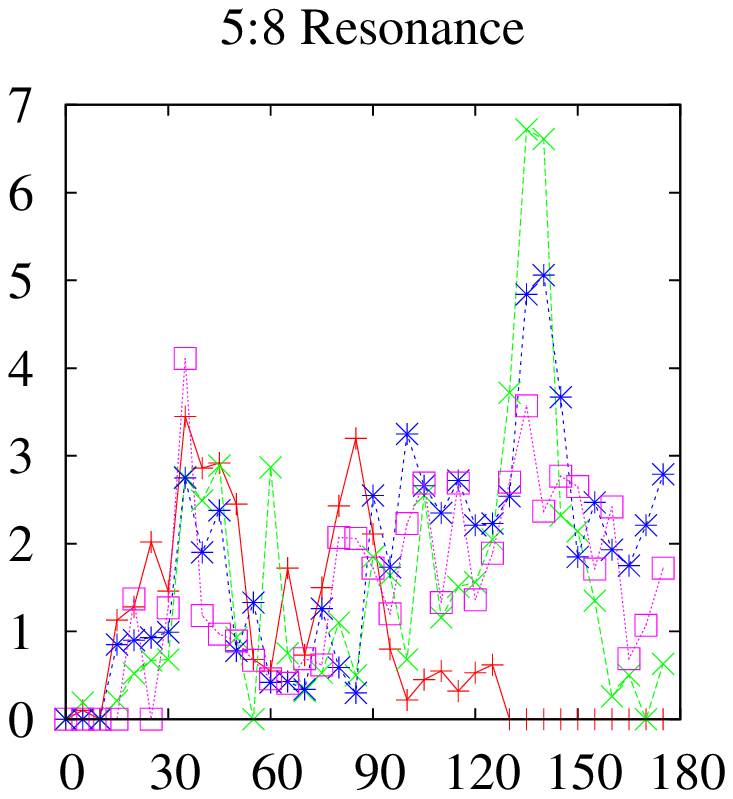}\hspace{-23mm}
\includegraphics[width=60mm]{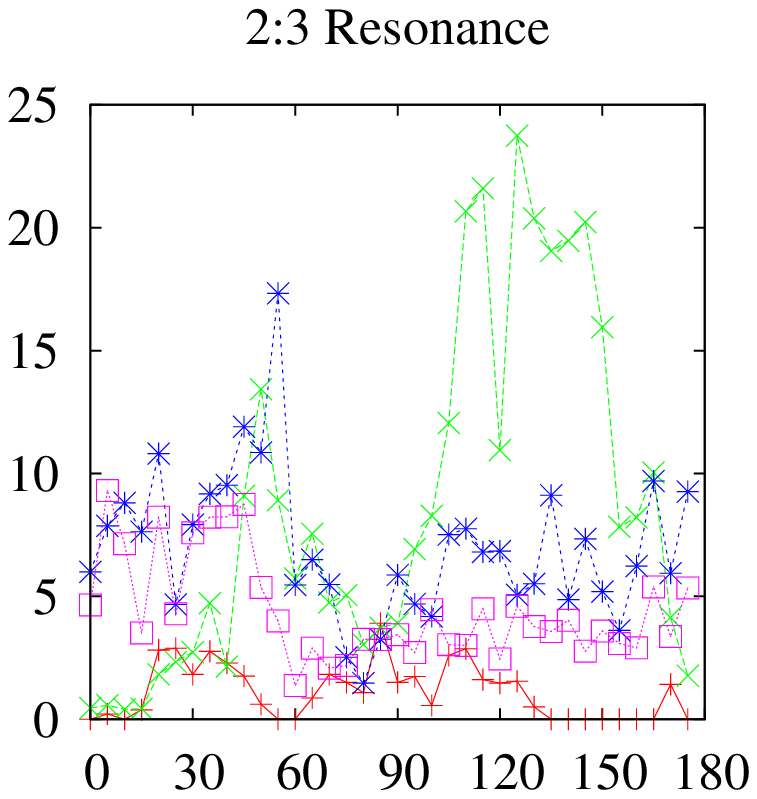}\\
\includegraphics[width=60mm]{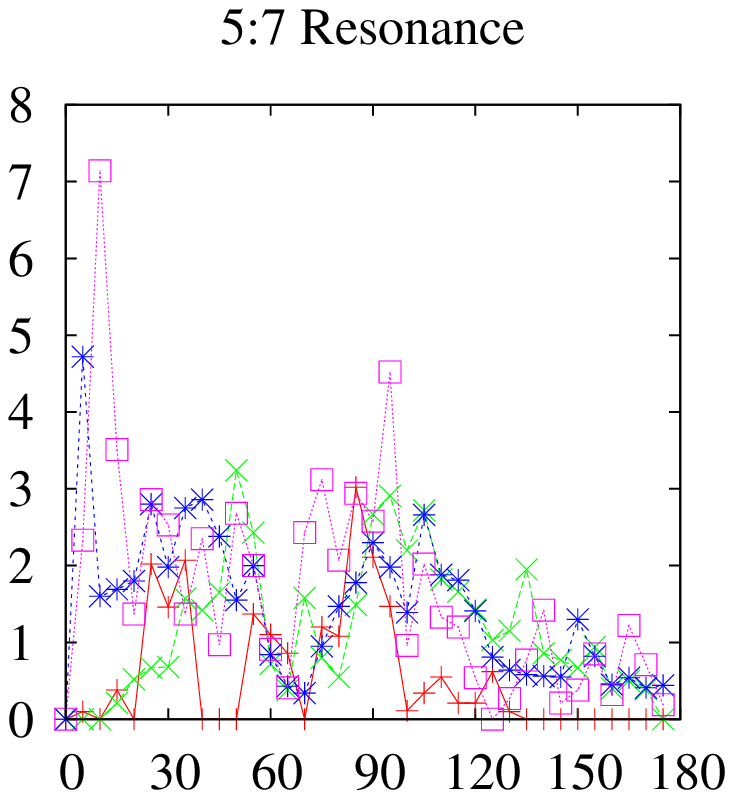}\hspace{-23mm}
\includegraphics[width=60mm]{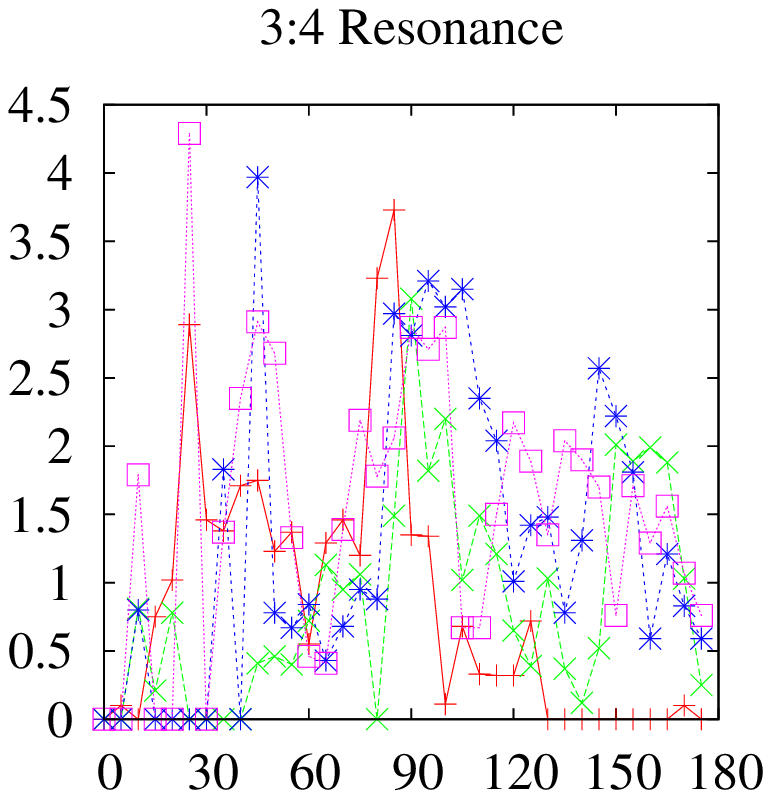}\hspace{-23mm}
\includegraphics[width=60mm]{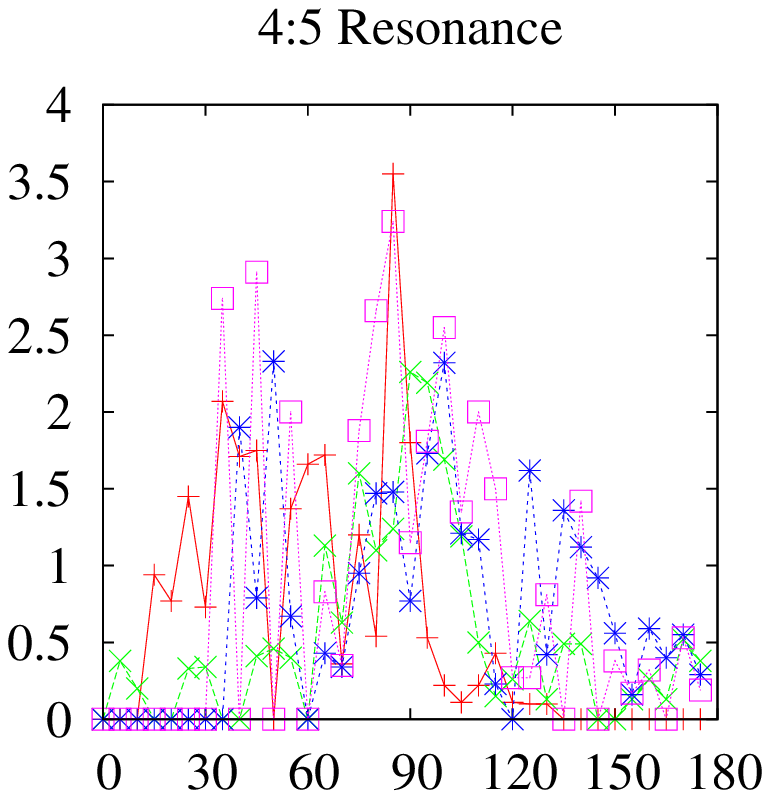}\hspace{-23mm}
\includegraphics[width=60mm]{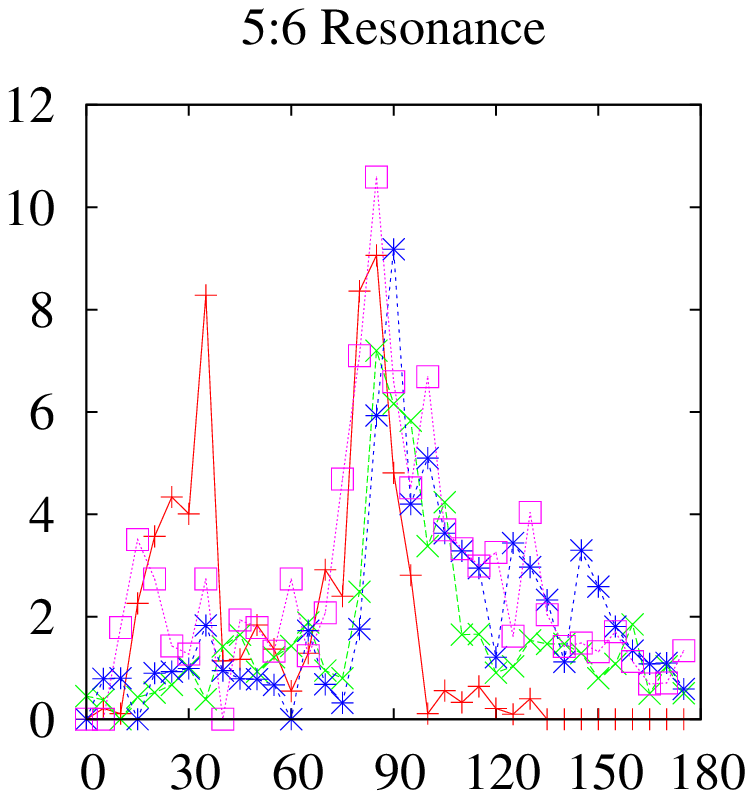}
\end{center}
\caption{Total capture fraction and its breakdown for outer resonance in order of encounter by the planet as a function of initial inclination, $I_0$. The planet has a circular orbit. Four curves are plotted corresponding to the four eccentricity standard deviations. Red line with cross sign for $\sigma_e=0.01$,  green line with times sign for $\sigma_e=0.10$, blue line with star sign for $\sigma_e=0.30$ and purple line with a box sign for $\sigma_e=0.70$.    {The total capture fraction for each $\sigma_e$ and every $I_0$ is the ratio of the total number of particle captured in any resonance to 1008. The capture fraction in a specific resonance is the ratio of the number of particles captured in that resonance to the total number of particles captured in any resonance.}}
\end{figure*}

\newpage

\begin{figure*}
\begin{center}
\includegraphics[width=60mm]{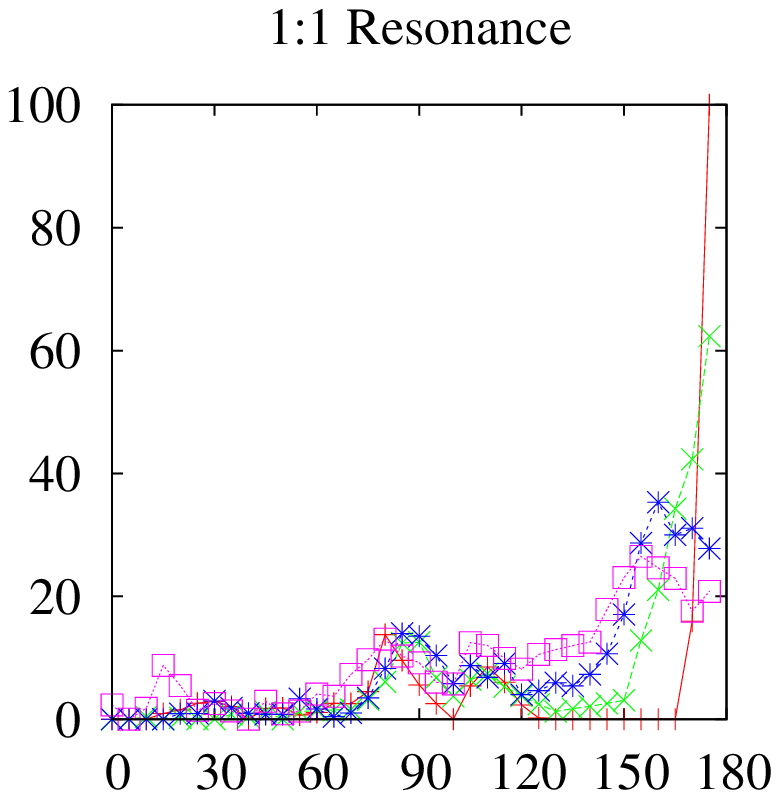}\hspace{-23mm}
\includegraphics[width=60mm]{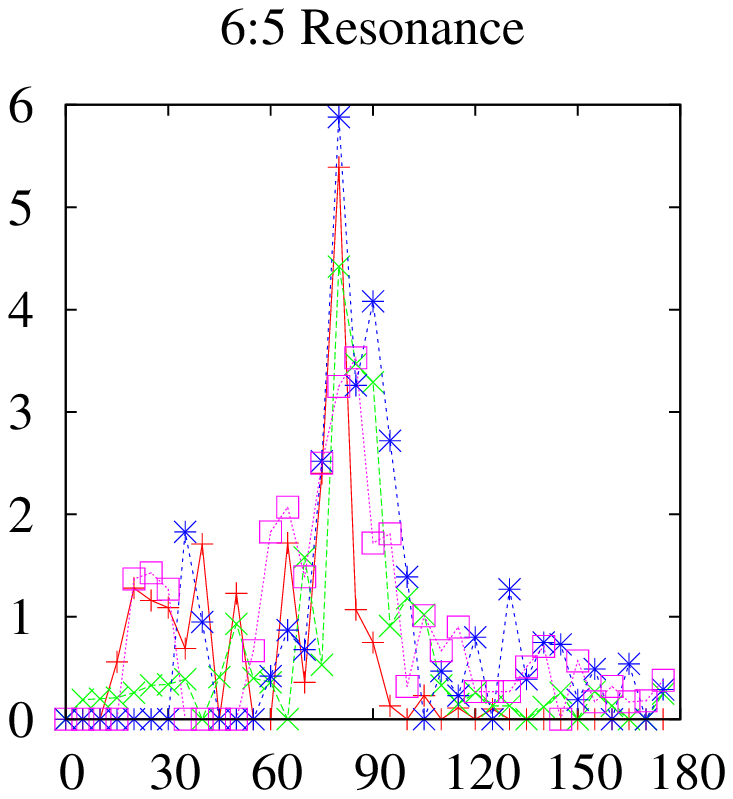}\hspace{-23mm}
\includegraphics[width=60mm]{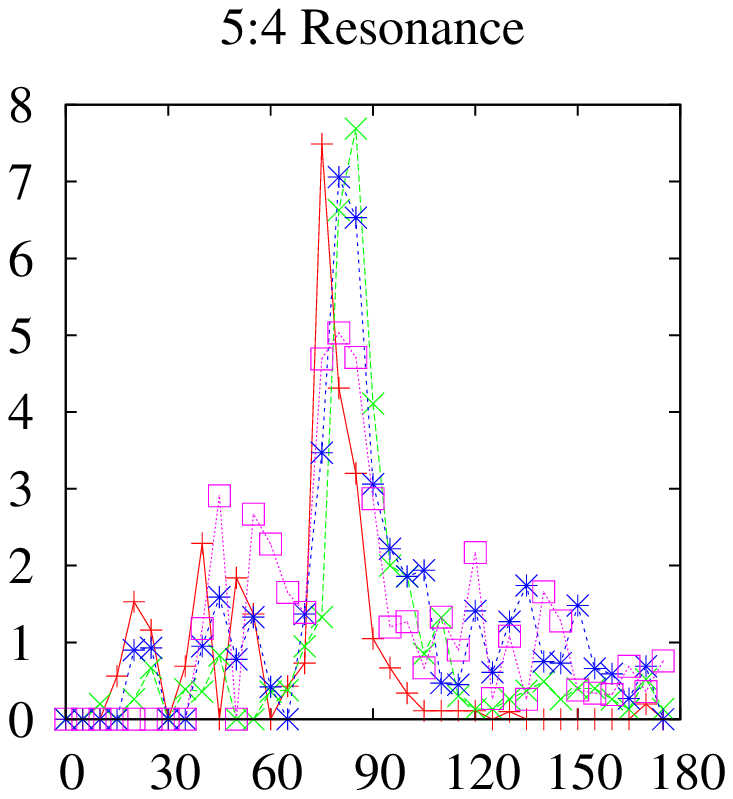}\hspace{-23mm}
\includegraphics[width=60mm]{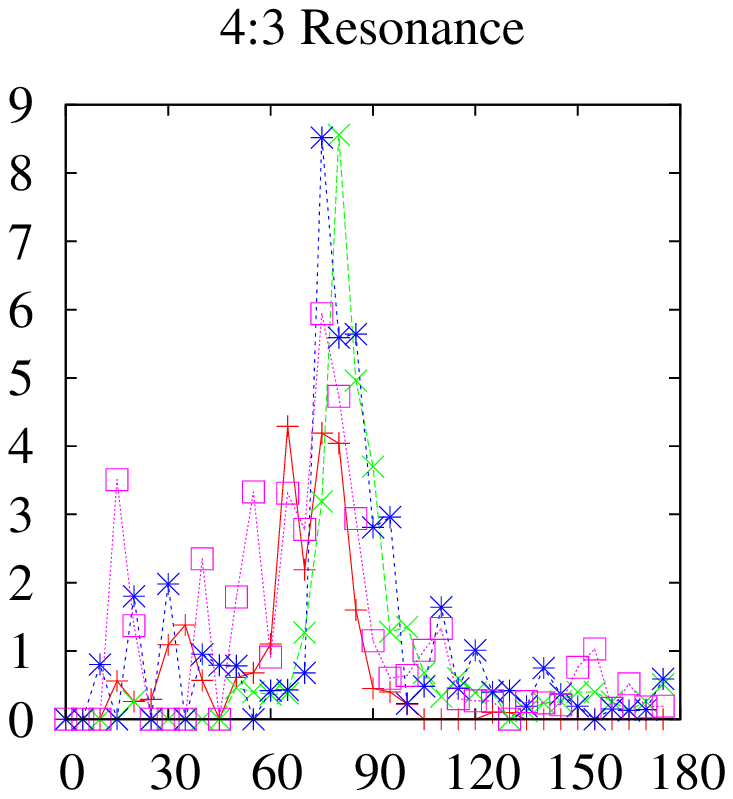}\\
\includegraphics[width=60mm]{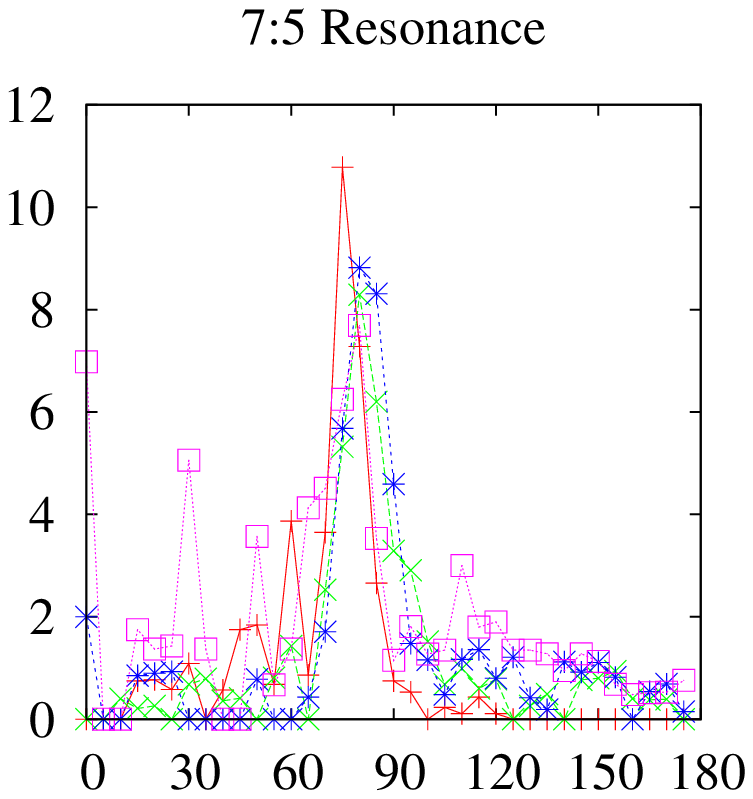}\hspace{-23mm}
\includegraphics[width=60mm]{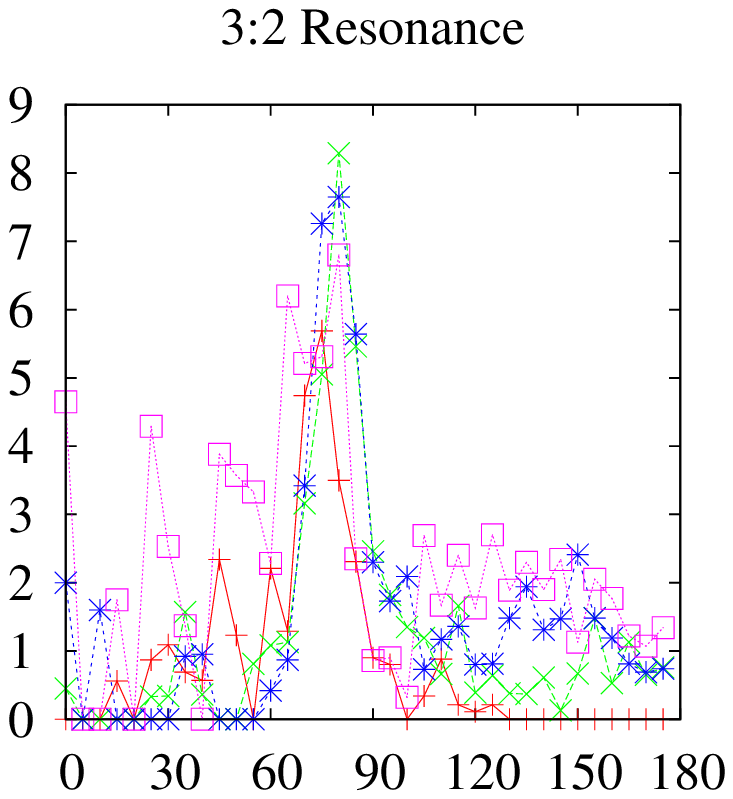}\hspace{-23mm}
\includegraphics[width=60mm]{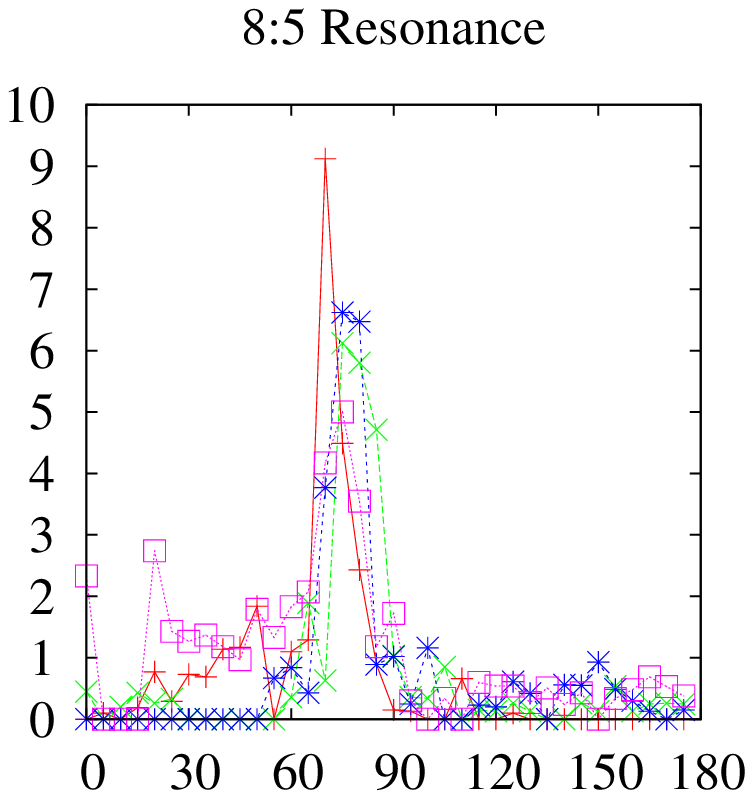}\hspace{-23mm}
\includegraphics[width=60mm]{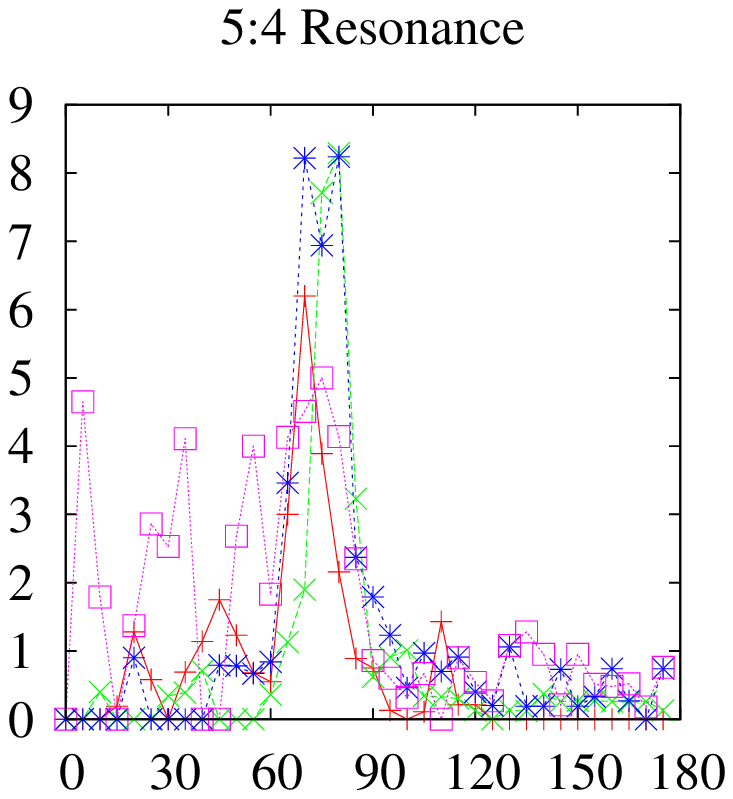}\\ 
\includegraphics[width=60mm]{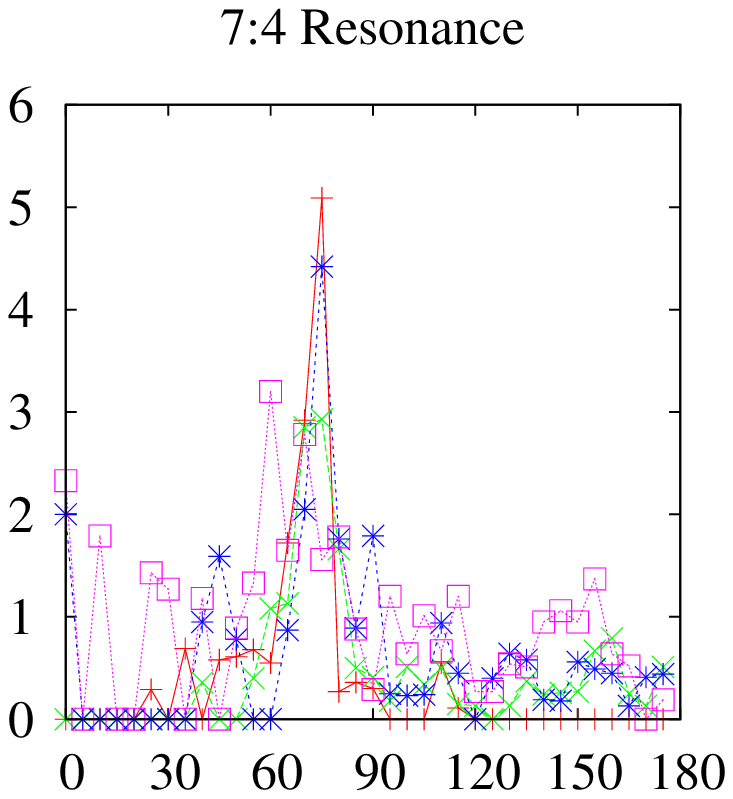}\hspace{-23mm}
\includegraphics[width=60mm]{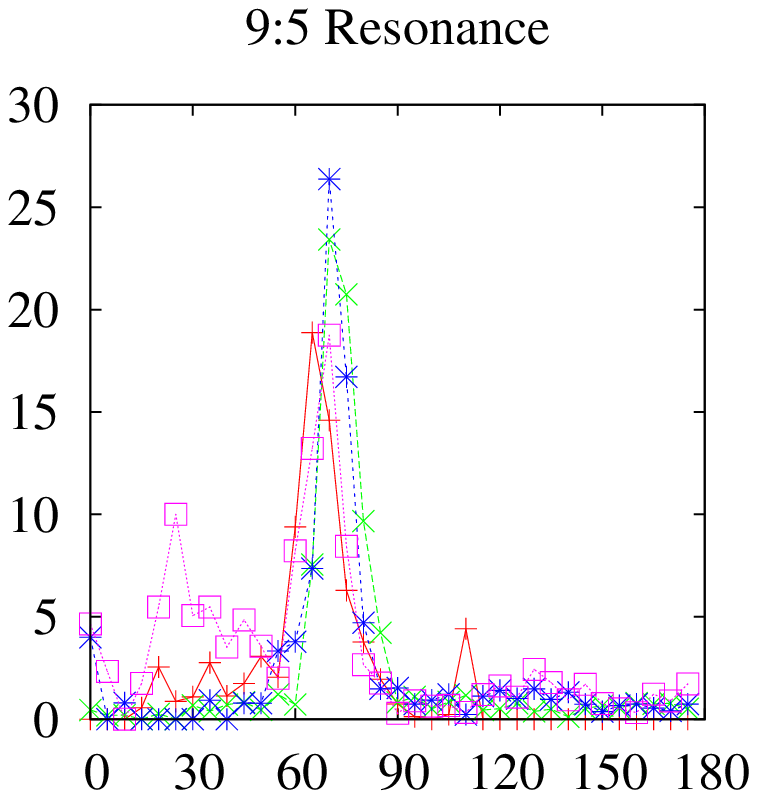}\hspace{-23mm}
\includegraphics[width=60mm]{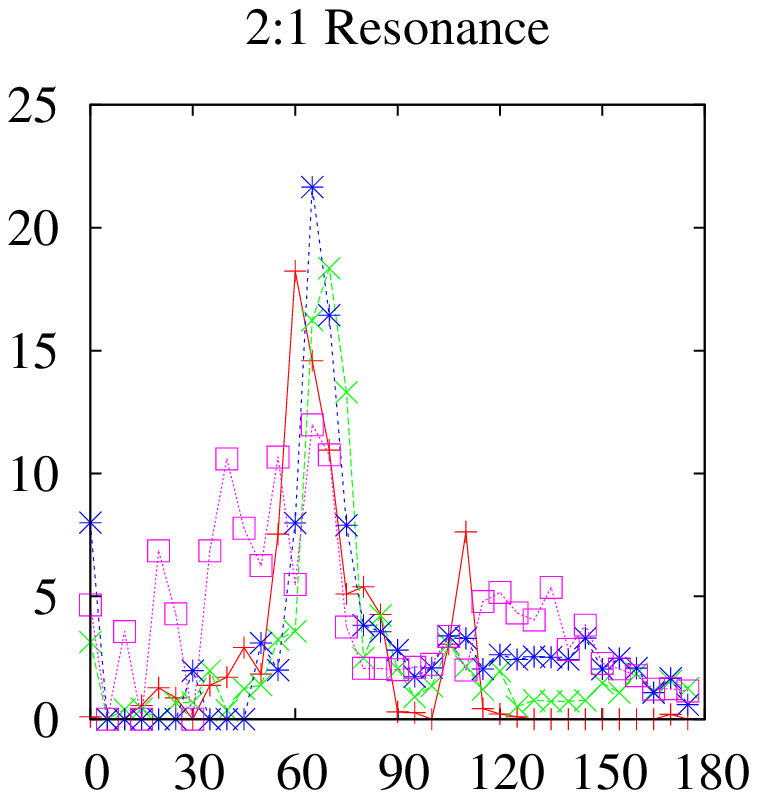}\hspace{-23mm}
\includegraphics[width=60mm]{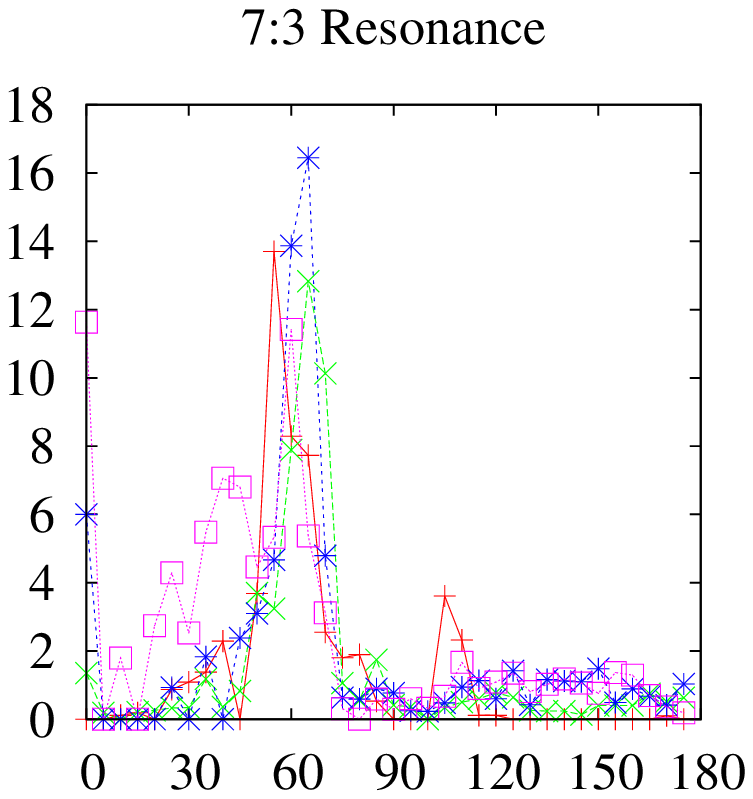}\\
\includegraphics[width=60mm]{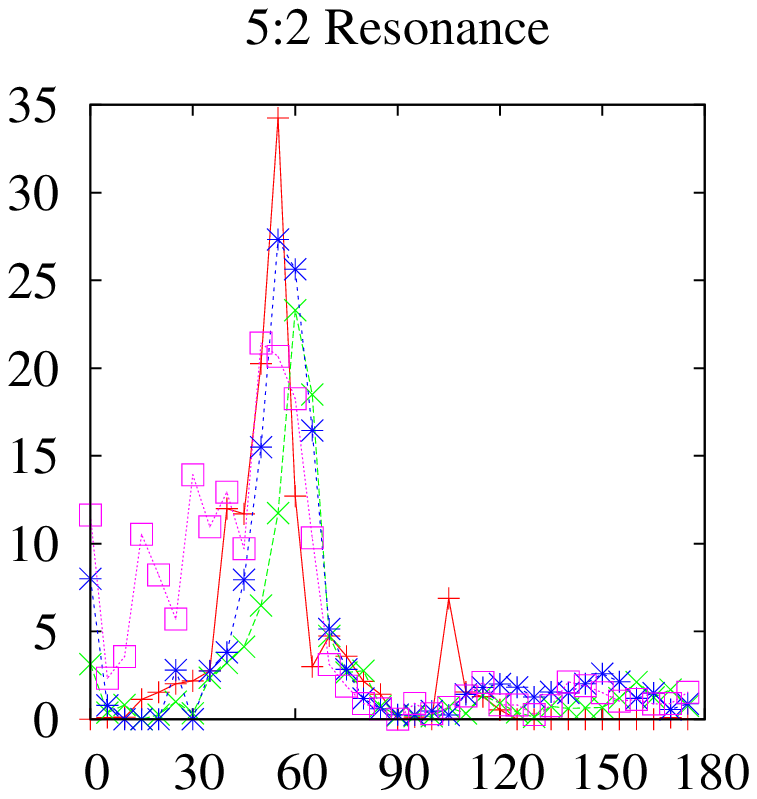}\hspace{-23mm}
\includegraphics[width=60mm]{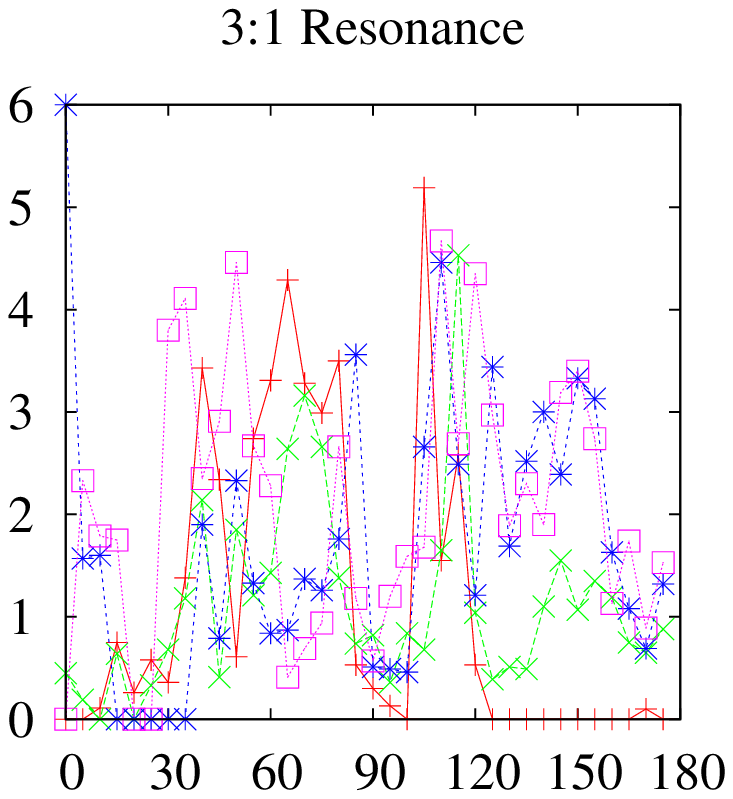}\hspace{-23mm}
\includegraphics[width=60mm]{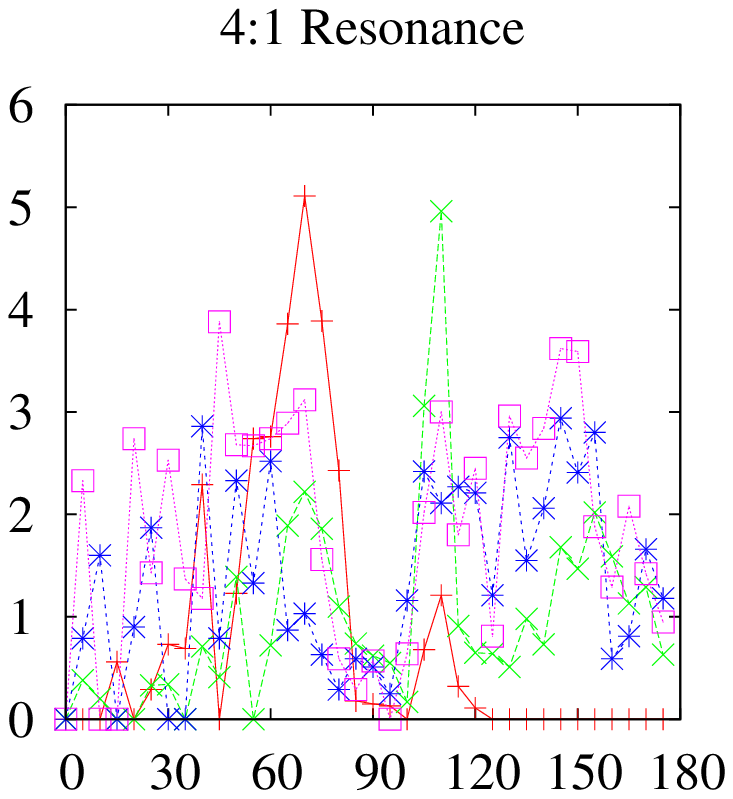}
\caption{Capture fraction for coorbital and inner resonances in order of encounter by the planet as a function of initial inclination. The planet has a circular orbit. 
Four curves are plotted  corresponding to the four eccentricity standard deviations. Red line with cross sign for $\sigma_e=0.01$,  green line with times sign for $\sigma_e=0.10$, blue line with star sign for $\sigma_e=0.30$ and purple line with a box sign for $\sigma_e=0.70$. {The capture fraction is defined in the previous figure caption.}}
\end{center}
\end{figure*}

\begin{figure*}
\begin{center}
\includegraphics[width=58mm]{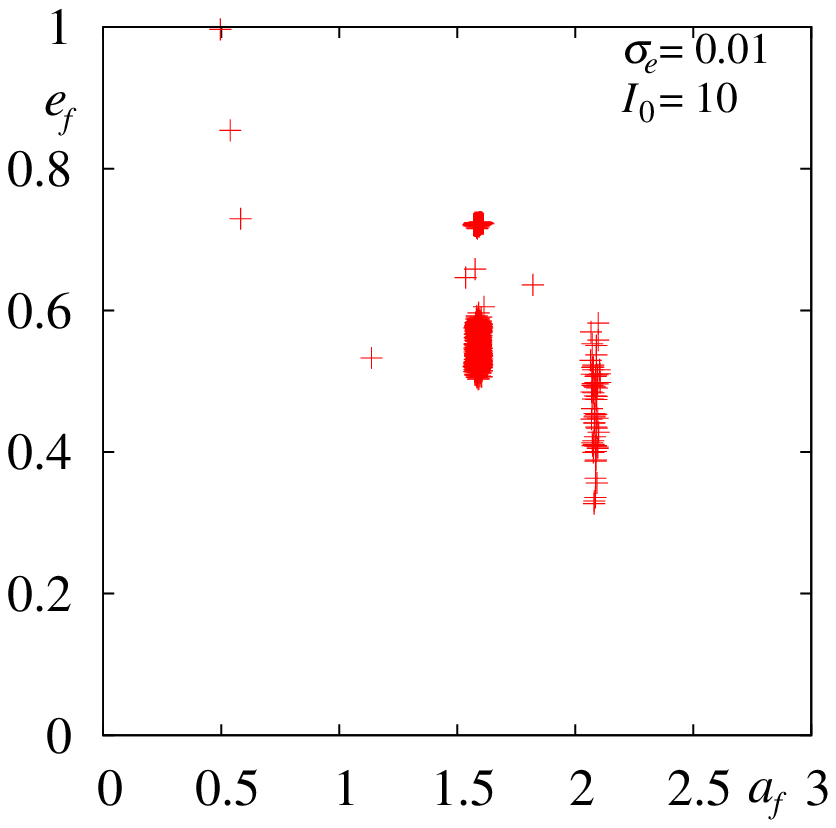}\hspace{-20mm}
\includegraphics[width=58mm]{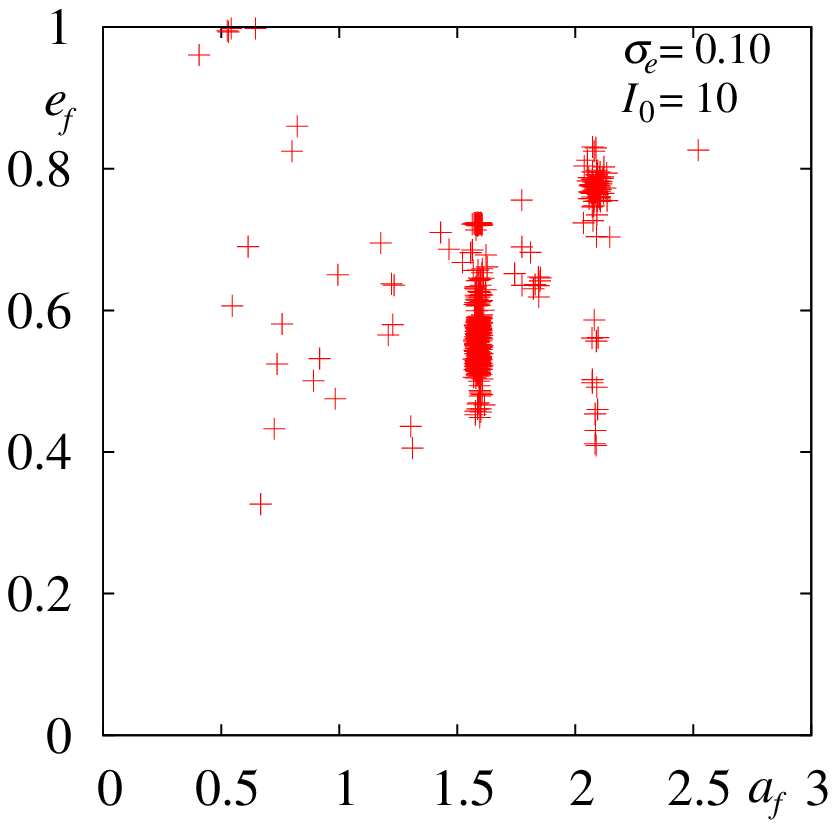}\hspace{-20mm}
\includegraphics[width=58mm]{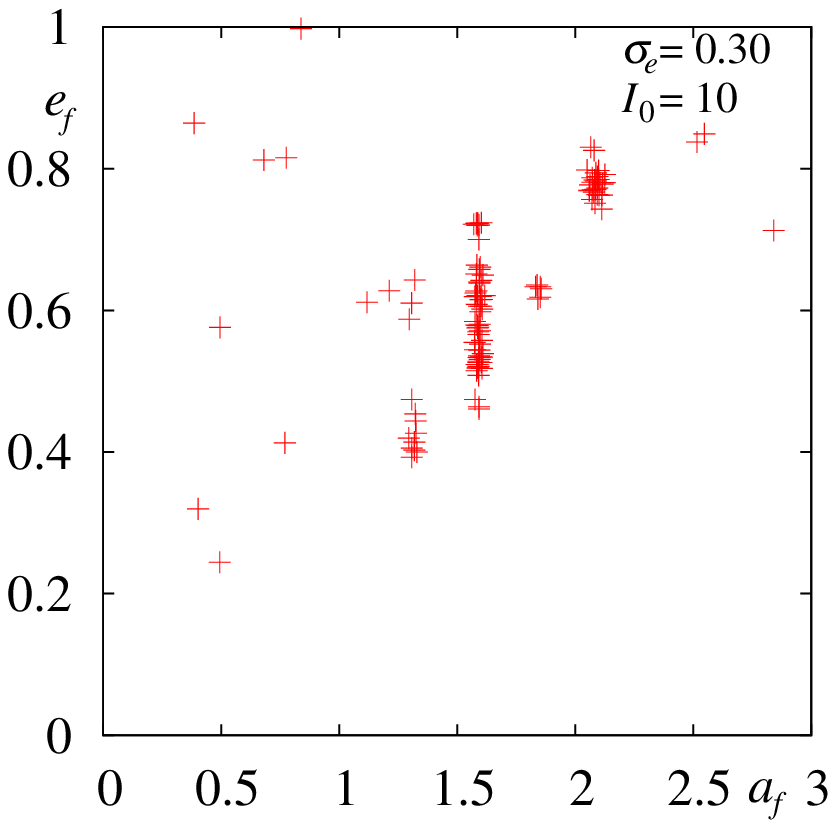}\hspace{-20mm}
\includegraphics[width=58mm]{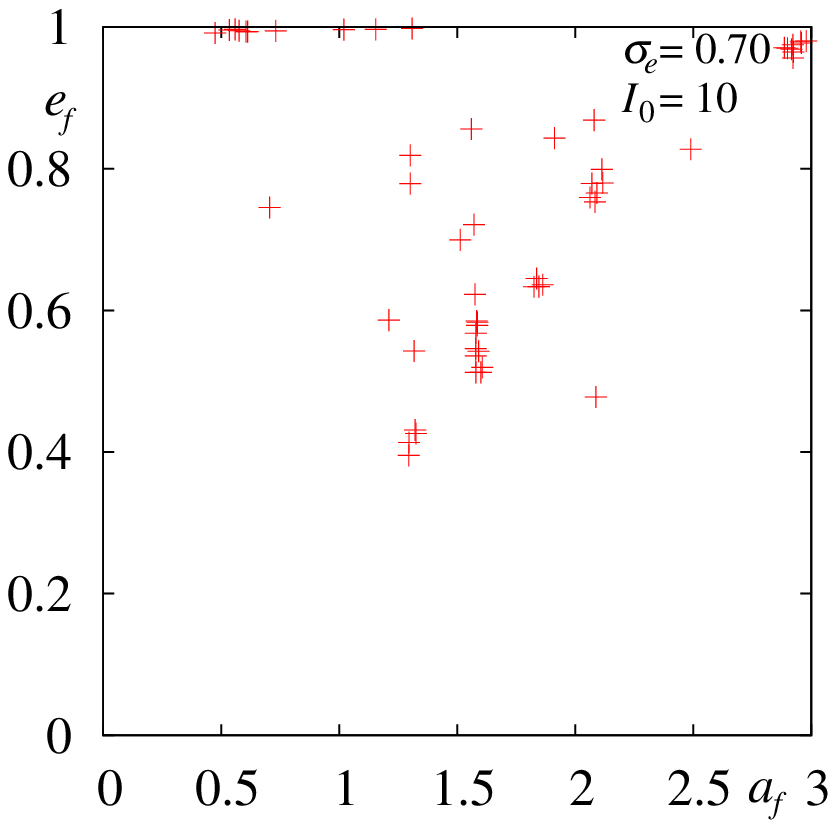}\\
\includegraphics[width=58mm]{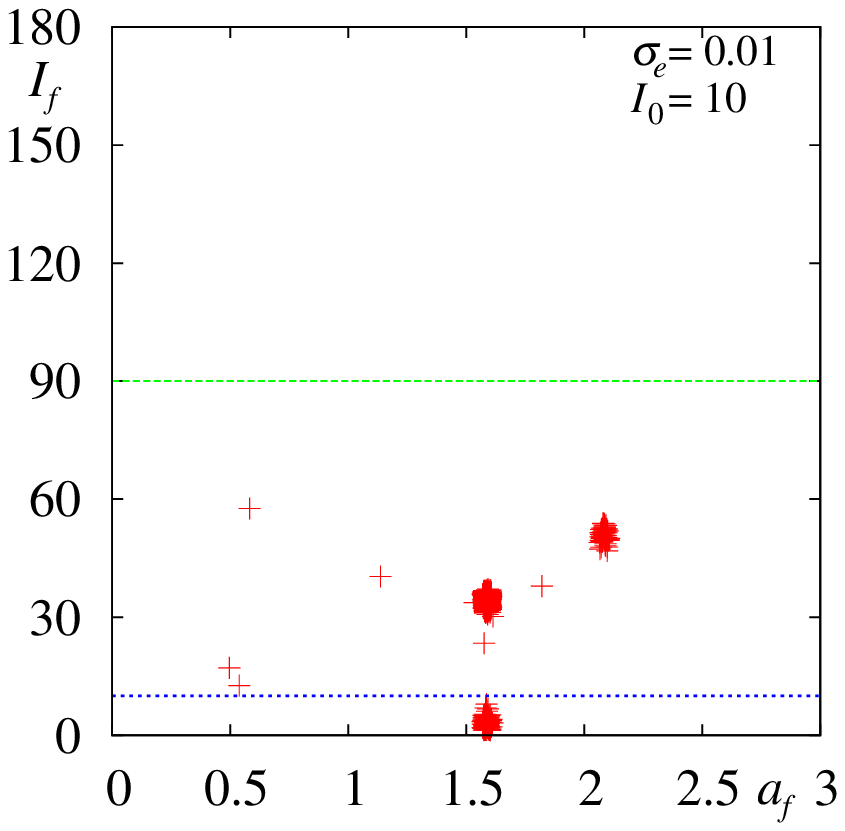}\hspace{-20mm}
\includegraphics[width=58mm]{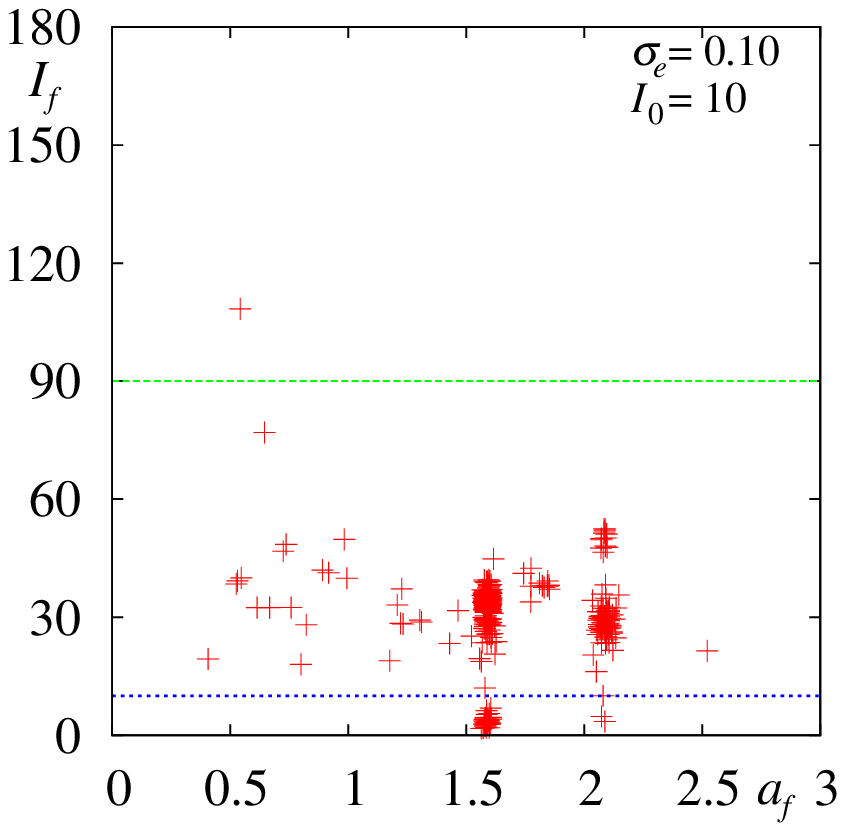}\hspace{-20mm}
\includegraphics[width=58mm]{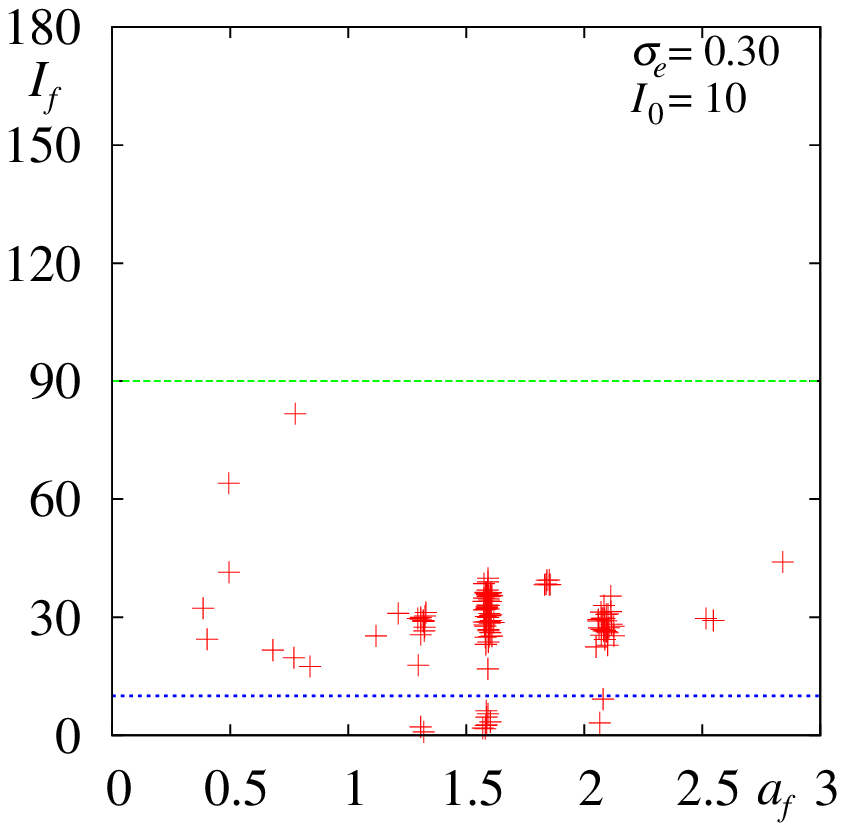}\hspace{-20mm}
\includegraphics[width=58mm]{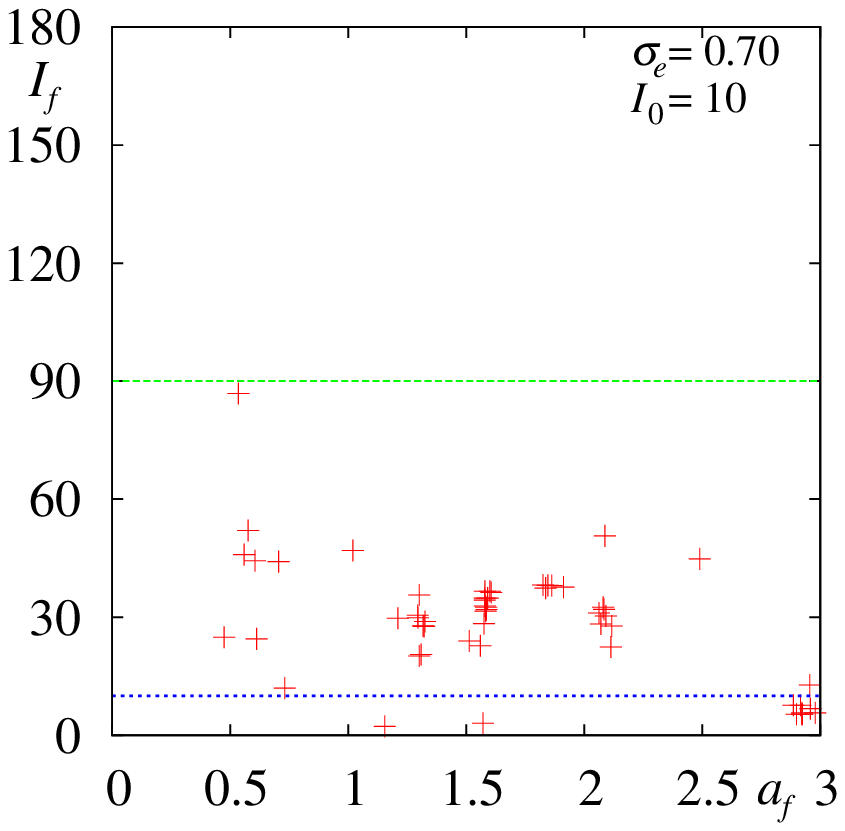}\\
\includegraphics[width=58mm]{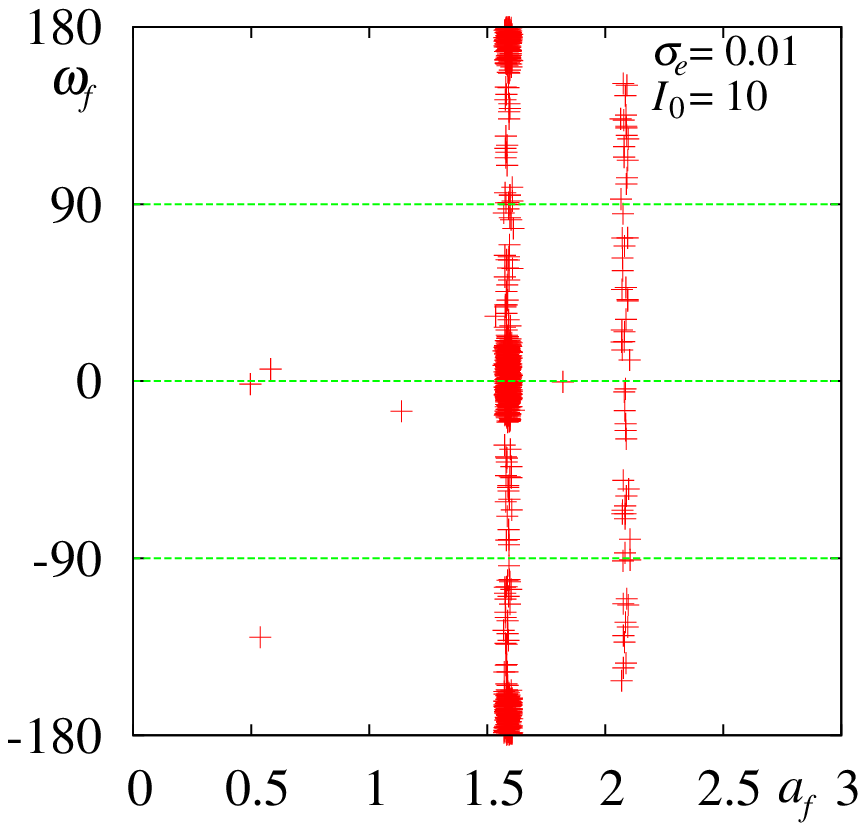}\hspace{-20mm}
\includegraphics[width=58mm]{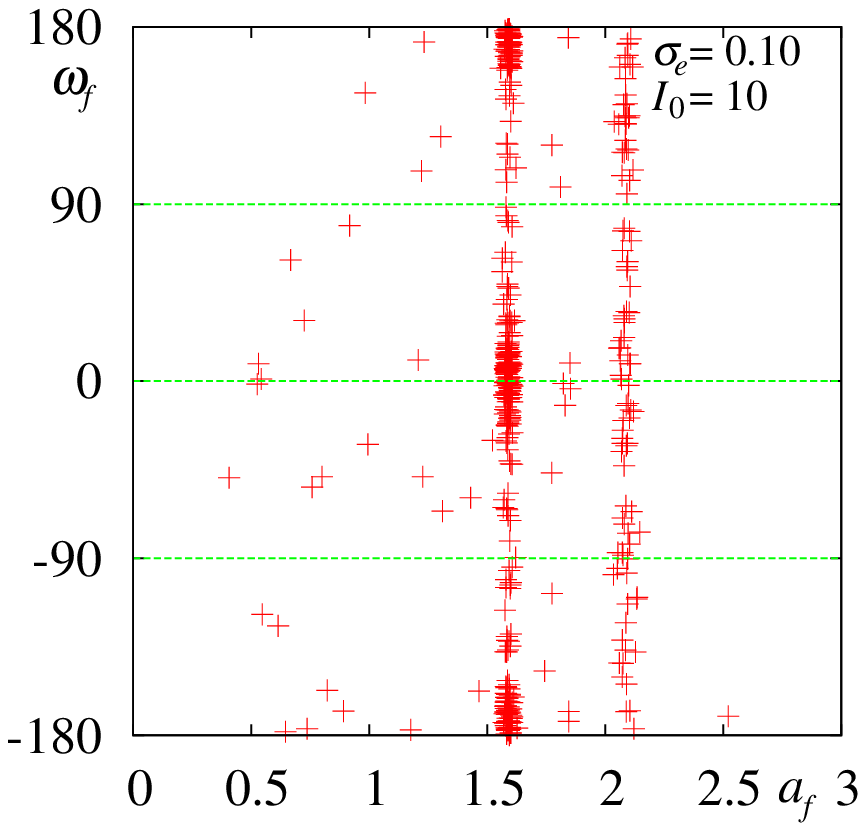}\hspace{-20mm}
\includegraphics[width=58mm]{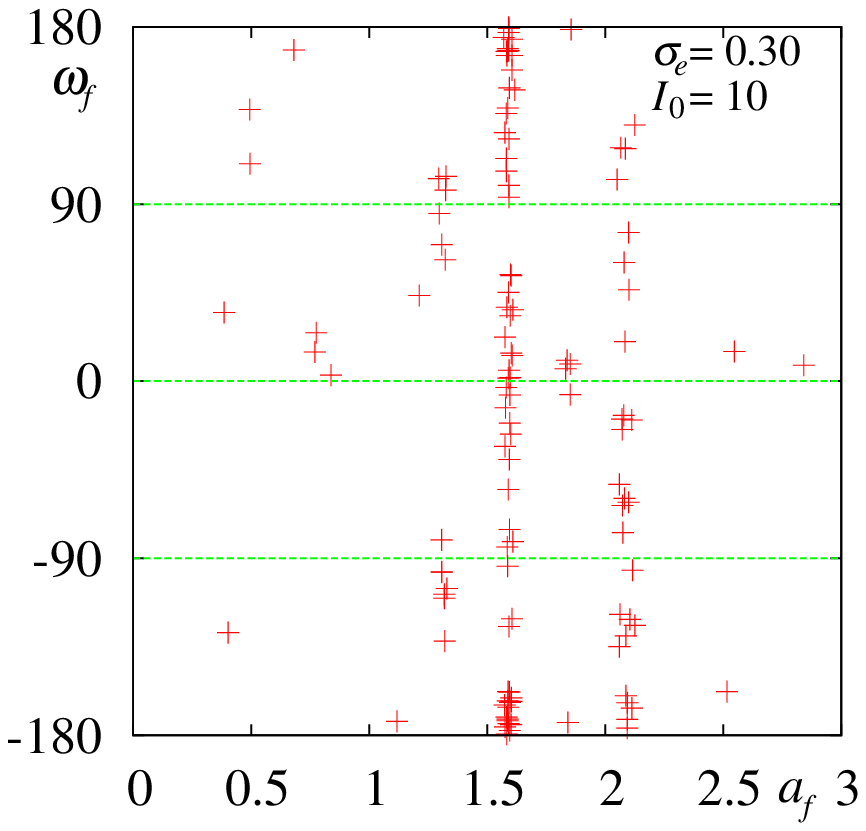}\hspace{-20mm}
\includegraphics[width=58mm]{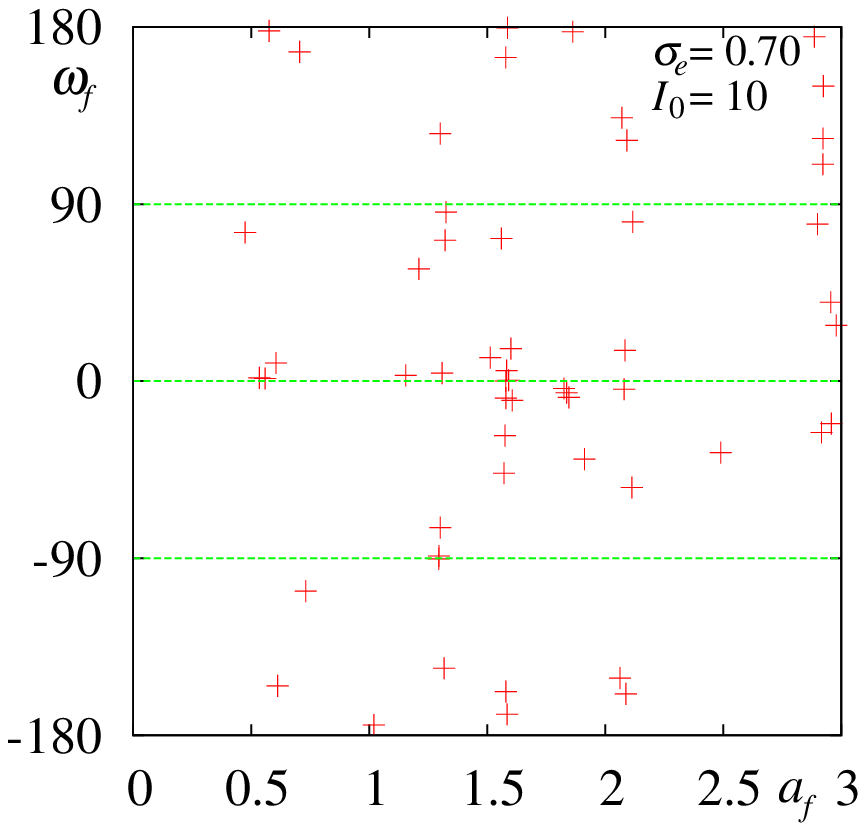}
\caption{Final eccentricity (upper row), inclination (middle row) and argument of pericenter (bottom row) as  functions of the final semi-major axis for an initial inclination $I_0=10^\circ$ and the four initial eccentricity standard deviations, $\sigma_e=0.01, \ 0.10,\ 0.30$ and $0.70$. In the inclination plots, the green dashed line indicates polar orbits and the blue dotted line, the initial inclination. In the argument of pericenter plots, the green dashed lines at $0$ and $\pm 90^\circ$ indicate the possible locations of the Kozai-Lidov resonances.}
\end{center}
\end{figure*}

\begin{figure*}
\begin{center}
\includegraphics[width=58mm]{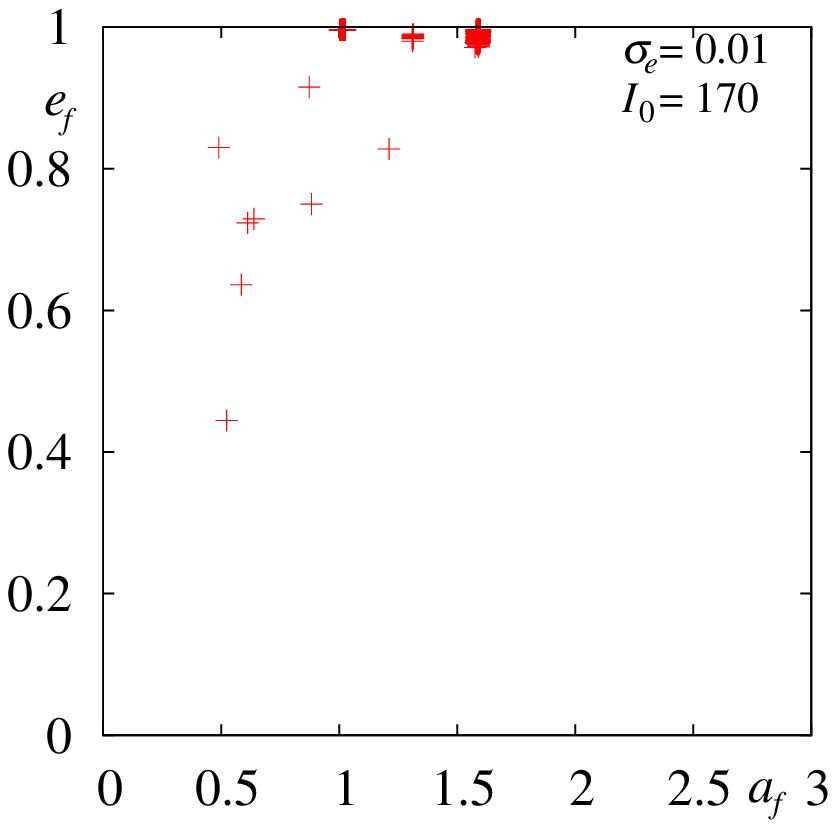}\hspace{-20mm}
\includegraphics[width=58mm]{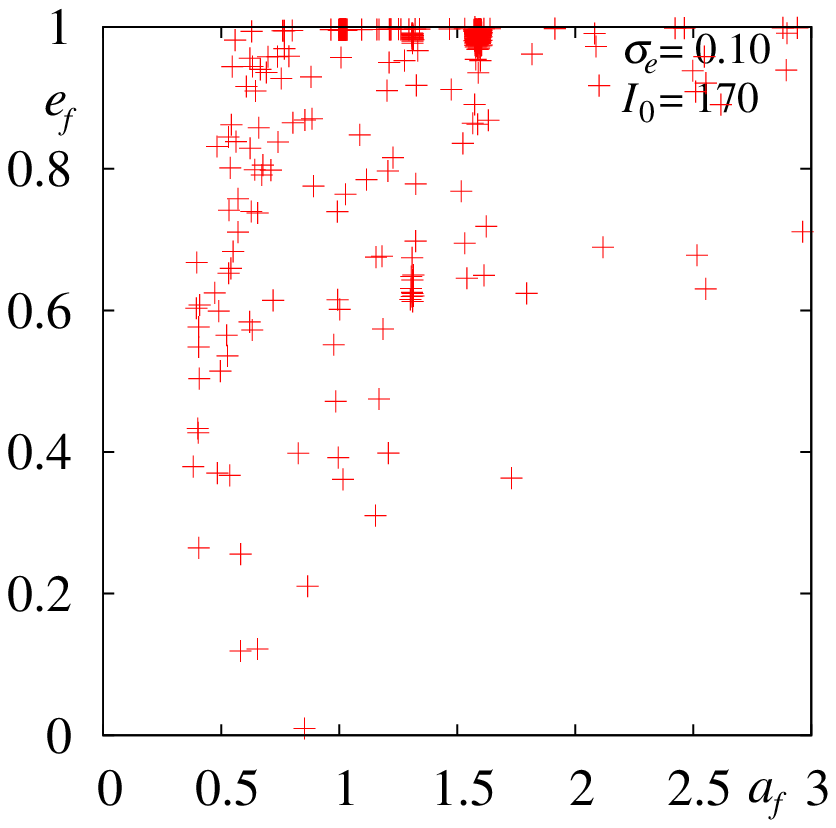}\hspace{-20mm}
\includegraphics[width=58mm]{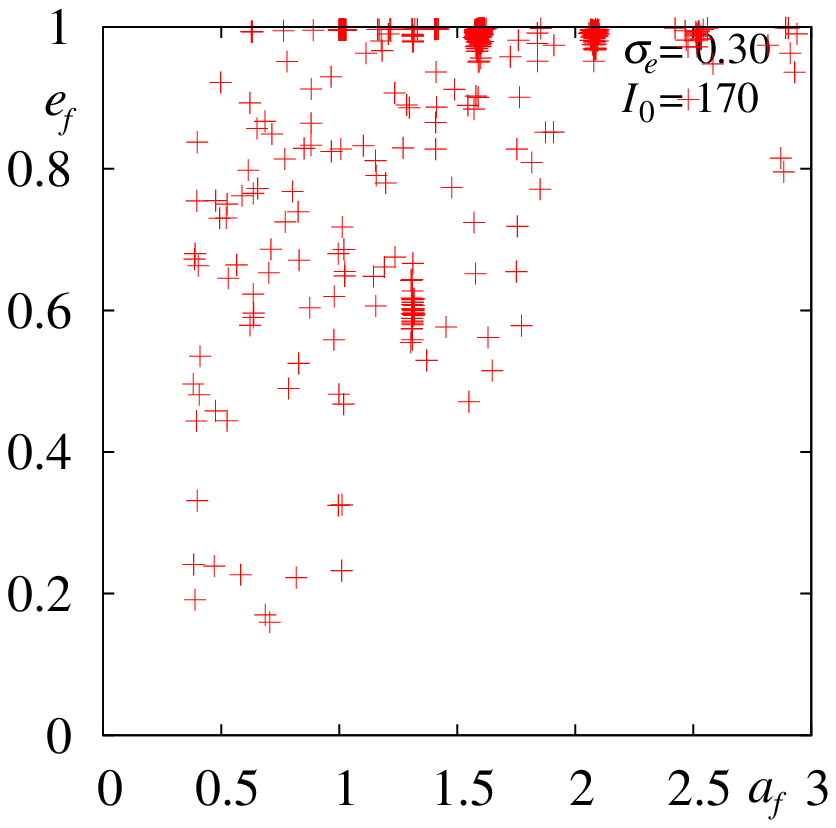}\hspace{-20mm}
\includegraphics[width=58mm]{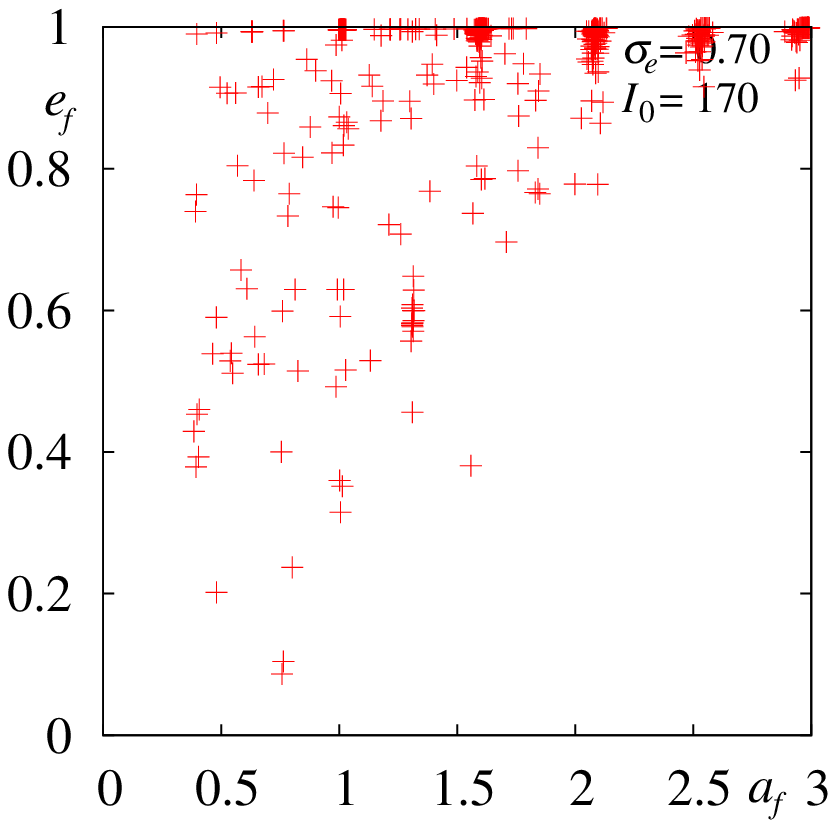}\\
\includegraphics[width=58mm]{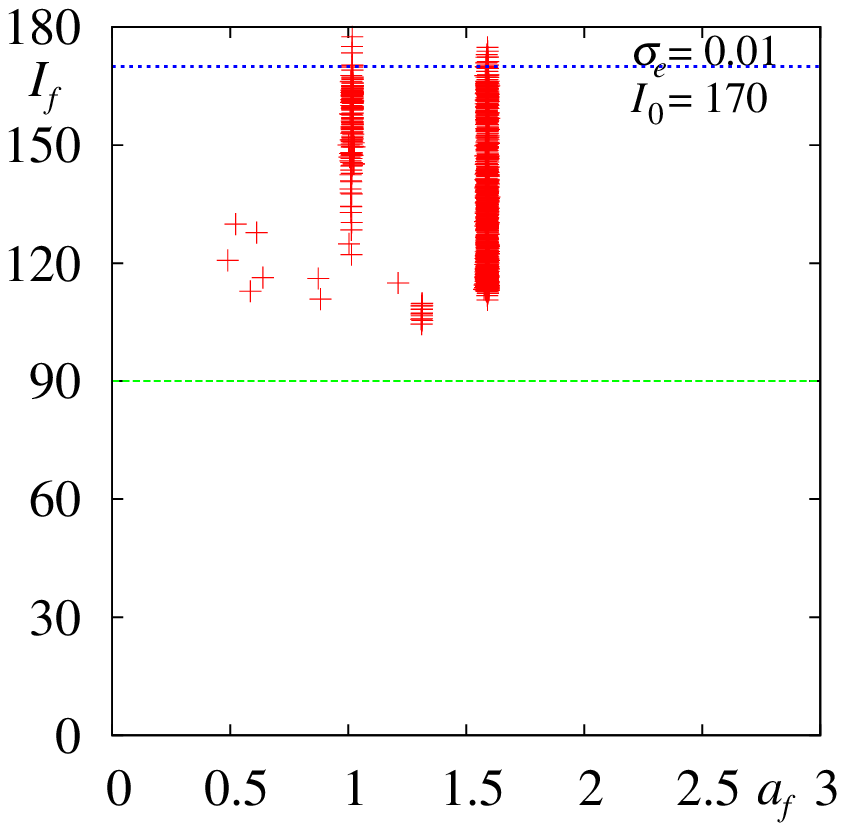}\hspace{-20mm}
\includegraphics[width=58mm]{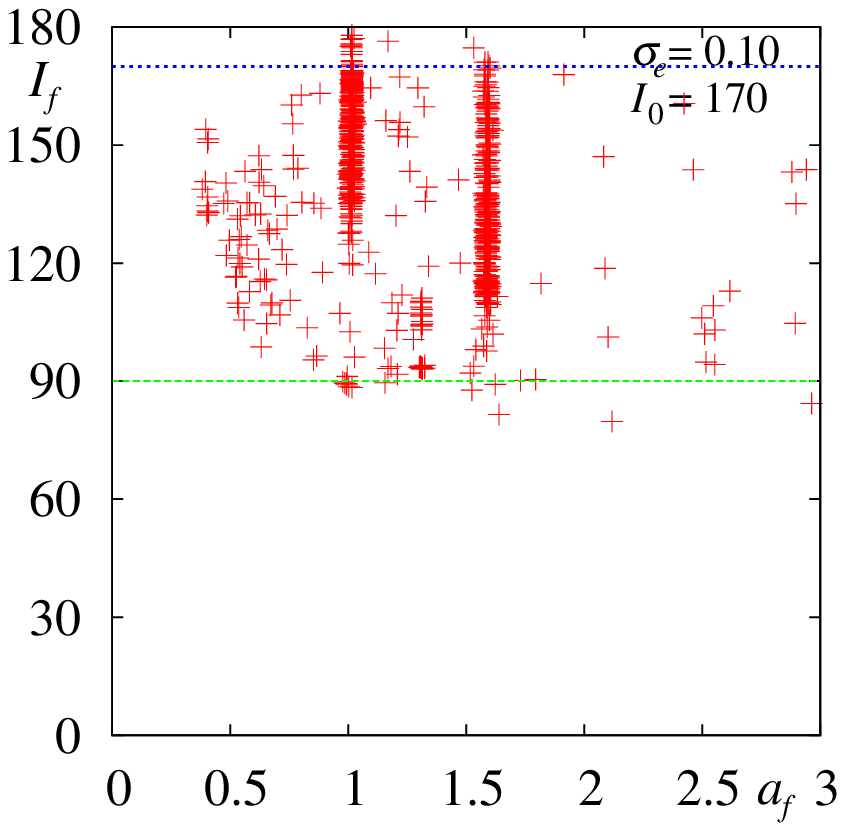}\hspace{-20mm}
\includegraphics[width=58mm]{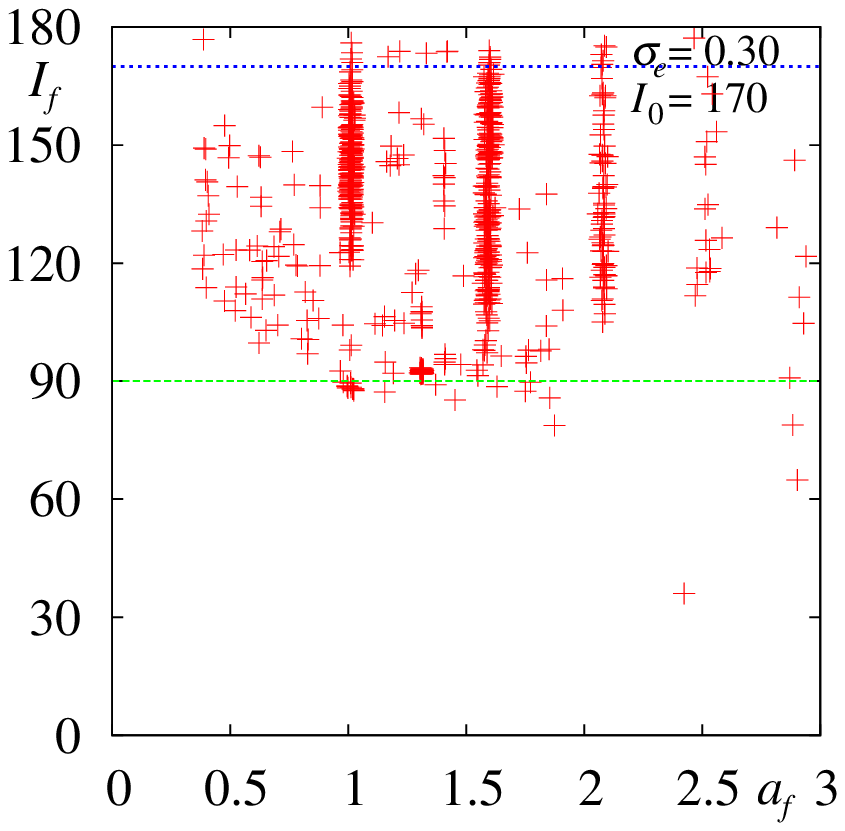}\hspace{-20mm}
\includegraphics[width=58mm]{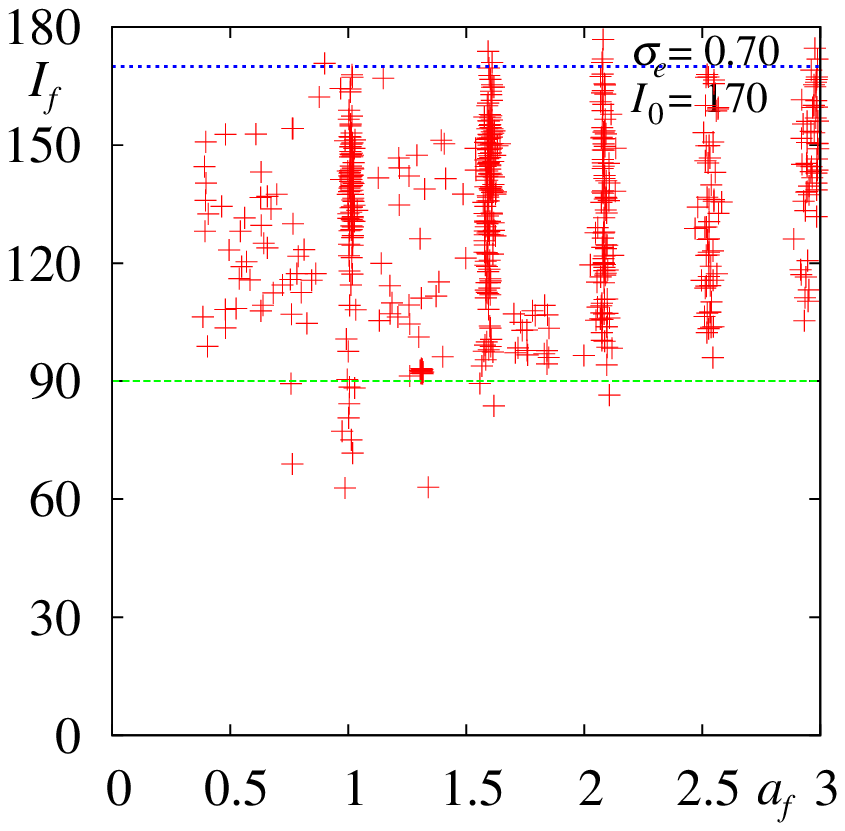}\\
\includegraphics[width=58mm]{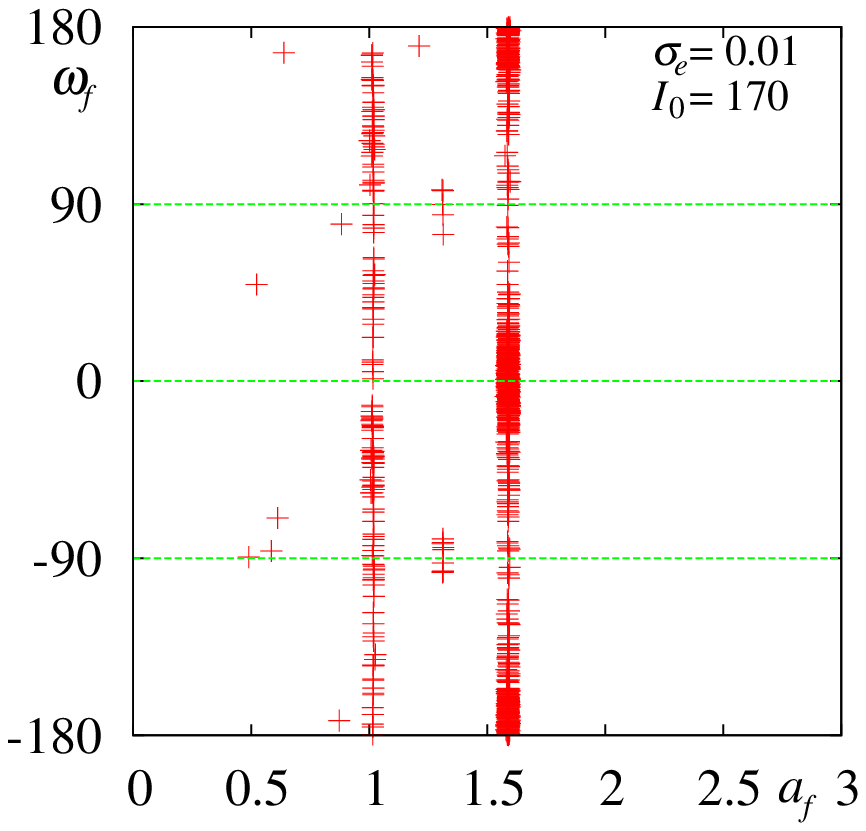}\hspace{-20mm}
\includegraphics[width=58mm]{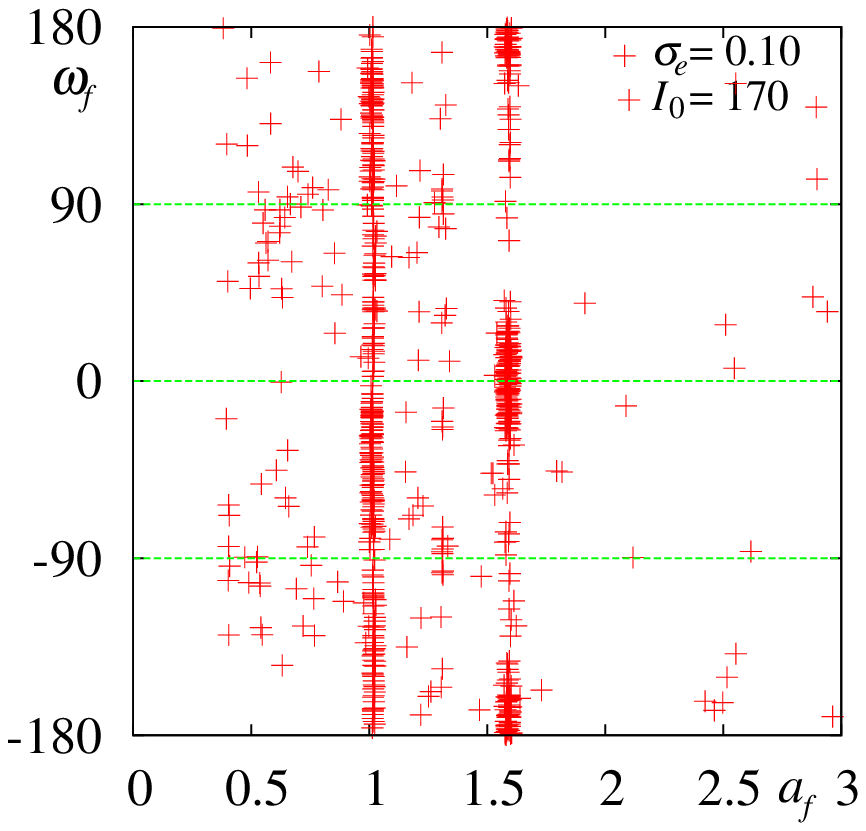}\hspace{-20mm}
\includegraphics[width=58mm]{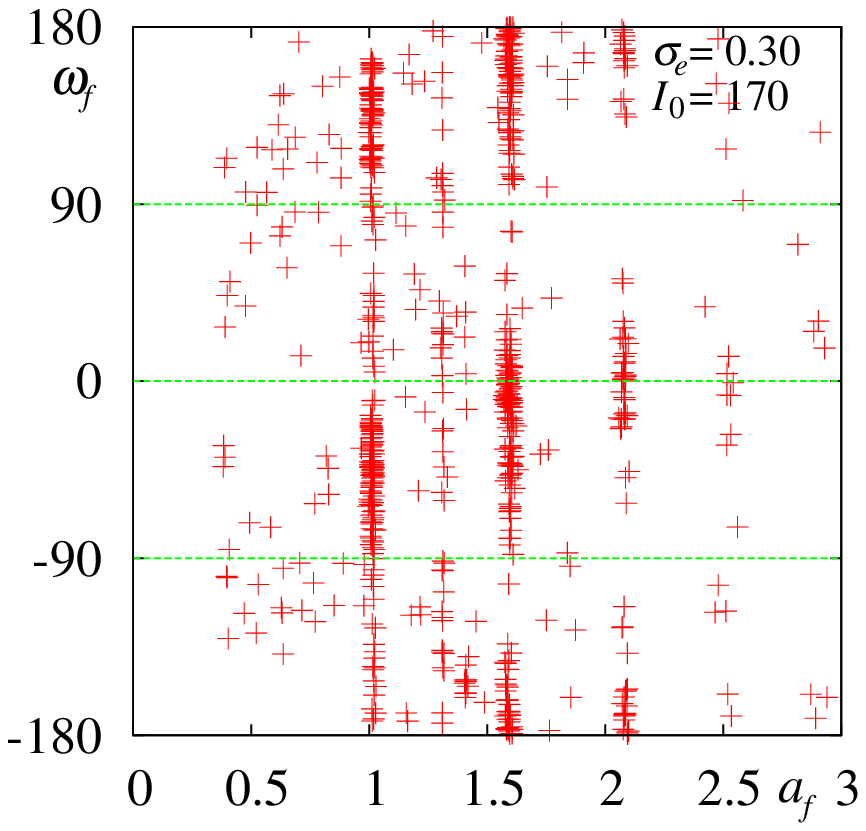}\hspace{-20mm}
\includegraphics[width=58mm]{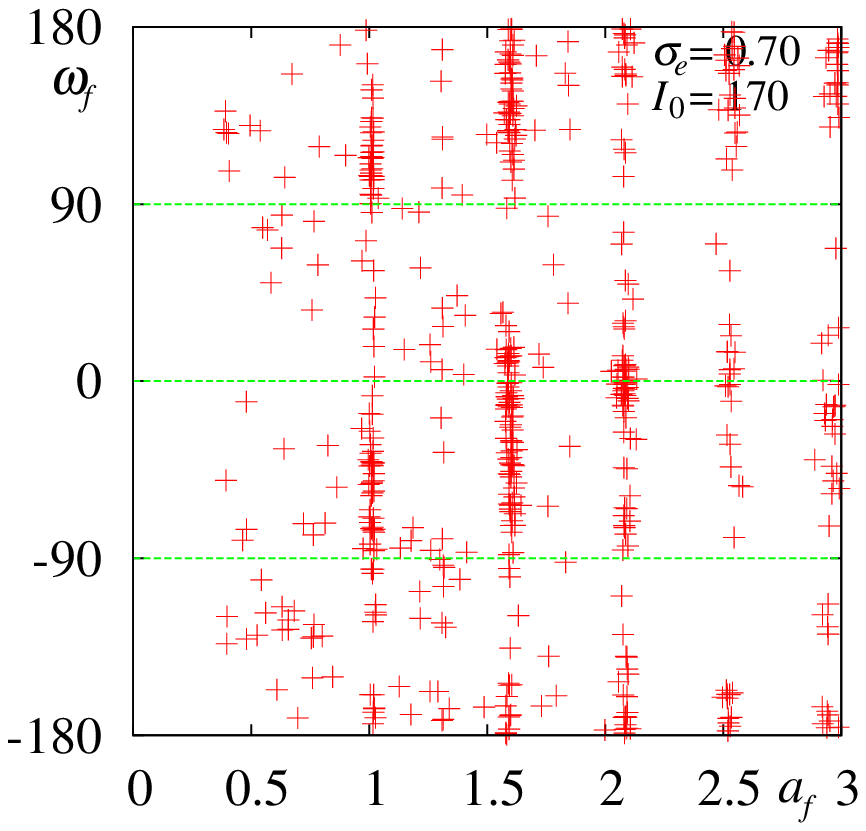}
\caption{Final eccentricity (upper row), inclination (middle row) and argument of pericenter (bottom row) as  functions of the final semi-major axis for an initial inclination $I_0=170^\circ$ and the four initial eccentricity standard deviations, $\sigma_e=0.01, \ 0.10,\ 0.30$ and $0.70$. In the inclination plots, the green dashed line indicates polar orbits and the blue dotted line, the initial inclination. In the argument of pericenter plots, the green dashed lines at $0$ and $\pm 90^\circ$ indicate the possible locations of the Kozai-Lidov resonances.}
\end{center}
\end{figure*}

\begin{figure*}
\begin{center}
\includegraphics[width=58mm]{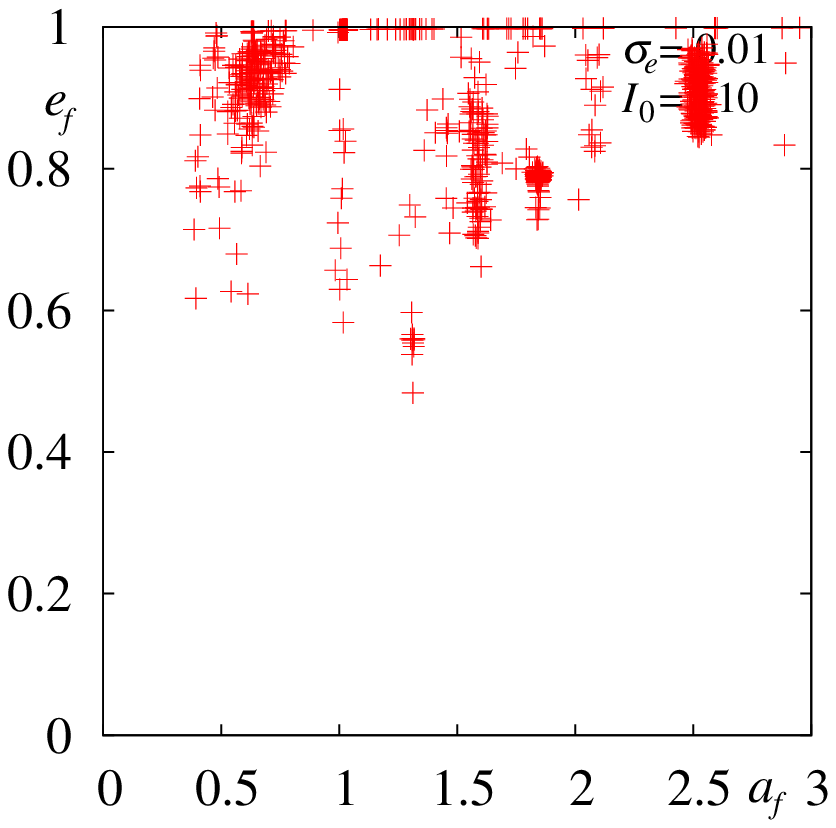}\hspace{-20mm}
\includegraphics[width=58mm]{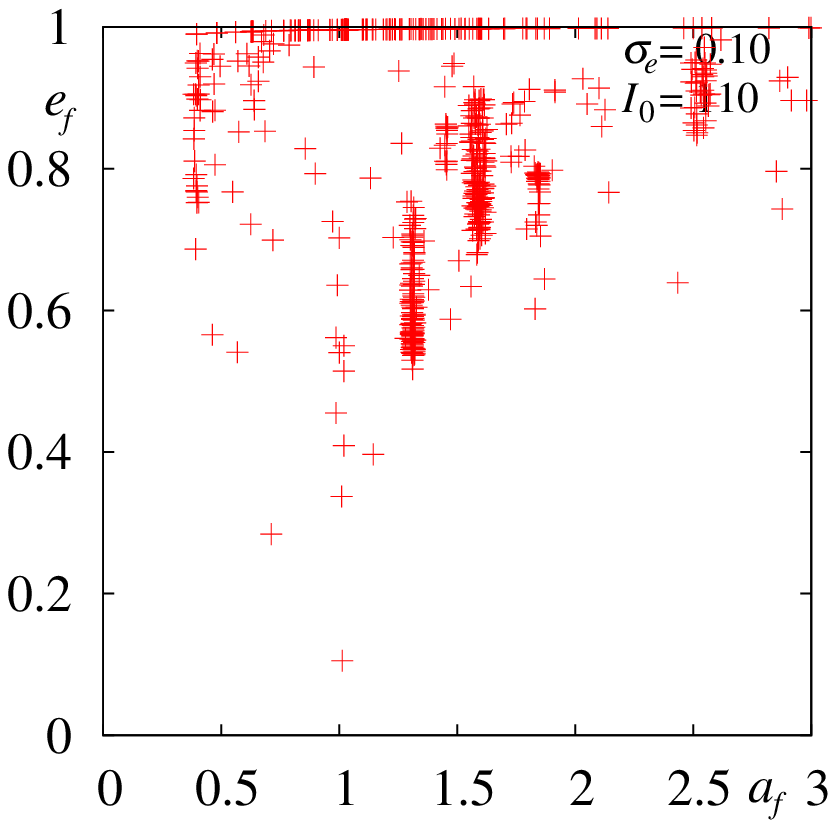}\hspace{-20mm}
\includegraphics[width=58mm]{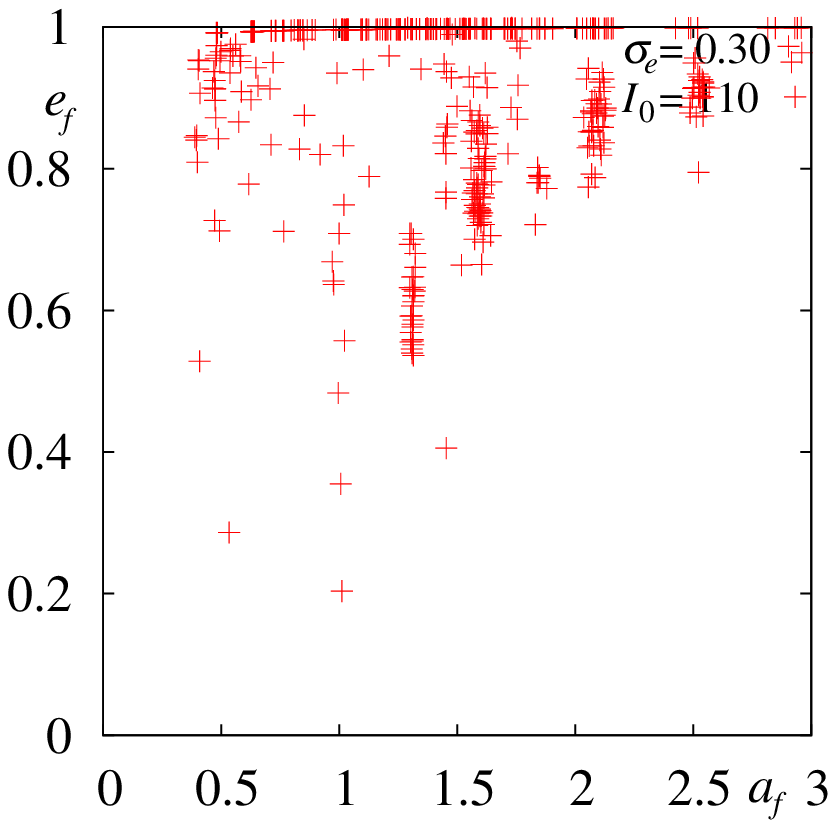}\hspace{-20mm}
\includegraphics[width=58mm]{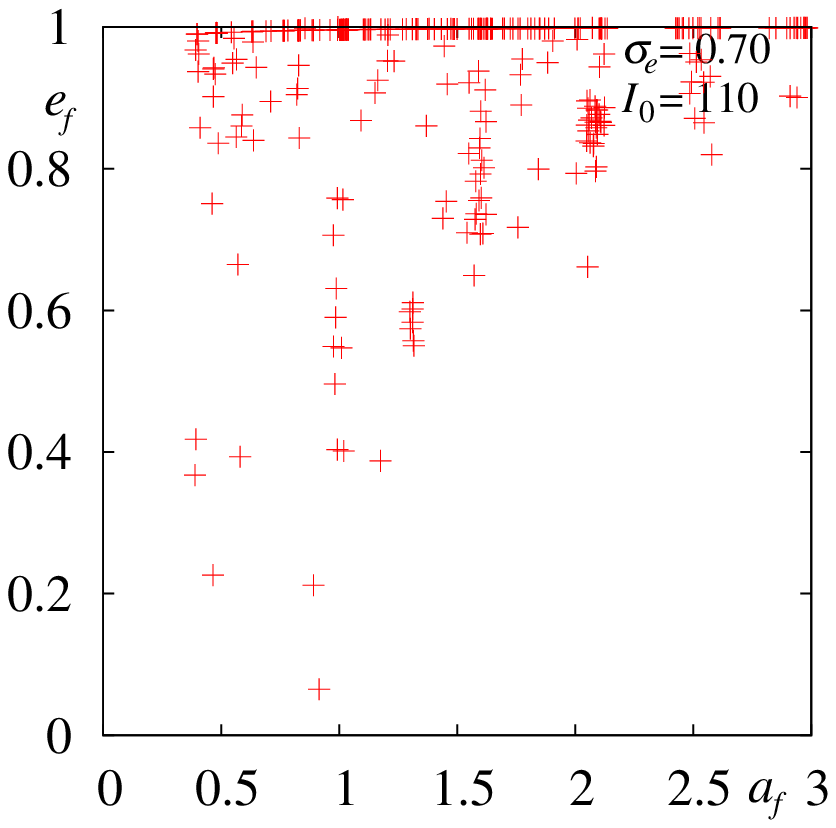}\\
\includegraphics[width=58mm]{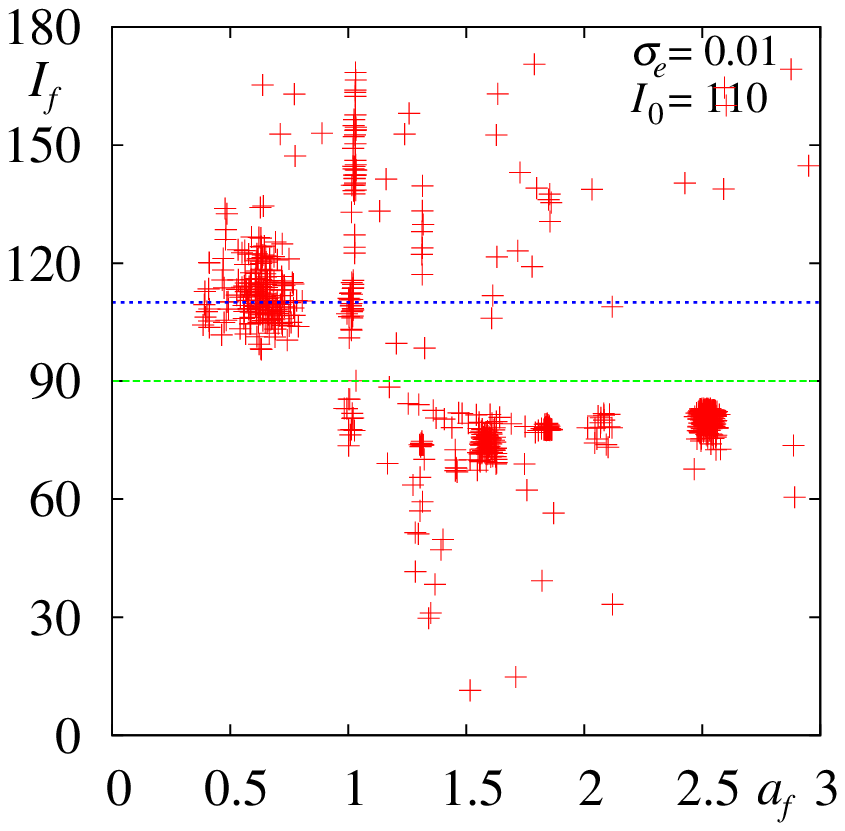}\hspace{-20mm}
\includegraphics[width=58mm]{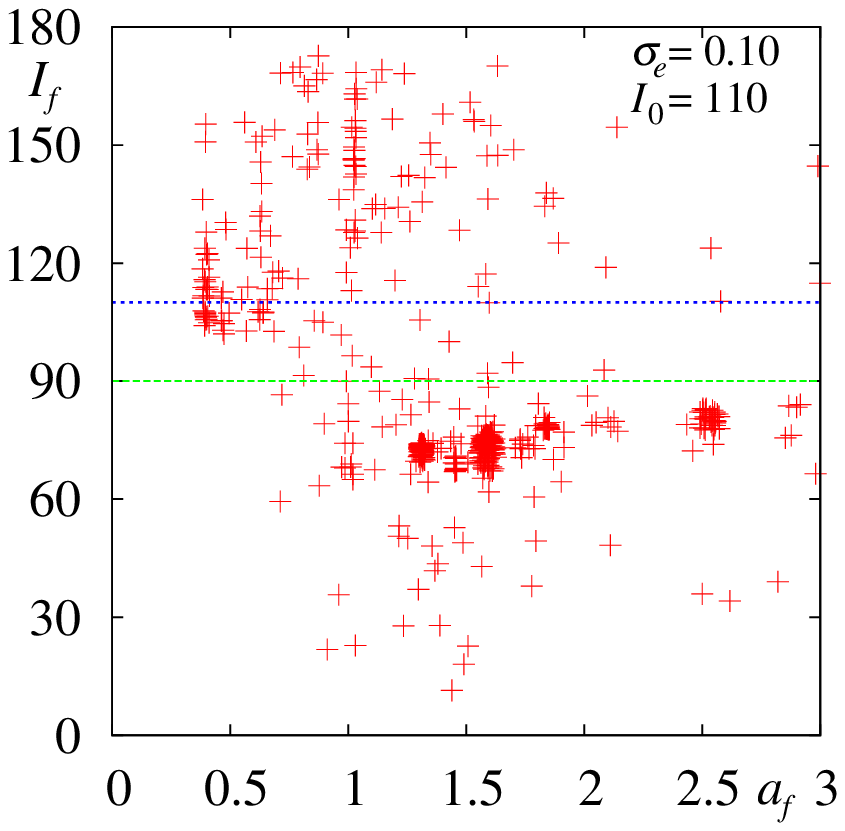}\hspace{-20mm}
\includegraphics[width=58mm]{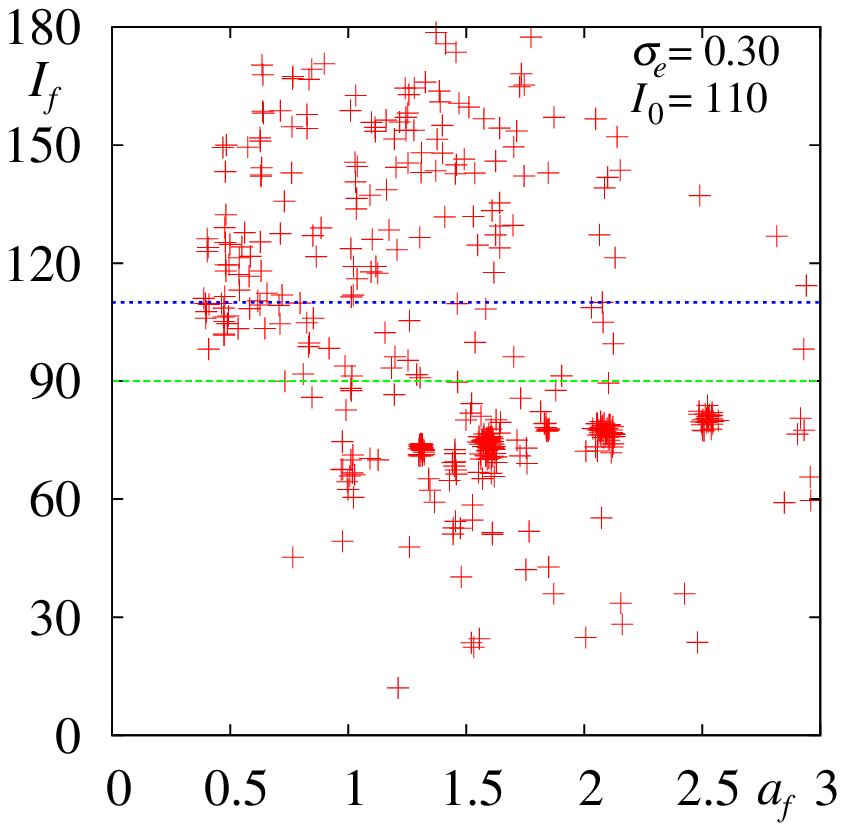}\hspace{-20mm}
\includegraphics[width=58mm]{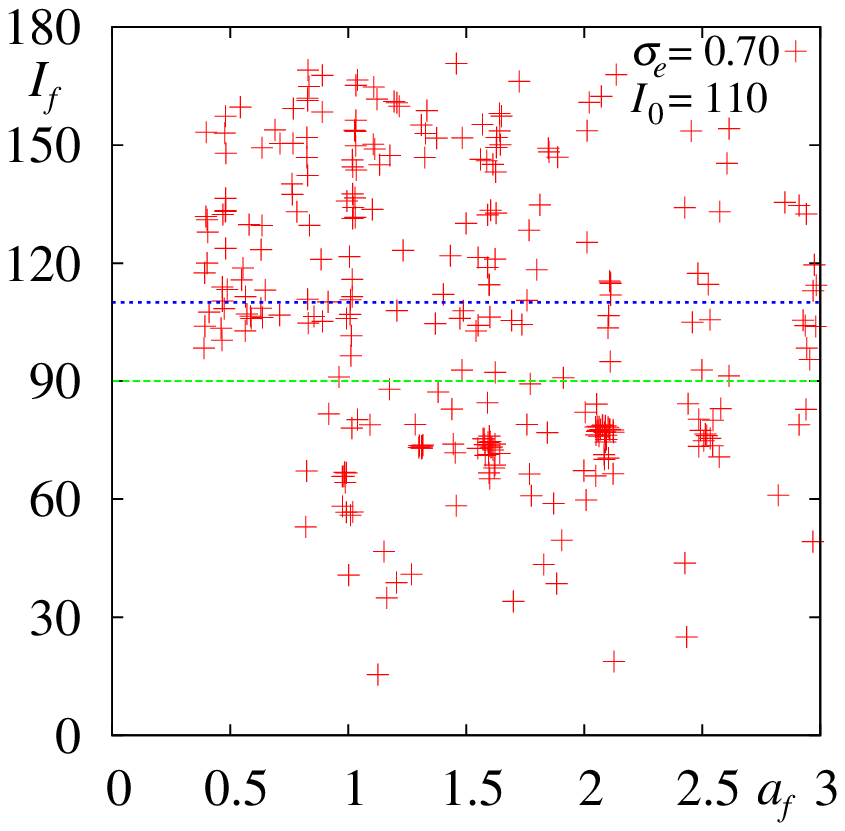}\\
\includegraphics[width=58mm]{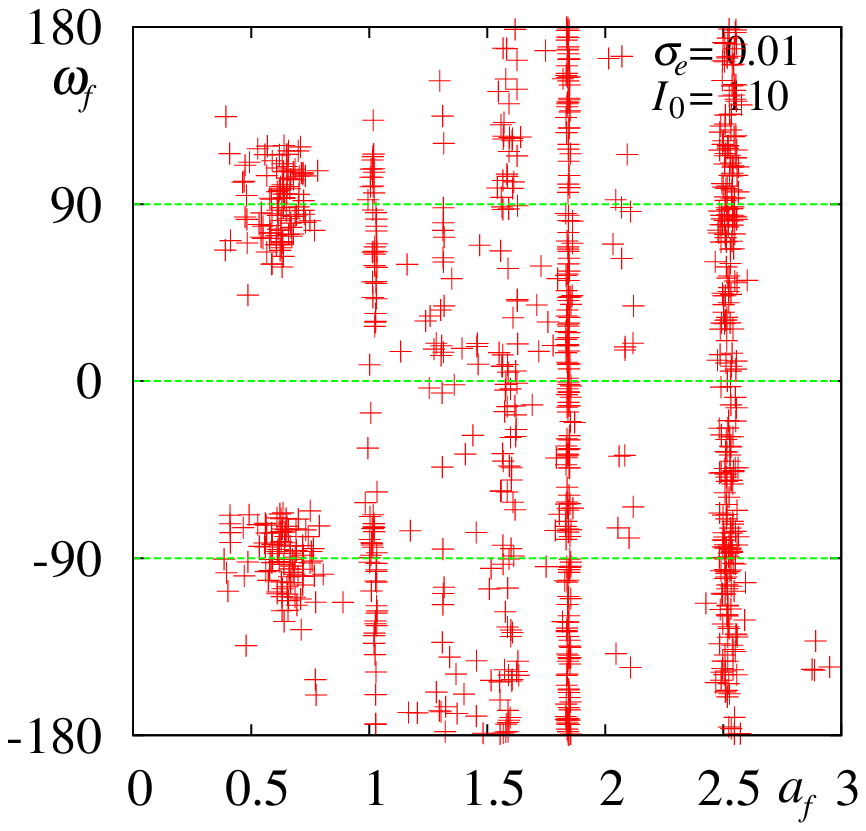}\hspace{-20mm}
\includegraphics[width=58mm]{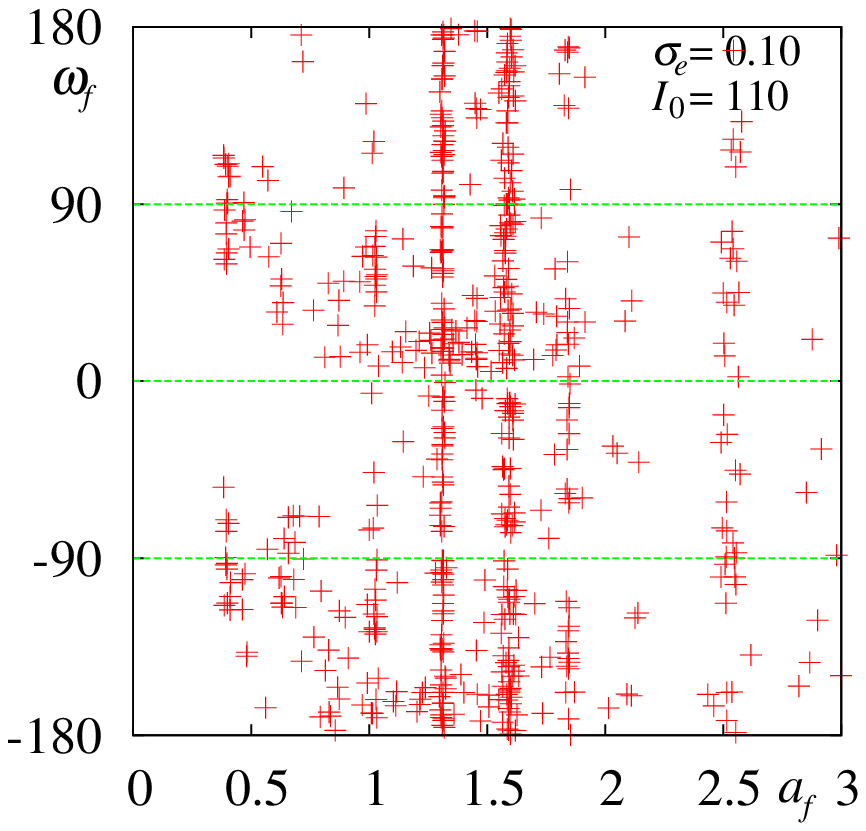}\hspace{-20mm}
\includegraphics[width=58mm]{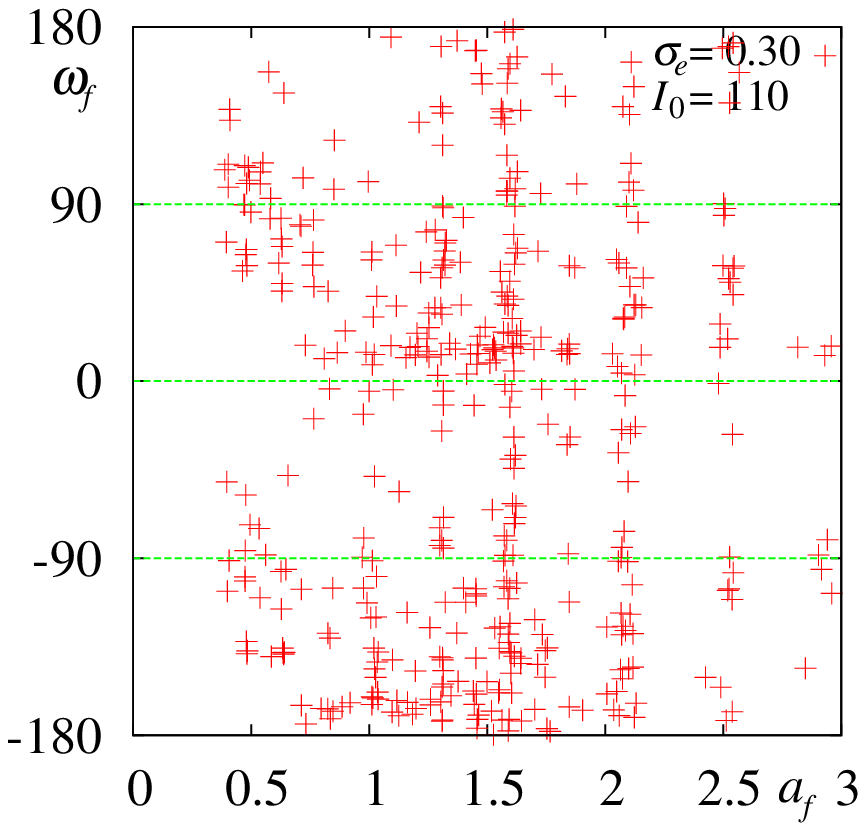}\hspace{-20mm}
\includegraphics[width=58mm]{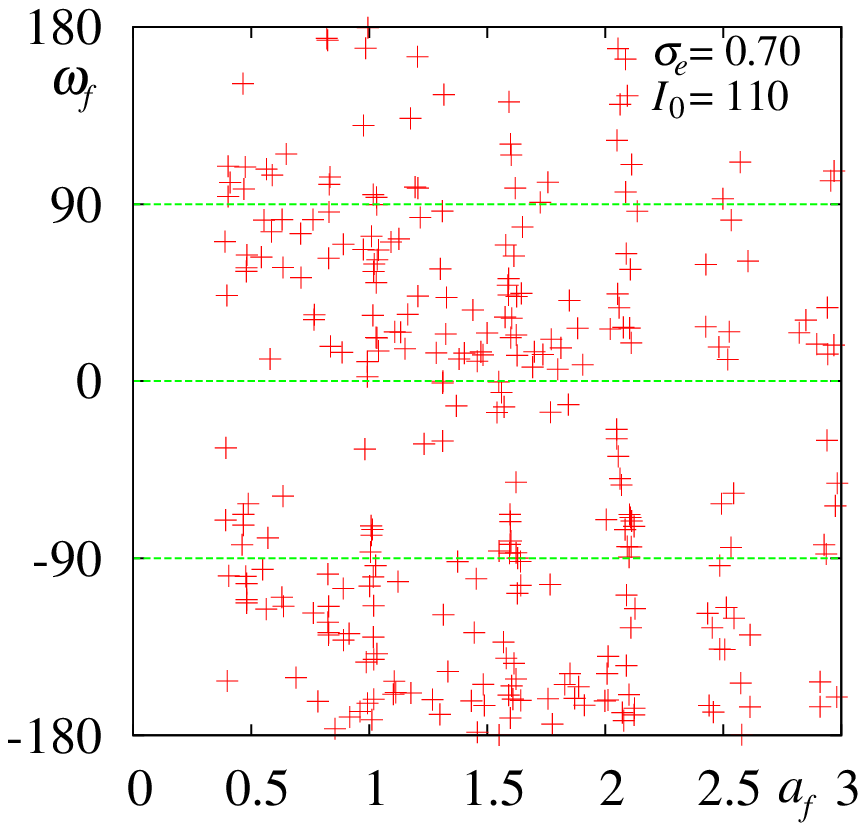}
\caption{Final eccentricity (upper row), inclination (middle row) and argument of pericenter (bottom row) as  functions of the final semi-major axis for an initial inclination $I_0=110^\circ$ and the four initial eccentricity standard deviations, $\sigma_e=0.01, \ 0.10,\ 0.30$ and $0.70$. In the inclination plots, the green dashed line indicates polar orbits and the blue dotted line, the initial inclination. In the argument of pericenter plots, the green dashed lines at $0$ and $\pm 90^\circ$ indicate the possible locations of the Kozai-Lidov resonances.}
\end{center}
\end{figure*}

\begin{figure*}
\begin{center}
\includegraphics[width=60mm]{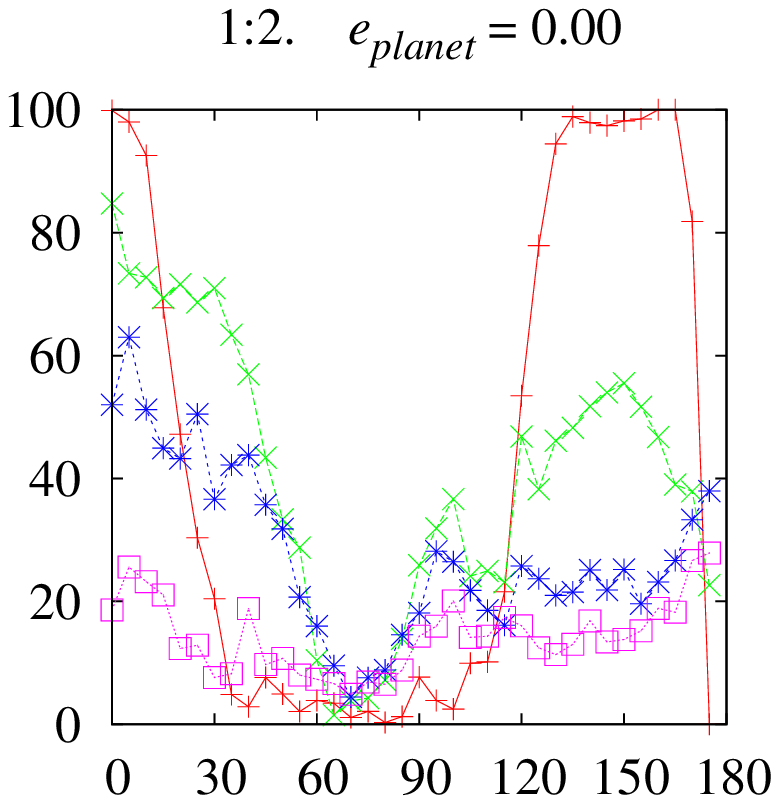}\hspace{-23mm}
\includegraphics[width=60mm]{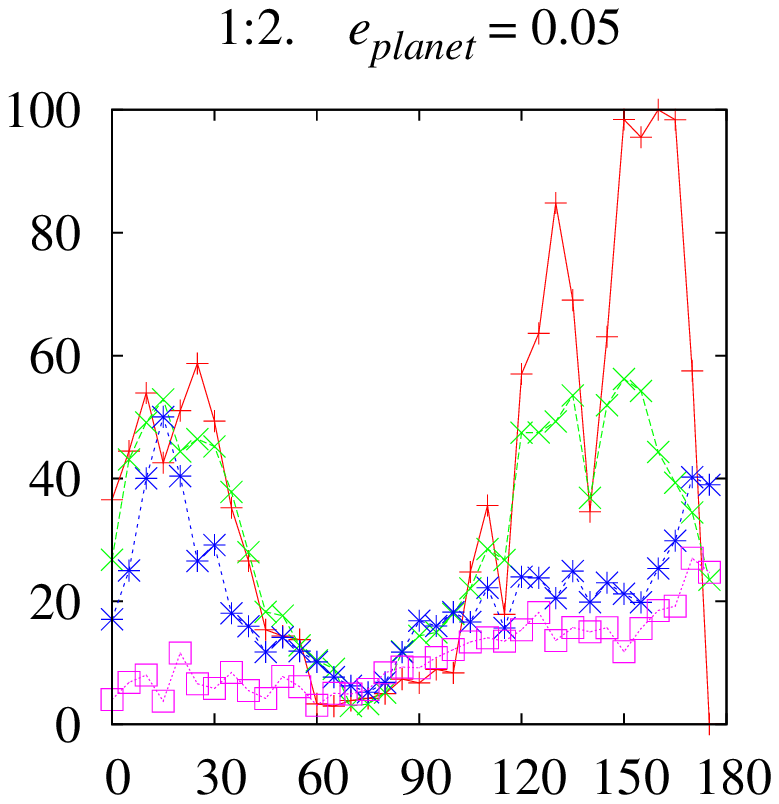}\hspace{-23mm}
\includegraphics[width=60mm]{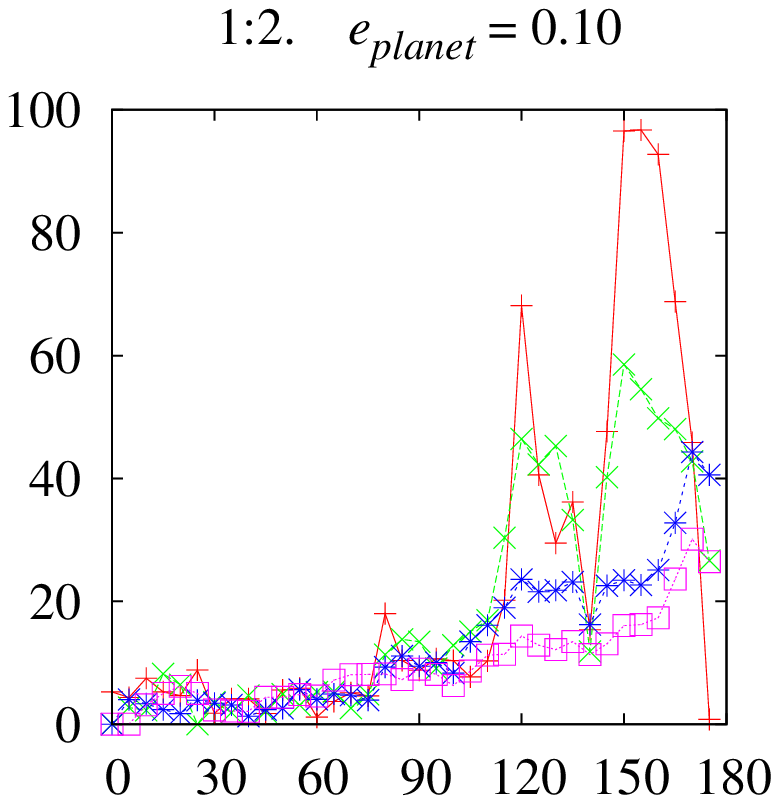}\hspace{-23mm}
\includegraphics[width=60mm]{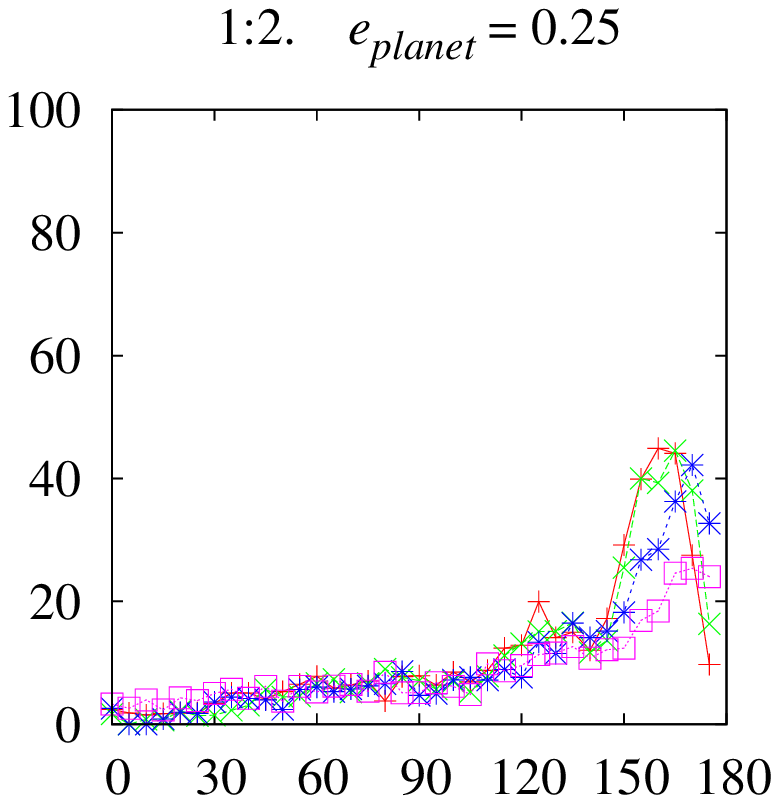}\\
\includegraphics[width=60mm]{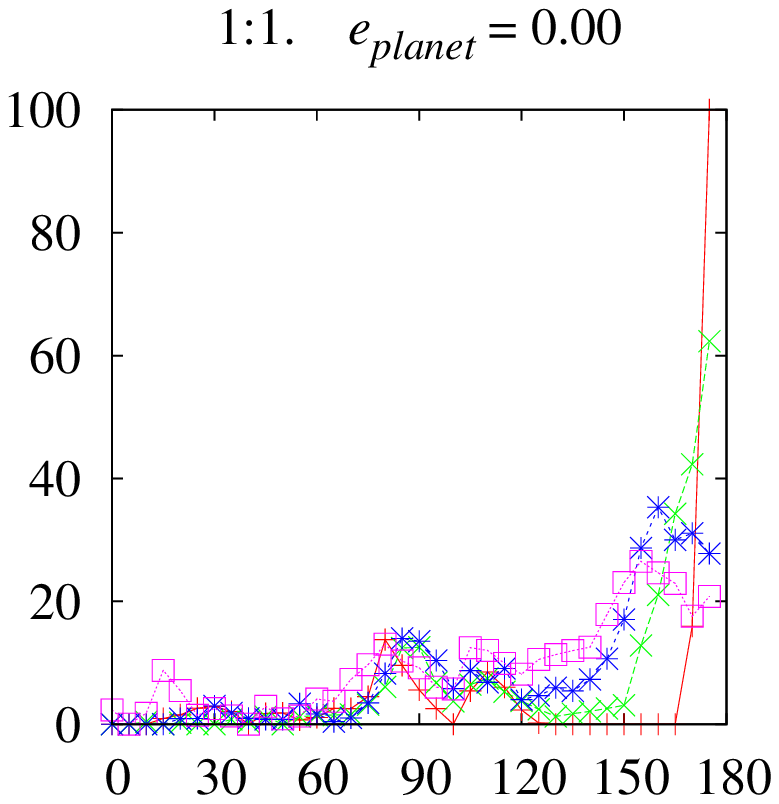}\hspace{-23mm}
\includegraphics[width=60mm]{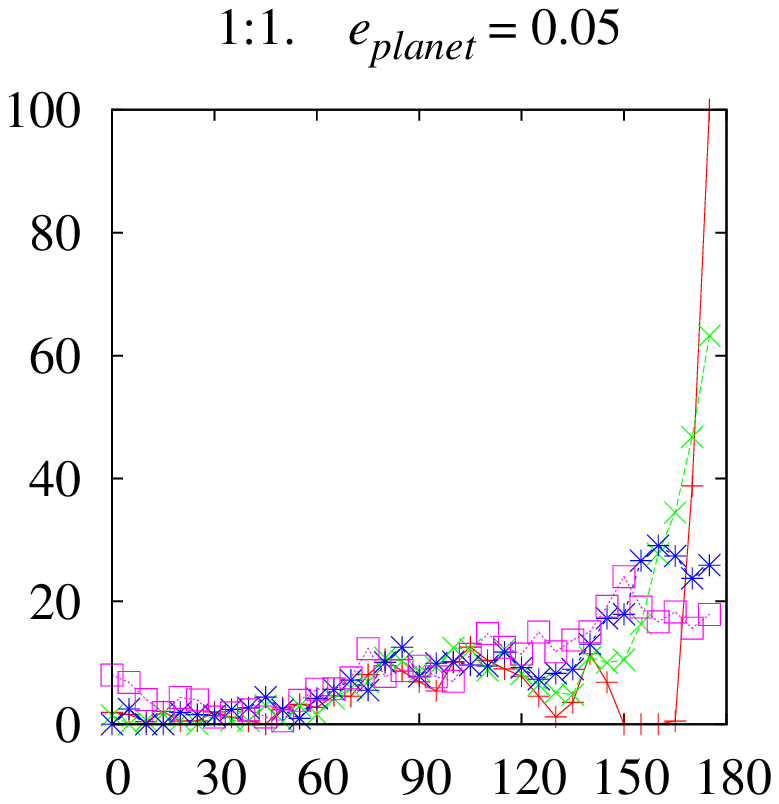}\hspace{-23mm}
\includegraphics[width=60mm]{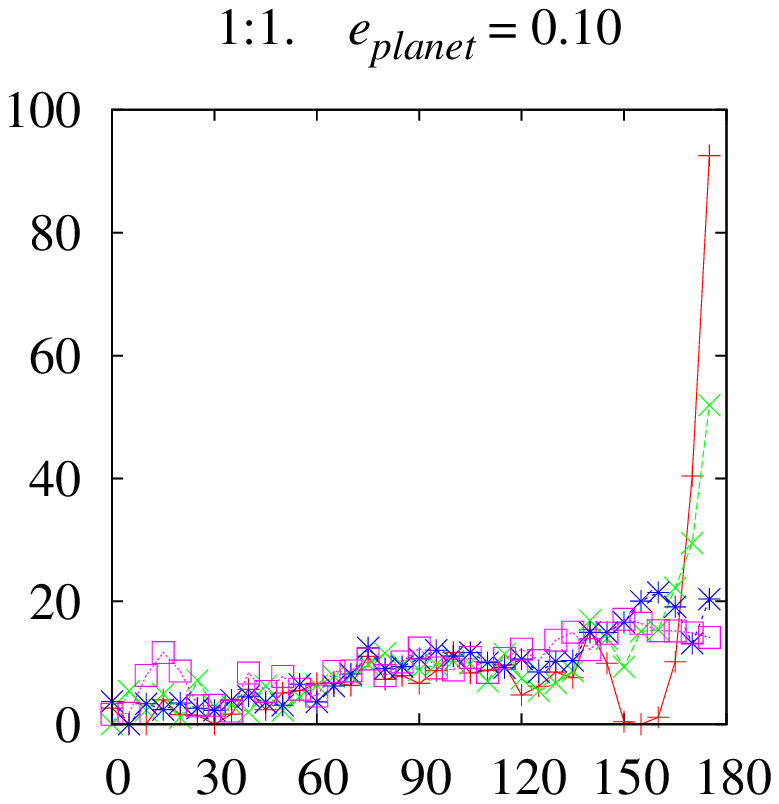}\hspace{-23mm}
\includegraphics[width=60mm]{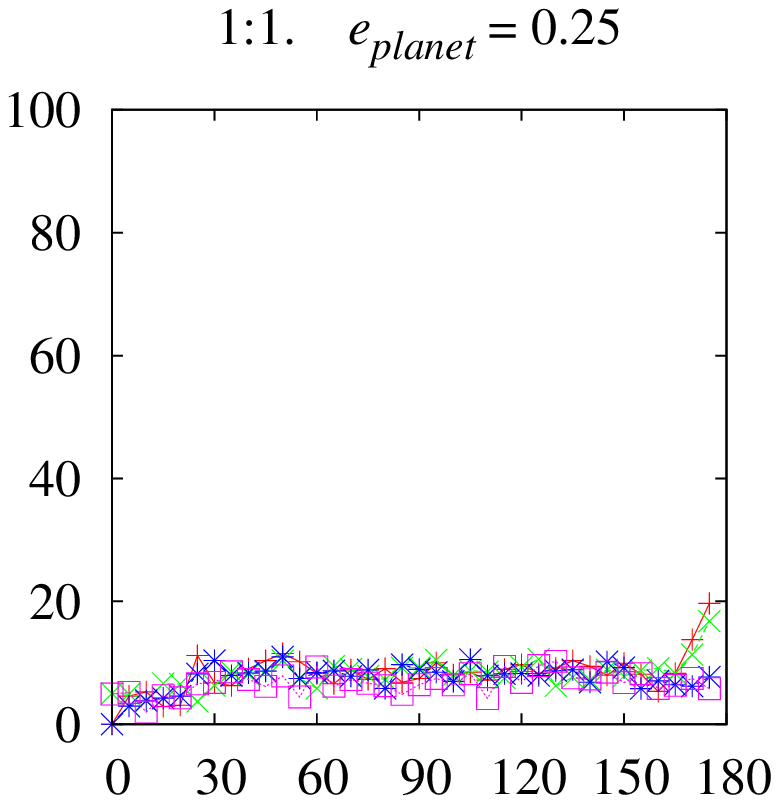}\\
\includegraphics[width=60mm]{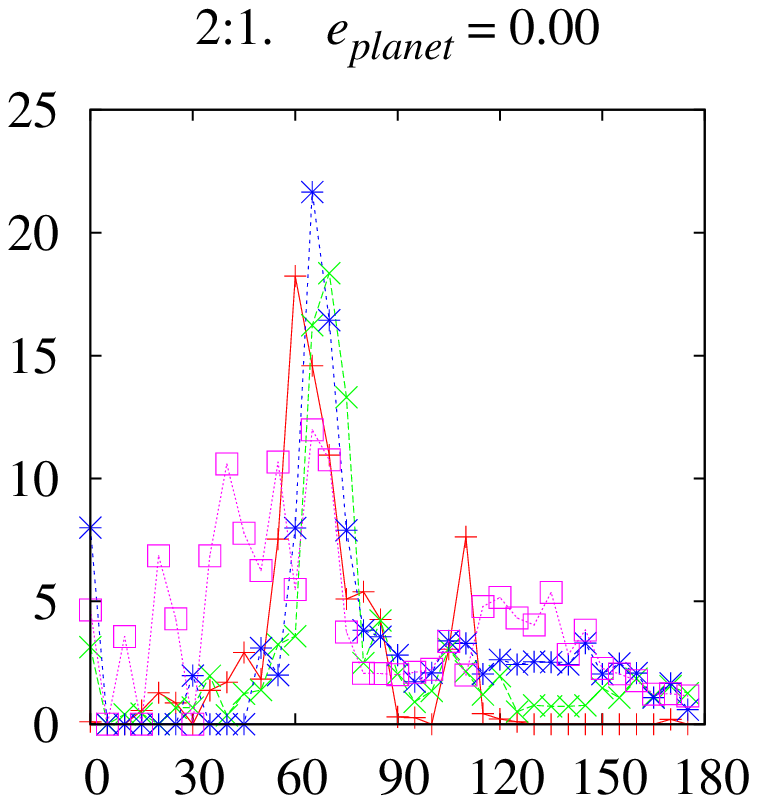}\hspace{-23mm}
\includegraphics[width=60mm]{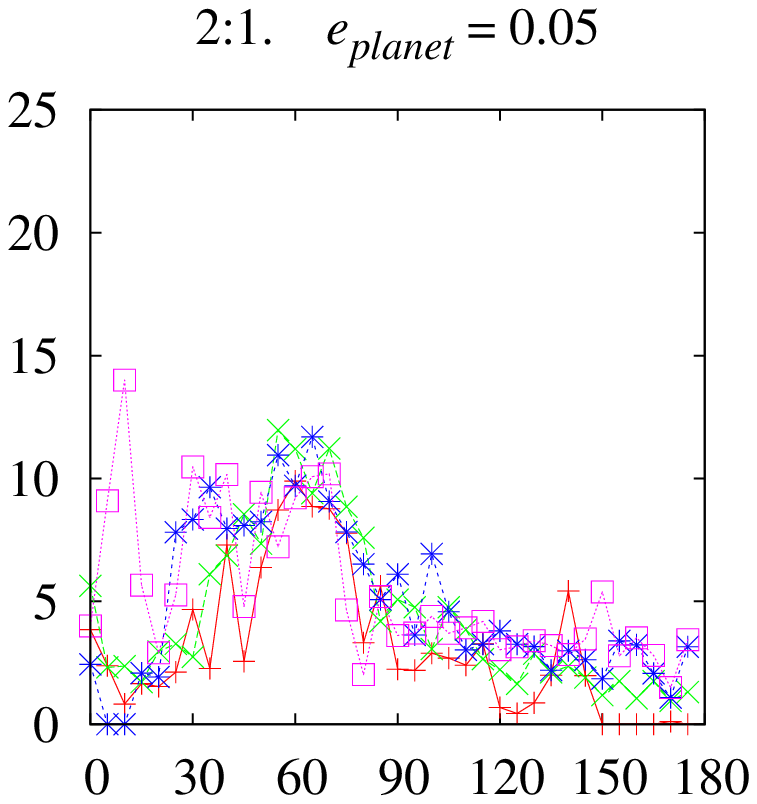}\hspace{-23mm}
\includegraphics[width=60mm]{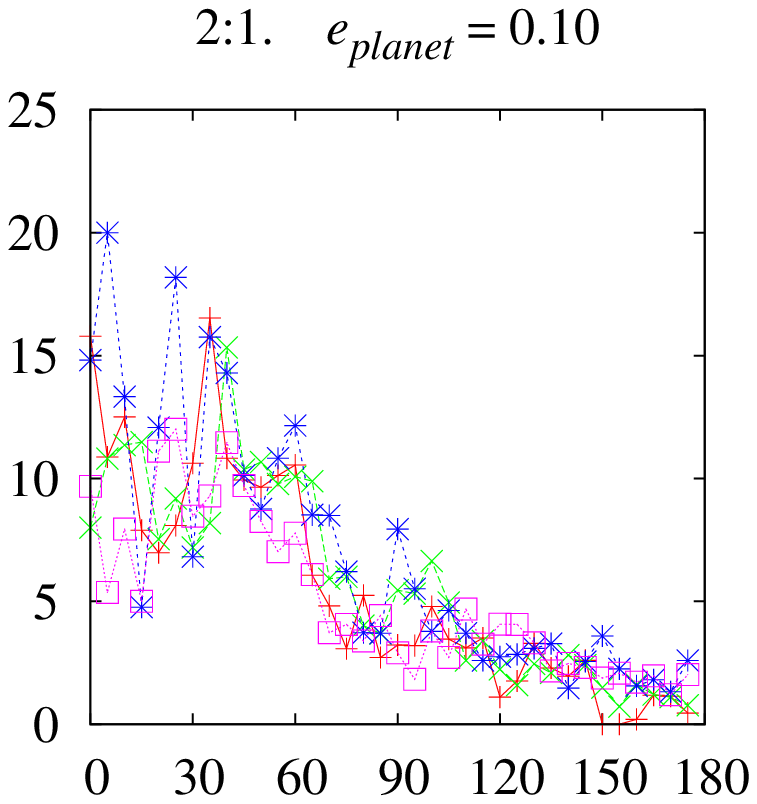}\hspace{-23mm}
\includegraphics[width=60mm]{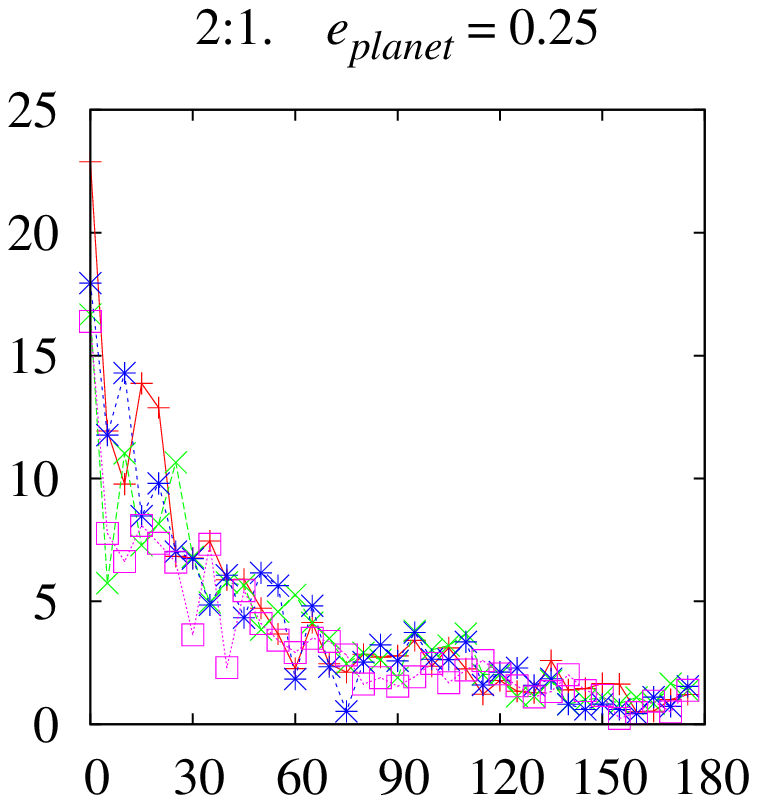}\\
\caption{Capture fraction for different planet eccentricities as functions of initial inclination for the 1:2, 1:1 and 2:1 resonances.
Four curves are plotted  corresponding to the four eccentricity standard deviations. Red line with cross sign for $\sigma_e=0.01$,  green line with times sign for $\sigma_e=0.10$, blue line with star sign for $\sigma_e=0.30$ and purple line with a box sign for $\sigma_e=0.70$.}
\end{center}
\end{figure*}

\begin{figure*}
\begin{center}
\includegraphics[width=60mm]{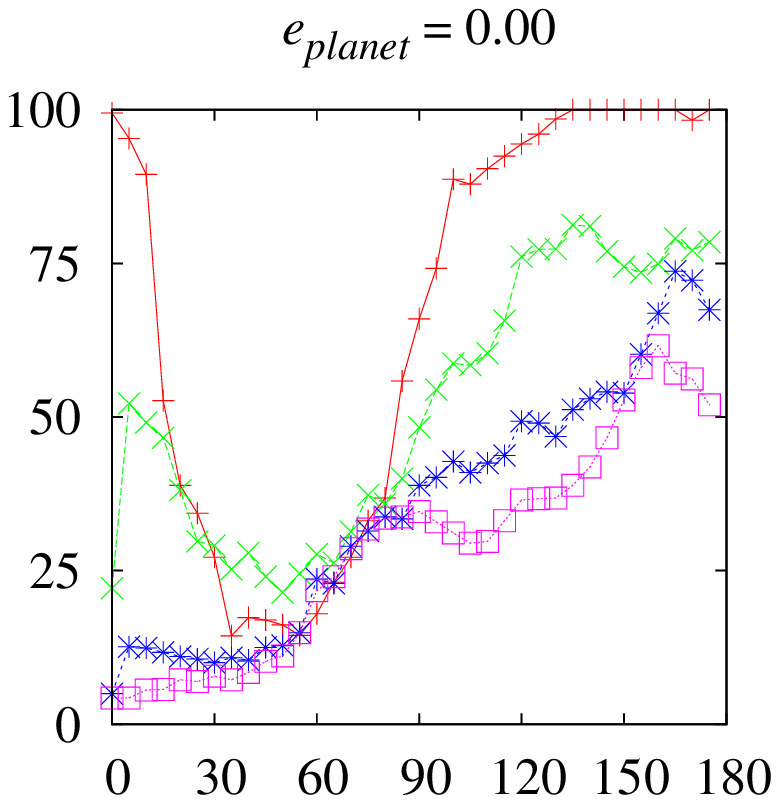}\hspace{-23mm}
\includegraphics[width=60mm]{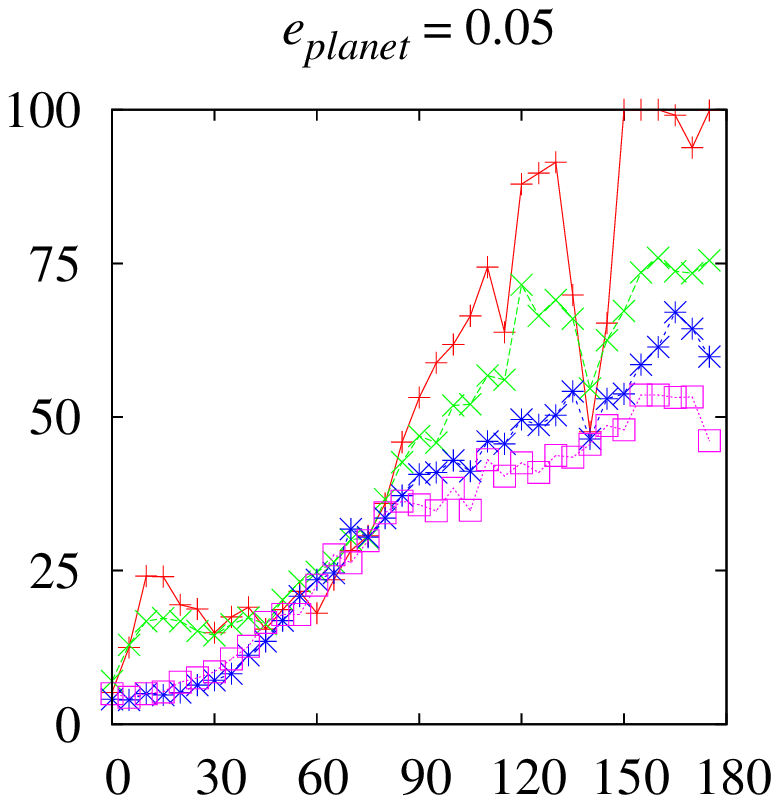}\hspace{-23mm}
\includegraphics[width=60mm]{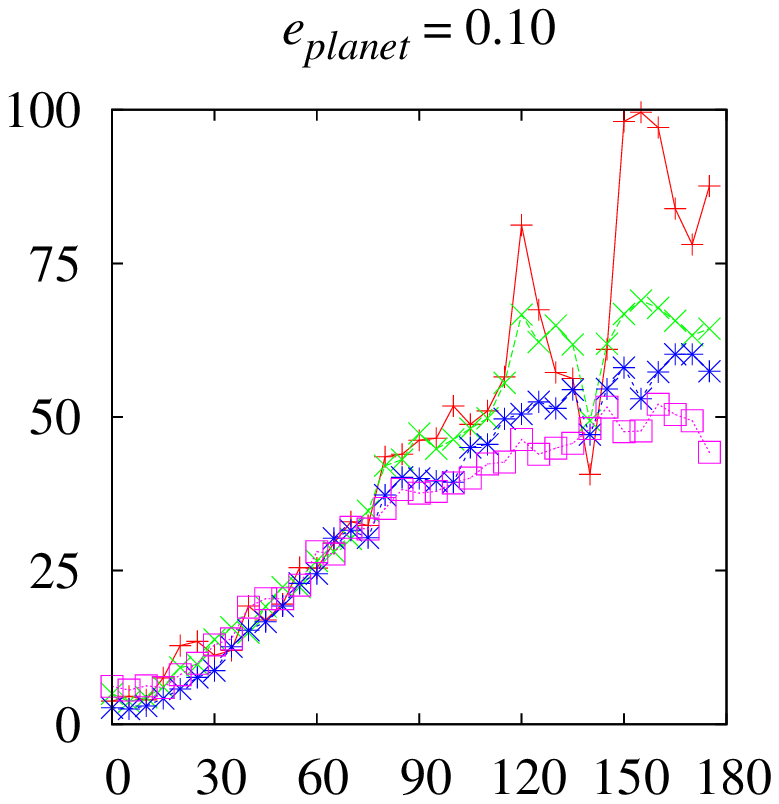}\hspace{-23mm}
\includegraphics[width=60mm]{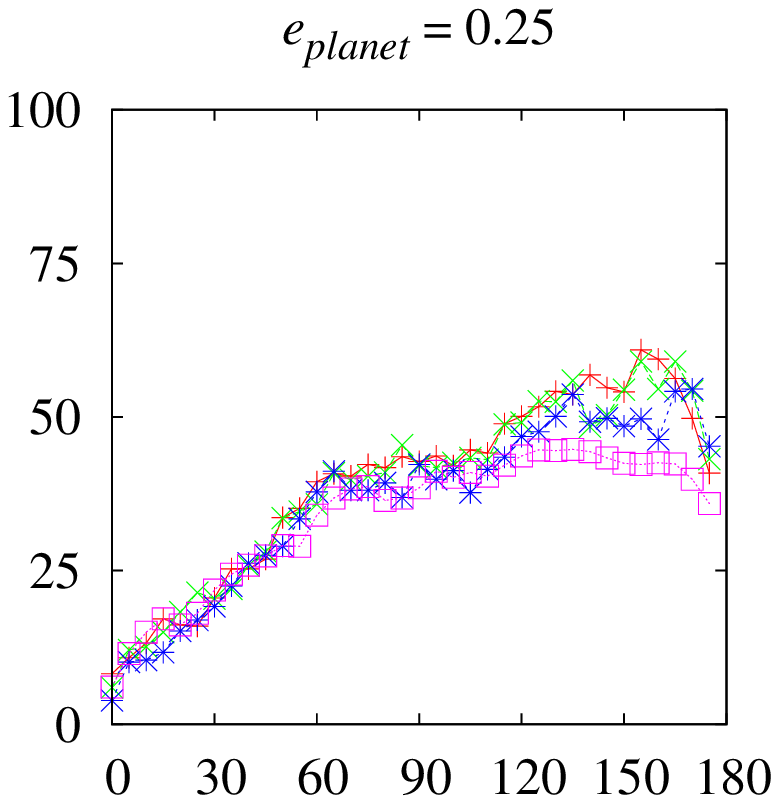}\caption{Total capture fraction for different planet eccentricities as functions of initial inclination.
Four curves are plotted  corresponding to the four eccentricity standard deviations. Red line with cross sign for $\sigma_e=0.01$,  green line with times sign for $\sigma_e=0.10$, blue line with star sign for $\sigma_e=0.30$ and purple line with a box sign for $\sigma_e=0.70$.}
\end{center}
\end{figure*}

\begin{figure*}
\begin{center}
\includegraphics[width=70mm]{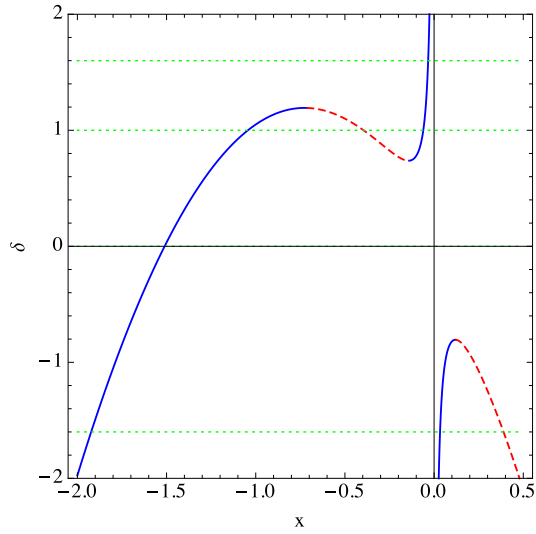}
\caption{Critical points of the Hamiltonian $H$ for a positive value of $T$ ($0.1$). The solid blue and dashed red  curves denote stable and unstable equilibria respectively. The green dotted curves denote the value of $\delta$ taken for the contour plots of Figure (12). For $T>0$ only the (CP1) equilibria are present. }
\end{center}
\end{figure*}

\begin{figure*}
\begin{center}
\includegraphics[width=70mm]{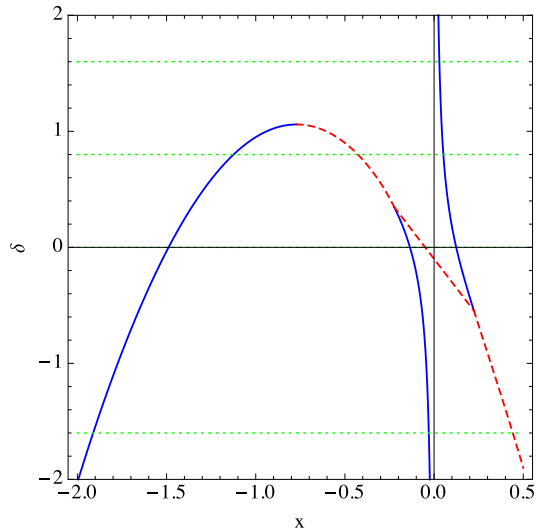}
\caption{Critical points of the Hamiltonian $H$ for a negative value of $T$ ($-0.1$). The solid blue and dashed red  curves denote stable and unstable equilibria respectively. The green dotted curves denote the value of $\delta$ taken for the contour plots of Figure (12). The line joining the (CP1) curves across $x=0$ corresponds to the (CP2) unstable equilibria.}
\end{center}
\end{figure*}

\begin{figure*}
\begin{center}
\includegraphics[width=40mm]{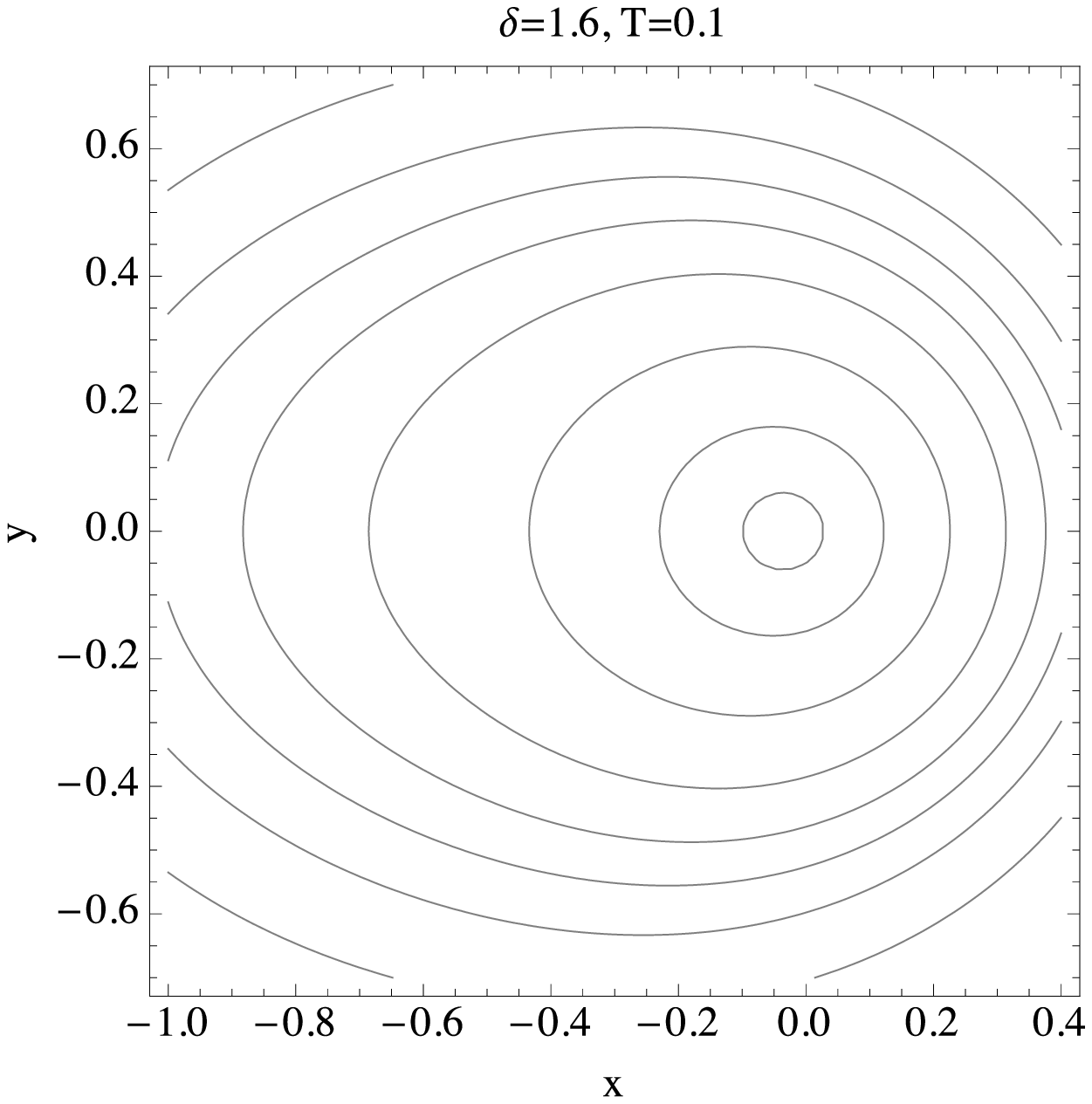}
\includegraphics[width=40mm]{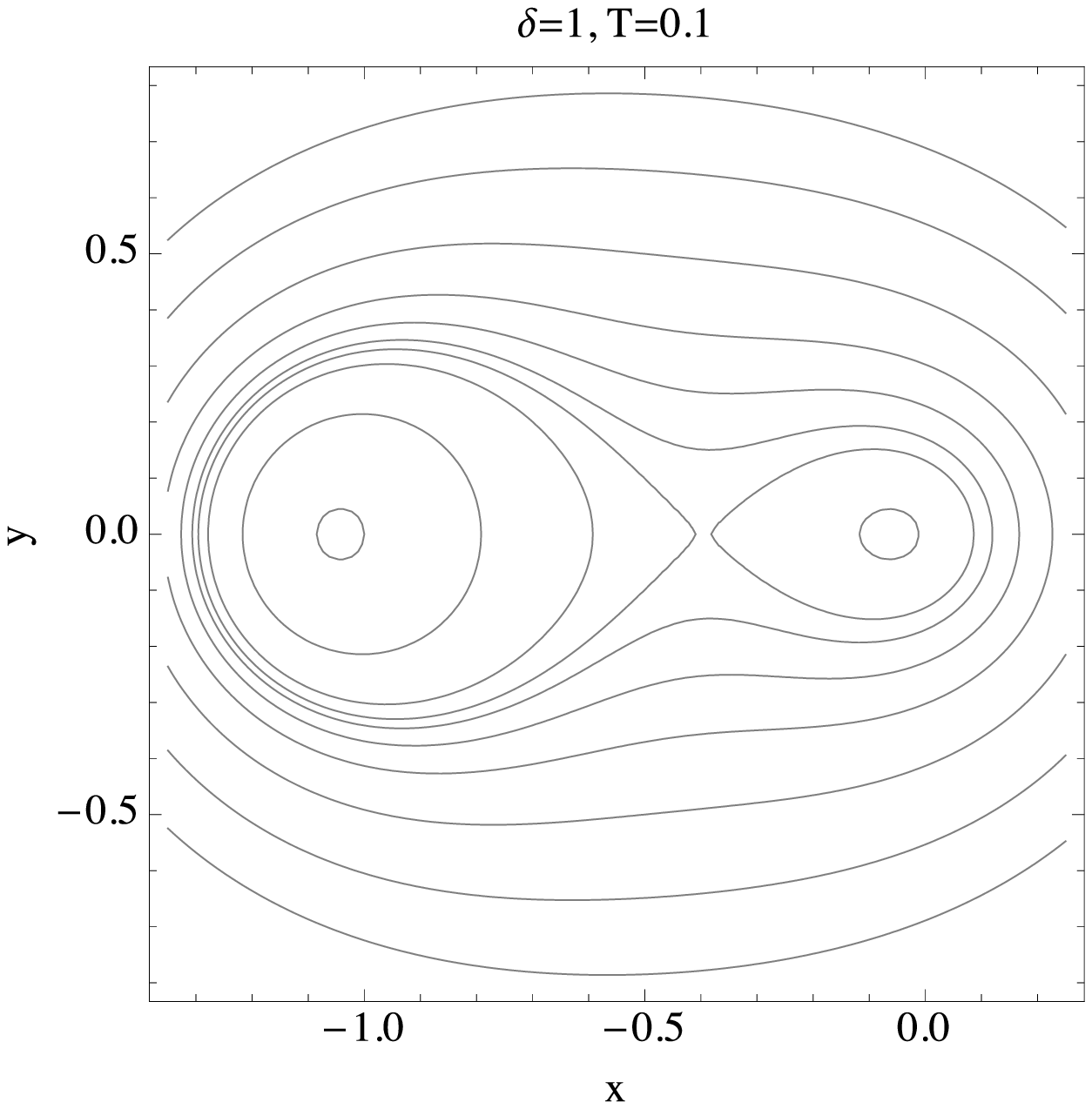}
\includegraphics[width=40mm]{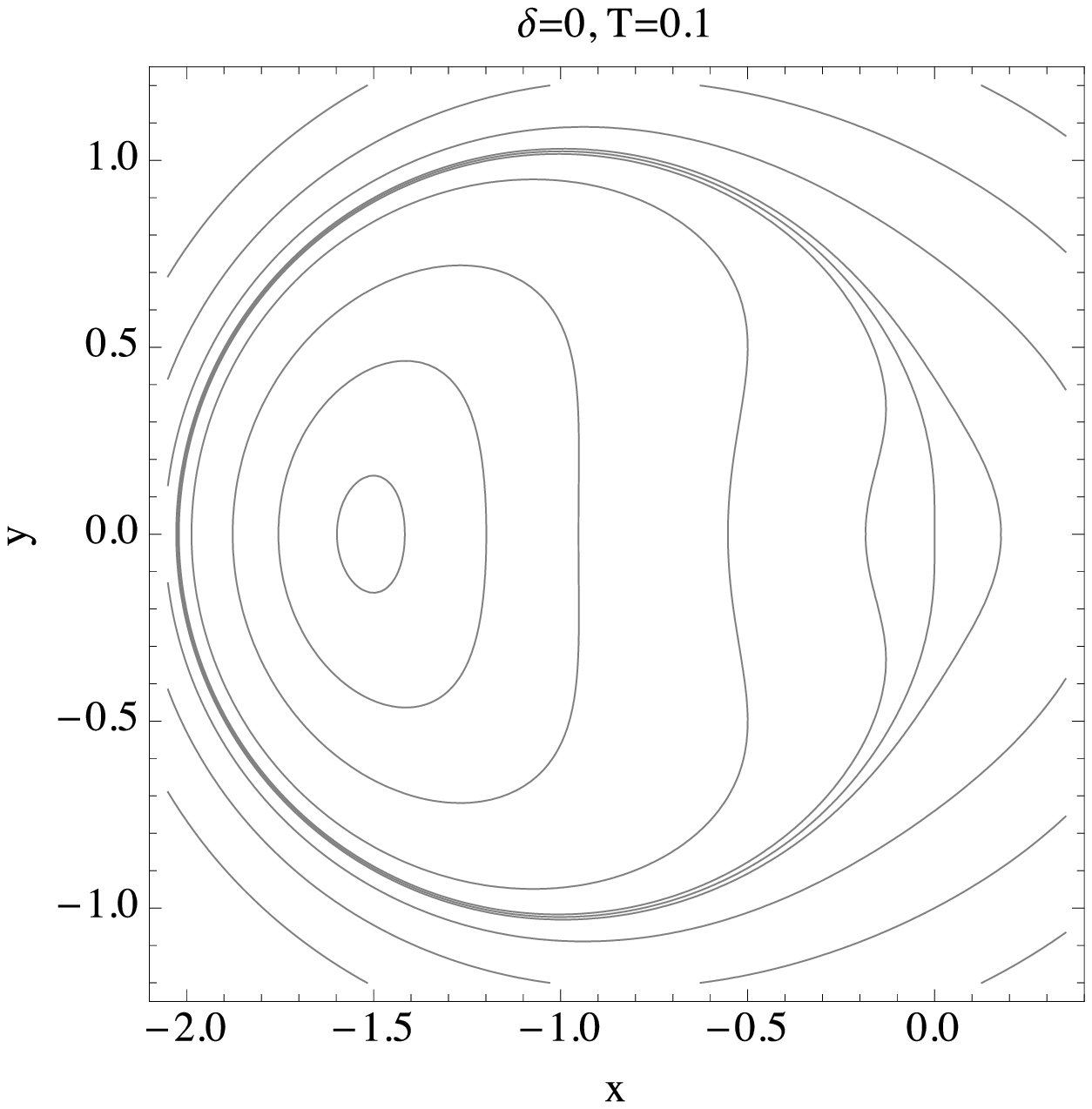}
\includegraphics[width=40mm]{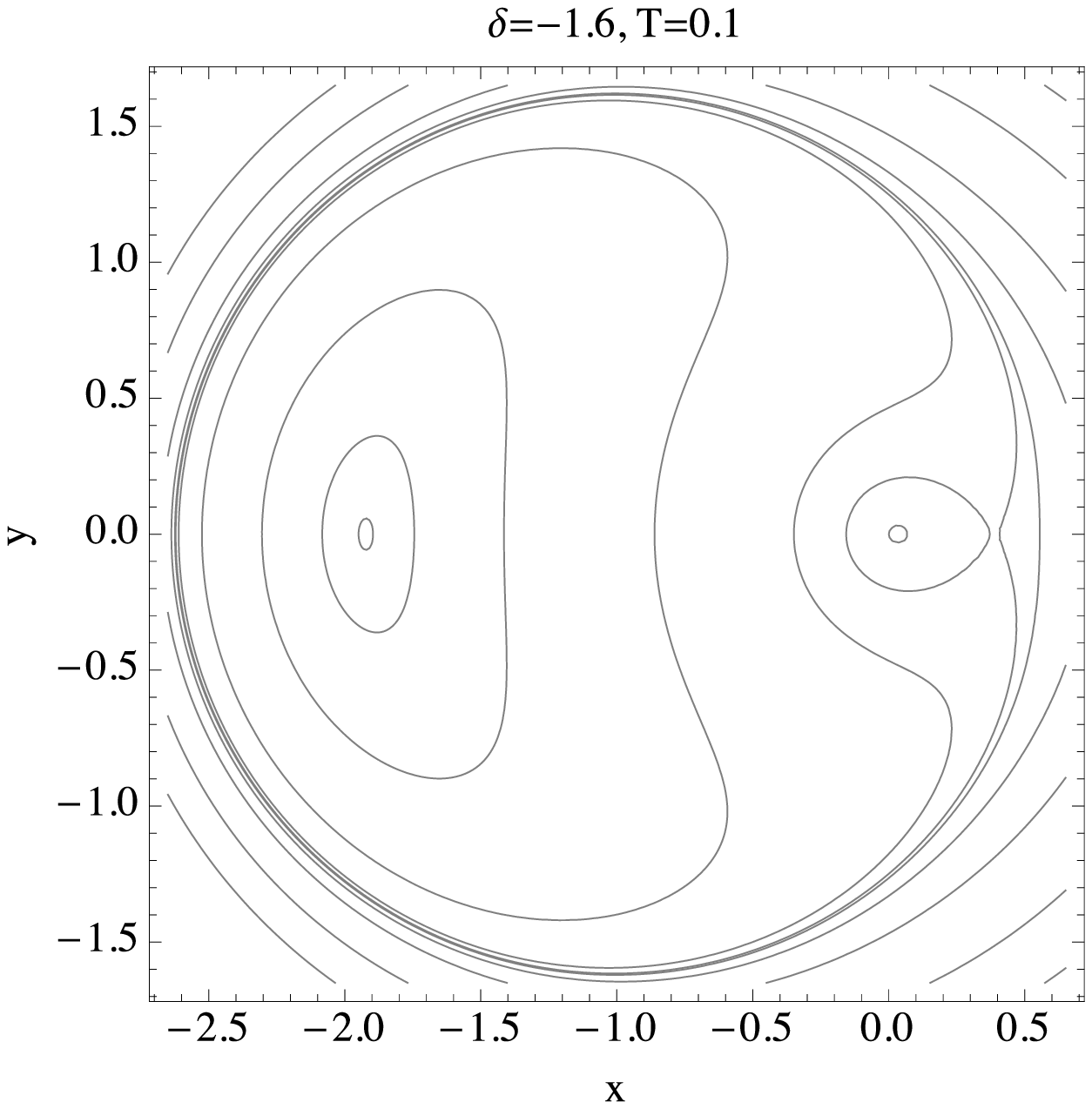}\\
\includegraphics[width=40mm]{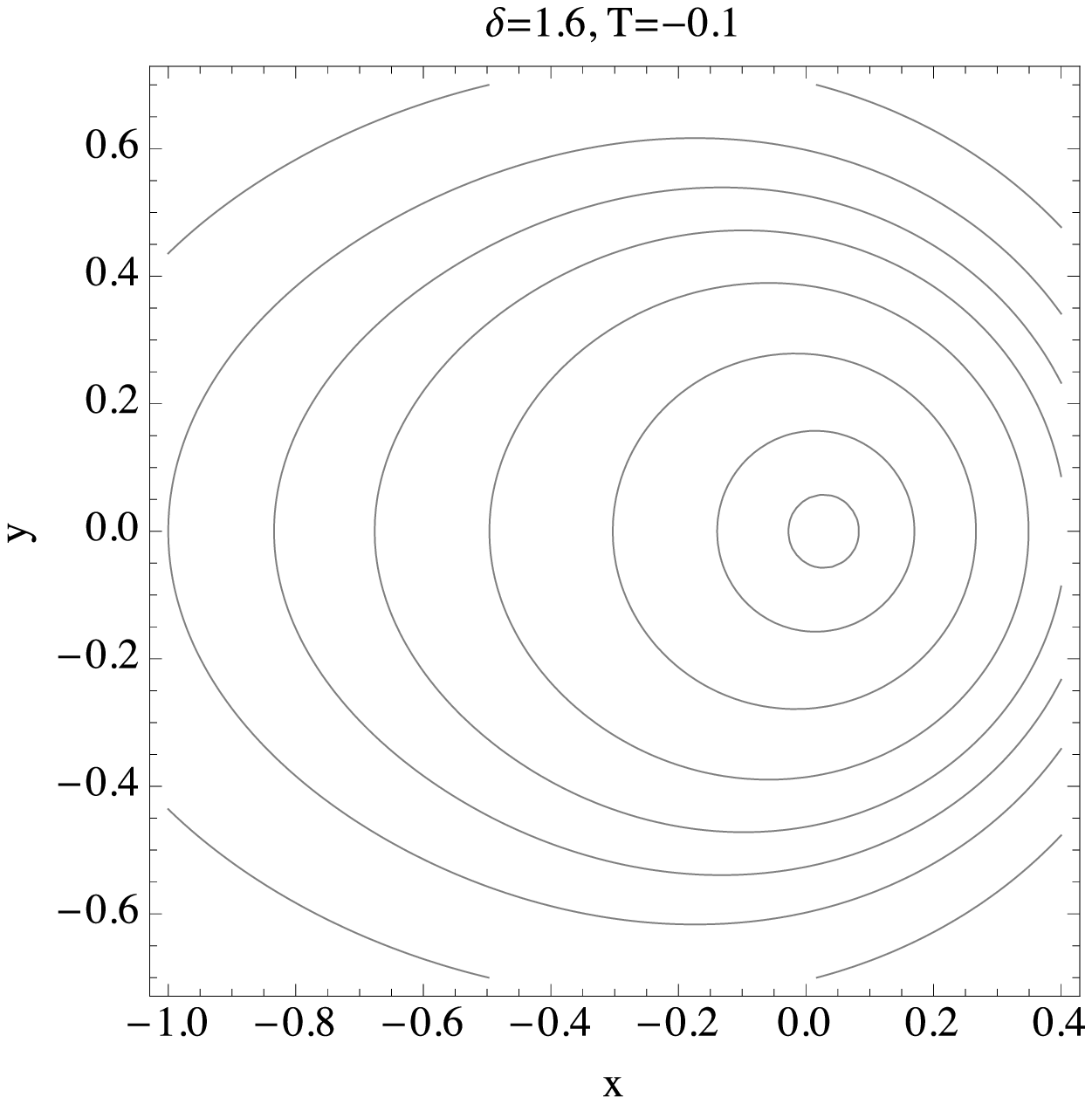}
\includegraphics[width=40mm]{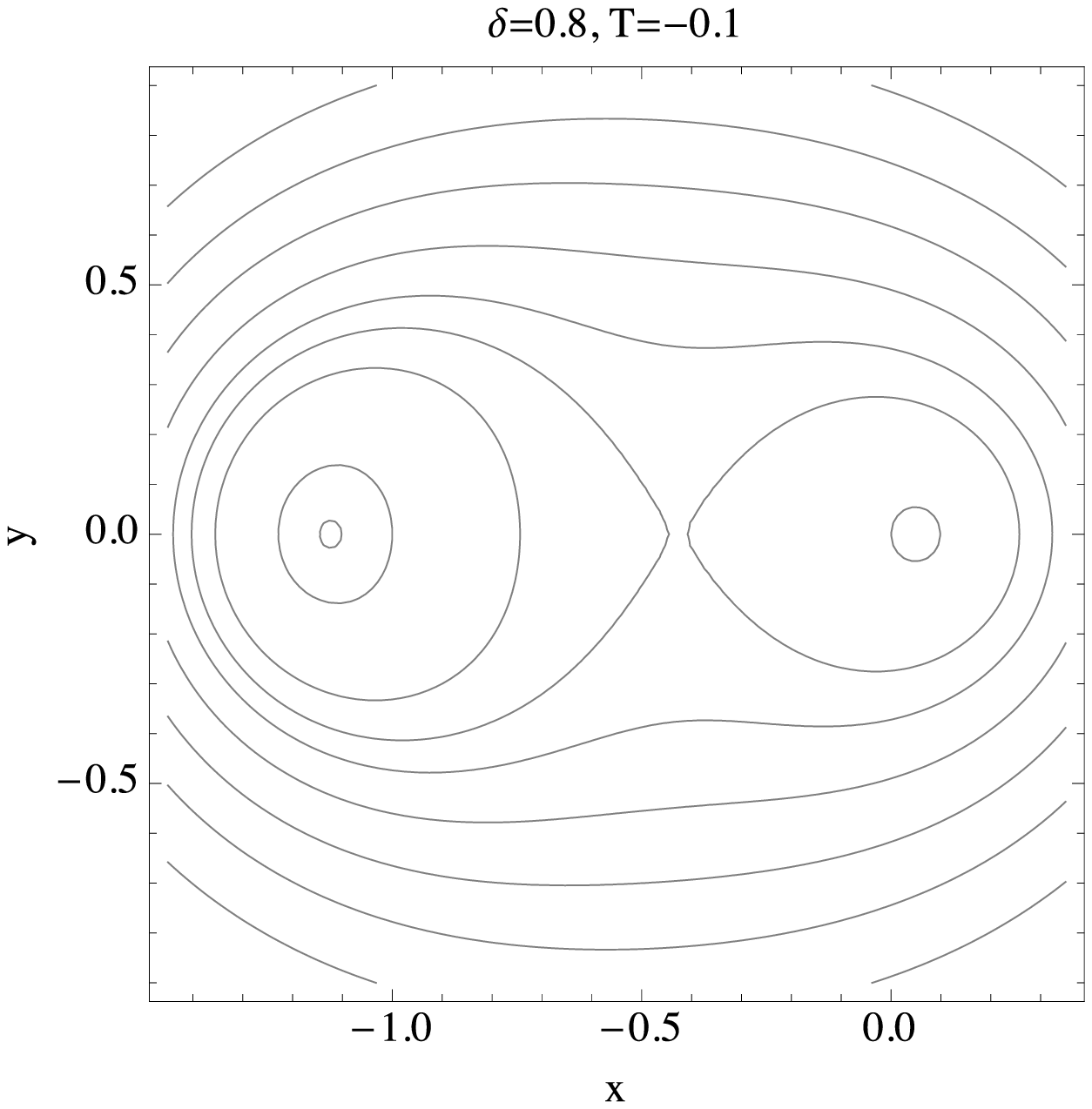}
\includegraphics[width=40mm]{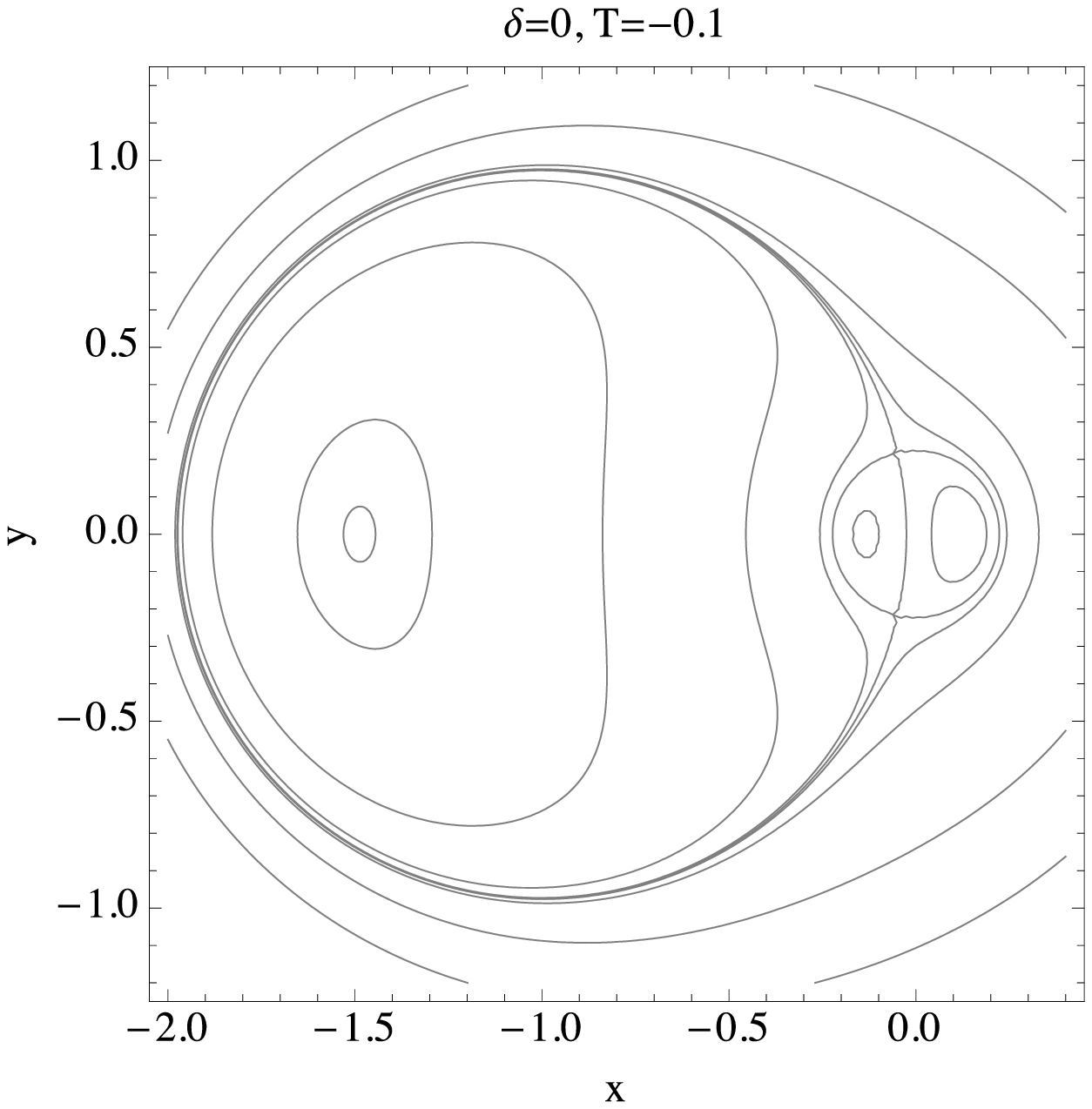}
\includegraphics[width=40mm]{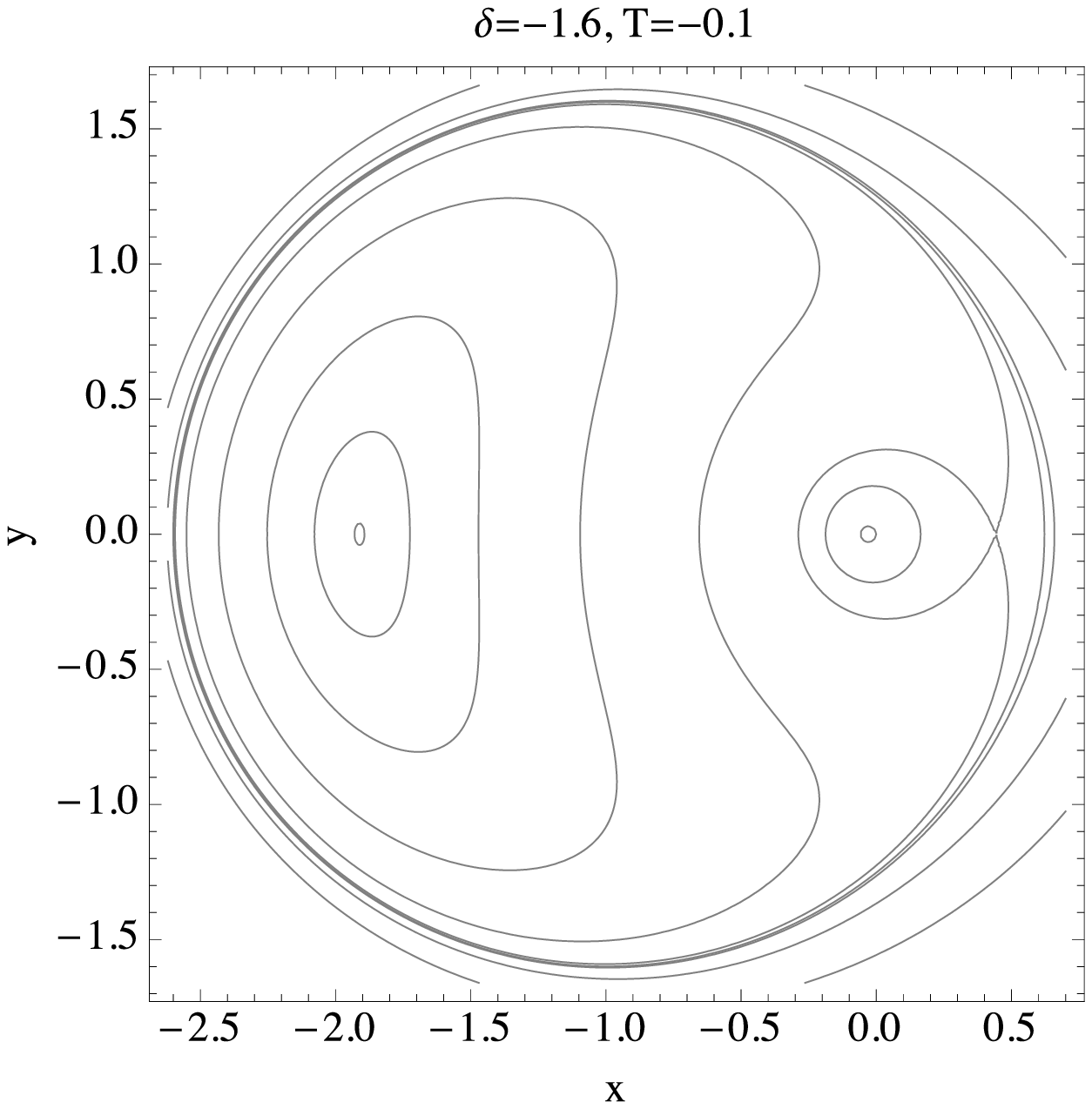}\caption{Coutour plots of the Hamiltonian $H$ (\ref{ham}) in the $xy$-plane for $T=0.1$ and $-0.1$.}
\end{center}
\end{figure*}

\newpage

\bibliographystyle{mn2e}

\bibliography{retrograde}

\end{document}